\pdfoutput=1
\documentclass[11pt,twoside,a4paper,cmspaper,final,collab]{cms-tdr}
\def\svnVersion{0552811}\def\svnDate{2024/05/20}\def\cmsCernNoTag{CERN-EP-2023-282}\def\cmsCernDate{\today}\def\cmsMessage{Published in the Journal of High Energy Physics as \href{http://dx.doi.org/10.1007/JHEP05(2024)116}{\doi{10.1007/JHEP05(2024)116}.}}
\begin{document}\cmsNoteHeader{SMP-22-007}

\newcommand{\lnDeltaR}{\ensuremath{\ln(R/\Delta R)}\xspace}
\newcommand{\lnkT}{\ensuremath{\ln(\kt/\GeV)}\xspace}
\newcommand{\ktFSRcutoff}{\ensuremath{\kt^\text{FSR\,cutoff}}\xspace}
\newcommand{\alpSFSR}{\ensuremath{\alpS^\mathrm{FSR}(m_\PZ)}\xspace}
\newcommand{\nJets}{\ensuremath{N_\text{jets}}\xspace}
\newcommand{\nEmissions}{\ensuremath{N_\text{emissions}}\xspace}
\newcommand{\DIRE} {{\textsc{dire}}\xspace}
\newcommand{\HeHEPData} {{\textsc{HEPData}}\xspace}
\newcommand{\VINCIA} {{\textsc{vincia}}\xspace}
\newcommand{\DeltaR}{\ensuremath{\Delta R}\xspace}
\newcommand{\CA}{\ensuremath{C_{\mathrm{A}}}\xspace}
\newcommand{\CF}{\ensuremath{C_{\mathrm{F}}}\xspace}
\newcommand{\CR}{\ensuremath{C_{\mathrm{R}}}\xspace}
\newcommand{\CReff}{\ensuremath{C_{\mathrm{R}}^{\text{eff}}}\xspace}

\cmsNoteHeader{SMP-22-007}
\title{Measurement of the primary Lund jet plane density in proton-proton collisions at \texorpdfstring{$\sqrt{s} = 13\TeV$}{sqrt(s) = 13 TeV}}

\date{\today}

\abstract{
A measurement is presented of the primary Lund jet plane (LJP) density in inclusive jet production in proton-proton collisions. The analysis uses 138\fbinv of data collected by the CMS experiment at $\sqrt{s} = 13\TeV$. The LJP, a representation of the phase space of emissions inside jets, is constructed using iterative jet declustering. The transverse momentum \kt and the splitting angle \DeltaR of an emission relative to its emitter are measured at each step of the jet declustering process. The average density of emissions as function of \lnkT and \lnDeltaR is measured for jets with distance parameters $R = 0.4$ or 0.8, transverse momentum $\pt >  700\GeV$, and rapidity $\abs{y} < 1.7$. The jet substructure is measured using the charged-particle tracks of the jet. The measured distributions, unfolded to the level of stable charged particles, are compared with theoretical predictions from simulations and with perturbative quantum chromodynamics calculations. Due to the ability of the LJP to factorize physical effects, these measurements can be used to improve different aspects of the physics modeling in event generators.
}
\hypersetup{
pdfauthor={CMS Collaboration},
pdftitle={CMS primary Lund jet plane density at 13 TeV},
pdfsubject={CMS},
pdfkeywords={CMS, measurement, jet substructure, quarks, gluons, jets, Lund jet plane}}

\maketitle 

\section{Introduction}\label{sec:introduction}

The production of jets, collimated sprays of particles that result from the fragmentation of energetic quarks and gluons, is one of the primary processes studied to deepen our understanding of the strong force, described by quantum chromodynamics (QCD). In recent years, it has become clear that valuable information about the formation of jets is contained in their substructure~\cite{Larkoski:2017jix,Kogler:2018hem, Marzani:2019hun, Kogler:2021kkw}. The phenomenological description of the substructure of jets requires an understanding of the cascade of partons produced by highly energetic quarks and gluons, a self-similar branching process referred to as parton showering, and the strongly coupled transition of partons into hadrons, known as hadronization. In a given hadronic collision, secondary parton-parton scatterings and beam-beam remnant interactions, collectively known as the underlying-event (UE) activity, also contribute to the substructure of jets. The separation of these effects in the jet substructure is challenging, since jet substructure observables are necessarily constructed using only the final-state hadrons, whose exact origin cannot be traced back easily. The modeling of jet formation in Monte Carlo (MC) event generators has a direct impact on the theoretical and experimental uncertainties associated with precision measurements of fundamental parameters of the standard model, such as the strong coupling \alpS and the mass of elementary particles, like the top quark. Thus, dedicated substructure measurements are beneficial, preferably with observables where the various physical effects contributing to the jet substructure can be separated in a transparent way.

\begin{figure}
  \centering
  \includegraphics[width = 0.55\textwidth]{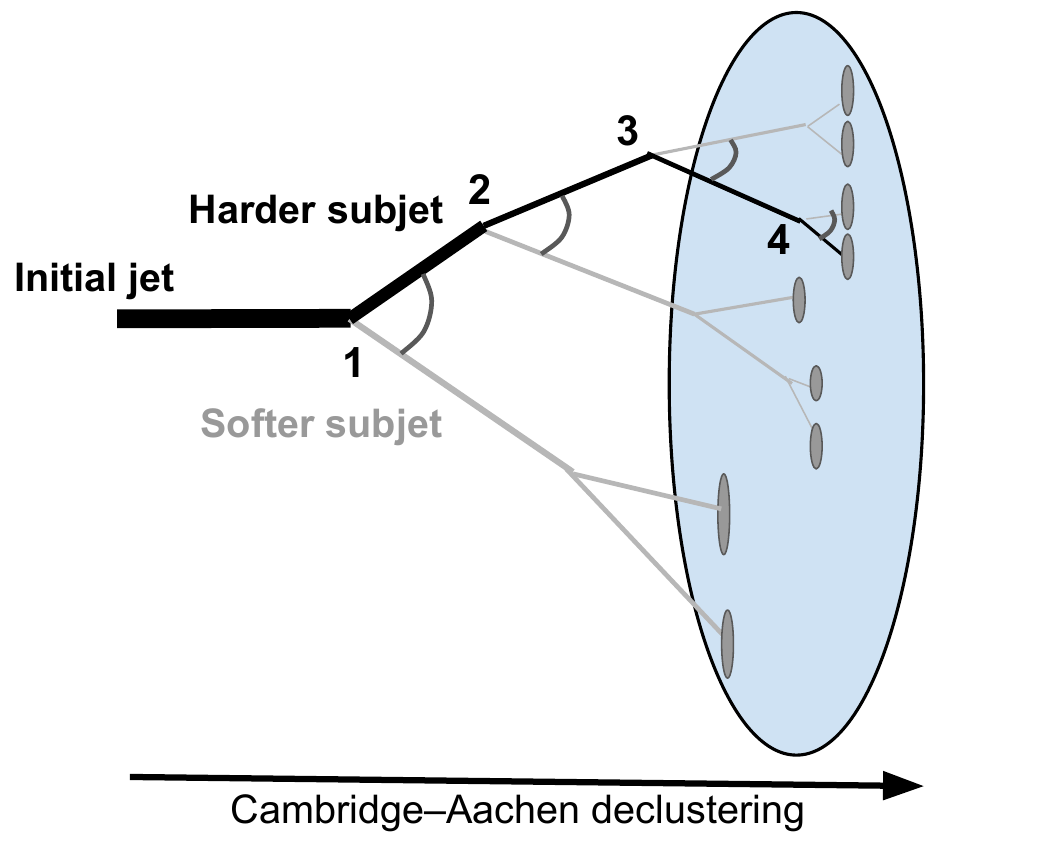}
  \hfill
  \includegraphics[width = 0.44\textwidth]{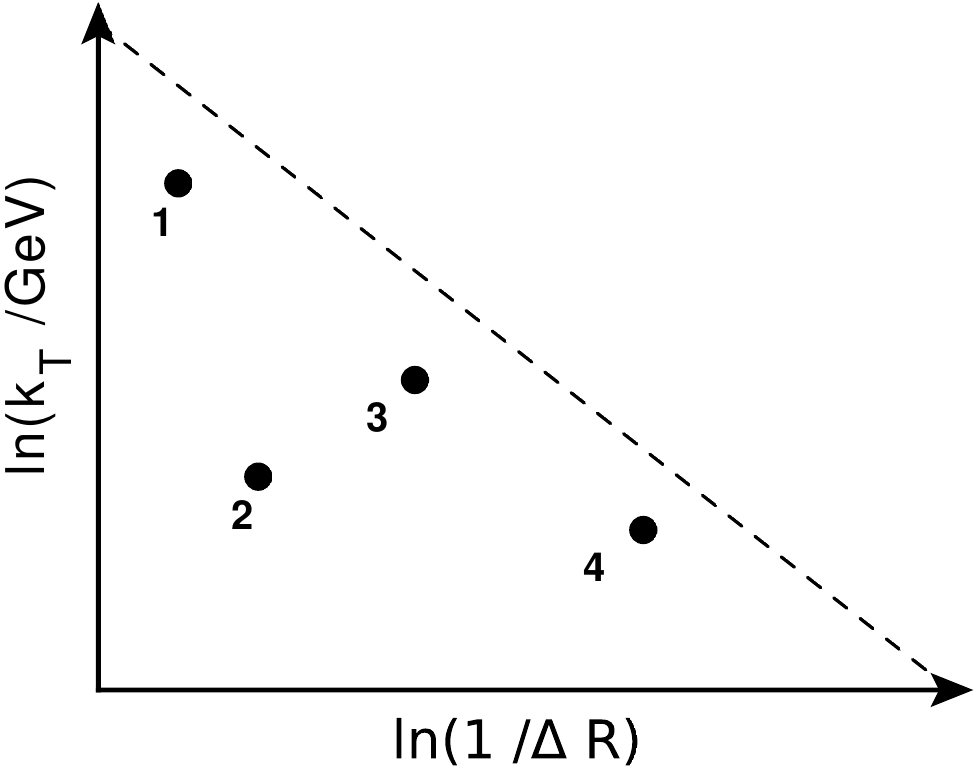}
  \caption{Left: schematic diagram of the Cambridge--Aachen primary declustering tree of a jet. The black lines represent the branch that follows the harder subjet at each step of the declustering tree. The softer subjet at each node is used as a proxy for an emission in the primary LJP. Right: schematic diagram of the primary emissions of a jet in the LJP, which is filled from left to right corresponding to emissions ordered from large to small angles. The numbers represent the order of appearance in the declustering tree. The dashed diagonal line represents the kinematical limit.}
  \label{fig:lp_cartoon}
\end{figure}

One way of representing radiation in QCD is via the Lund plane~\cite{Andersson:1988gp}, which is a two-dimensional representation of the phase space of $1\to 2$ partonic splittings (Fig.~\ref{fig:lp_cartoon}). The kinematics of a splitting can be described with two degrees of freedom: the splitting angle of the branching \DeltaR, and the transverse momentum \kt of the emission relative to its emitter. The logarithm of \kt and the logarithm of $1/\DeltaR$ are used for the vertical and horizontal axes of the Lund plane, respectively. This choice is based on the scaling given by the soft and collinear divergences of QCD, which is such that partons are emitted uniformly as a function of \lnkT and $\ln(1/\DeltaR)$. Historically, the Lund plane has been used for the development of parton showering algorithms and for calculations with resummation at all orders in \alpS.

References~\cite{Andrews:2018jcm, Dreyer:2018nbf} proposed to use iterative jet declustering using the Cambridge--Aachen (CA) algorithm \cite{Dokshitzer:1997in, Wobisch:1998wt} to represent the internal structure of the jet in the Lund plane. The CA algorithm clusters the pair of particles (and pairs of subjets thereafter) with the smallest separation in rapidity and azimuthal angle first, and the pair of subjets with the largest separation last. The strict angular ordering of the CA clustering algorithm resembles the angular ordering of $1\to 2$ splittings in QCD due to color coherence effects~\cite{Dokshitzer:1997in, Wobisch:1998wt} and at the same time it favors the collinear divergences of QCD. The different branches of the jet clustering tree can be used as proxies for emissions whose kinematics are mapped onto the Lund plane. This representation of the phase space of emissions is known as the Lund jet plane (LJP).

The constituents of a jet are reclustered using the CA algorithm, following the prescription in Ref.~\cite{Dreyer:2018nbf}. Then, the CA pairwise clustering history is followed in reverse, as shown schematically in Fig.~\ref{fig:lp_cartoon}. Starting with the reclustered jet ${j}$, the last step of the CA clustering is undone, such that the original jet is declustered into the two subjets of the previous clustering step $j \to j_1+j_2$, where $j_1$ and $j_2$ are the harder and softer subjets in the branching, respectively. The notation of $j$, $j_1$, and $j_2$ is used for subjets in the declustering process. The properties of the branching are recorded,
\begin{equation}
  \DeltaR = \sqrt{\smash[b]{(y^{j_1}-y^{j_2})^2+(\phi^{j_1}-\phi^{j_2})^2}}, \hspace{1cm} \kt = \pt^{j_2} \DeltaR
\end{equation}
where $y^{j_{1,2}}$ and $\phi^{j_{1,2}}$ represent the rapidity and azimuthal angle of the harder and softer subjets and $\pt^{j_{2}}$ the transverse momentum of the softer subjet relative to the beam axis. The radiation pattern of the jet can be explored in numerous ways depending on which branch is being followed in the CA declustering tree. The focus of this paper is specifically the primary LJP~\cite{Dreyer:2018nbf}, which corresponds to the LJP of emissions obtained by declustering the harder subjet at each step of the declustering process, as shown schematically in Fig.~\ref{fig:lp_cartoon} (left). The harder subjet is the one with the highest \pt at a given step of the CA declustering. The softer subjet is treated as an emission in the LJP. A given jet is then represented as a sequence of angular-ordered points in the LJP, as illustrated in Fig.~\ref{fig:lp_cartoon} (right). Due to the strict angular ordering of the CA clustering algorithm, the emissions in the primary LJP are always filled from large to small \DeltaR, from left to right on the LJP. The jet is declustered iteratively until the hardest subjet is composed of a single particle. The primary LJP contains a significant amount of information about the radiation pattern of the jet while allowing for a transparent physical interpretation.

The observable measured in this paper is the average density of emissions in the primary LJP~\cite{Lifson:2020gua},
\begin{equation}\label{eq:lundplanedensity}
 \rho(\kt, \DeltaR) \equiv \frac{1}{\nJets} \frac{\rd^2 \nEmissions}{\rd \lnkT \rd \lnDeltaR}, 
\end{equation}
where \nJets represents the total number of jets in a given fiducial region, \nEmissions the total number of emissions of such jets, and $R$ is the jet distance parameter under consideration. The per-jet normalization renders the observable insensitive to the inclusive jet cross section. The density is expressed double-differentially in \lnkT and \lnDeltaR, since this is the approximate momentum and angular scaling behavior of QCD radiation. Indeed, for independent emissions in the soft and collinear limit of perturbative QCD (pQCD), the LJP density is directly proportional to \alpS,
\begin{equation}\label{eq:lundplanedensityLO}
   \rho(\kt, \DeltaR) \approx \frac{2}{\pi} \CR \alpS (\kt),
\end{equation}
where \CR is the appropriate color factor of the emission, for instance $\CF = 4/3$ for gluon emissions from quarks and $\CA=3$ for gluon emissions from gluons. Equation~\ref{eq:lundplanedensityLO} does not account for parton flavor changes in the declustering history. The hard scale used in the evolution of \alpS is given by the \kt of an emission ~\cite{Dreyer:2018nbf}. The fact that the LJP density scales with $\alpS(\kt)$ means that the emission density is expected to be approximately uniform for large \kt values and to grow rapidly at small \kt following the running of $\alpS \propto 1/\ln(\kt/\Lambda_\mathrm{QCD})$, where $\Lambda_\mathrm{QCD}$ is the energy scale where the theory becomes strongly coupled. For \kt values of about 1\GeV, there is a transition towards the nonperturbative regime, dominated by hadronization effects.

\begin{figure}
  \centering
  \includegraphics[width = 0.8\textwidth]{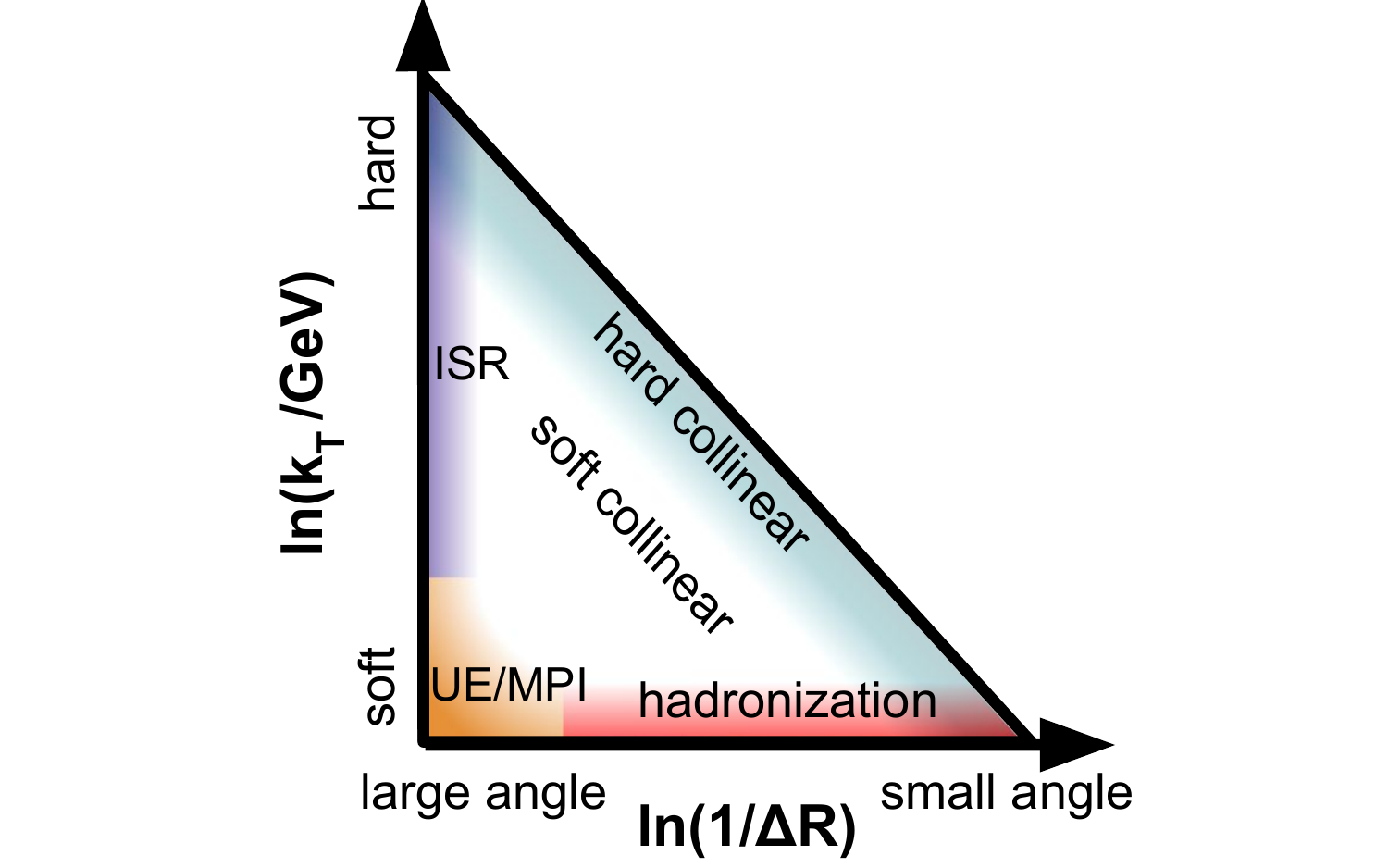}
  \caption{Schematic diagram of the mechanisms affecting different regions of the primary LJP in a given proton-proton collision. Initial-state radiation (ISR), the underlying event (UE) activity, and multiple-parton interactions (MPI) affect wide-angle radiation at $\DeltaR \sim R$, close to the boundary of the jet. In an experimental context, pileup contributes to the same region as the UE. Hadronization affects the low \lnkT region (below $\kt \sim 1\GeV$) at all angles. Soft and hard collinear parton splittings affect the rest of the LJP. The diagonal line represents the kinematical limit of the primary LJP, which corresponds to $\pt^{j_1} = \pt^{j_2}$.}
  \label{fig:lp_factorization}
\end{figure}

The primary LJP density provides detailed information about the radiation pattern of the jet, which branches out to numerous applications in high-energy physics. For instance, measurements of the primary LJP density can be used to improve the parton shower, hadronization, and UE activity modeling, since their effects approximately factorize in the LJP~\cite{Dreyer:2018nbf}, as illustrated in Fig.~\ref{fig:lp_factorization}. Precision measurements of the primary LJP density can be used to benchmark the next generation of general purpose parton showers with resummation at next-to-leading-logarithmic (NLL) accuracy \cite{Dasgupta:2018nvj, Dasgupta:2020fwr, Hamilton:2020rcu, Forshaw:2020wrq, Nagy:2020dvz, Herren:2022jej, vanBeekveld:2022zhl, vanBeekveld:2022ukn}. The LJP has been used to obtain the first direct evidence of the dead-cone effect in heavy-flavor jets \cite{ALICE:2021aqk}. Highly boosted color-singlet particles have unique signatures in the LJP, which can be exploited for jet flavor tagging~\cite{Dreyer:2018nbf, Dreyer:2020brq, Cavallini:2021vot}. The LJP can provide an effective space-time picture of the quark-gluon plasma created in ultrarelativistic heavy-ion collisions~\cite{Andrews:2018jcm}.

The primary LJP density can be calculated analytically in the framework of perturbation theory~\cite{Lifson:2020gua}. The most recent calculations include corrections at next-to-leading order (NLO) in \alpS for the fixed-order matrix element matched to an NLL resummation to all-orders in \alpS~\cite{Lifson:2020gua}. Substructure observables obtained with grooming techniques~\cite{Larkoski:2017jix,Kogler:2018hem, Marzani:2019hun, Kogler:2021kkw}, such as the groomed jet radius or groomed momentum fraction obtained with the soft-drop grooming algorithm~\cite{Dasgupta:2013ihk, Larkoski:2014wba}, effectively select a subset of emissions of the primary LJP. Measurements of groomed jet observables based on jet clustering have been reported in Refs.~\cite{Sirunyan:2017bsd, Sirunyan:2018xdh, Sirunyan:2018asm, ALICE:2017nij, ATLAS:2019kwg, ATLAS:2019mgf, ALICE:2019ykw, ALICE:2021obz, ALICE:2021njq, CMS:2021iwu, ATLAS:2022vii, CMS:2022kqg, ATLAS:2023hso}.

The primary LJP has been measured by the ATLAS Collaboration \cite{ATLAS:2020bbn} using jets initially clustered with the anti-\kt algorithm \cite{Cacciari:2008gp, Cacciari:2011ma} with a distance parameter $R = 0.4$. The two highest \pt jets in the event were selected, where the leading and subleading jet have $\pt^\text{jet1} > 675\GeV$ and $\pt^\text{jet2} > \frac{2}{3} \pt^\text{jet1} $. They reported the LJP density as a function of the momentum fraction shared between the softer and harder subjets $z = \pt^{j2}/(\pt^{j_1}+\pt^{j_2})$ and as a function of \lnDeltaR. The \kt variable is related to $z$ via $\kt = z \DeltaR \pt^j$, where $\pt^j$ is the transverse momentum of the parent subjet of the splitting. Although a relation between \kt and $z$ exists for a given branching, this requires knowing the \pt of the parent subjet $j$ associated to the CA declustering step $j\to j_1 + j_2$. The momentum fraction $z$ carries information of the momentum balance of $j_1$ and $j_2$, but it is less straightforward to identify the \kt of the emission based only on $z$ without making additional assumptions, as done in Ref.~\cite{Lifson:2020gua}. The broad features of the LJP density, such as the running of \alpS and the separation of physical mechanisms, are also exposed in this representation of the primary LJP.

In this paper, a measurement of the primary LJP density using the CMS detector is presented. The analysis uses anti-\kt jets with $\pt^\text{jet} > 700\GeV$ and $\abs{y} < 1.7$ using proton-proton ($\Pp\Pp$) data collected at $\sqrt{s} = 13\TeV$ during Run 2 (2016--2018), corresponding to an integrated luminosity of 138\fbinv. The charged particles of the anti-\kt jet are reclustered with the CA algorithm to construct the LJP. The present measurement is reported for two distance parameters: $R=0.4$ and $R=0.8$. The value of 0.4 is the default choice in most measurements at 13\TeV at the LHC, whereas the 0.8 value, which has not been used in previous measurements of the LJP, allows us to extend the available phase space for wide-angle, hard emissions. Large $R$ jets are typically used to search for highly-boosted particles that decay hadronically. Identification algorithms, which are optimized using MC-simulated events, are used to distinguish such boosted particles from the QCD background. Thus, precision jet substructure measurements of large $R$ jets can help to improve the physics modeling in the simulation and reduce possible biases in the performance of such algorithms. For this measurement, the \lnkT versus \lnDeltaR representation of the primary LJP proposed in Ref.~\cite{Dreyer:2018nbf} is used. The choice of \kt has the advantage that one can unambiguously identify the absolute momentum scale of the emissions in the LJP, which in turn enables a clean separation of the mechanisms contributing to the strongly and weakly coupled regions of phase space.

The measurement is performed using data recorded by the CMS detector, whose components are described in Section~\ref{sec:cms_detector}, which also describes the algorithms used to reconstruct events are described. Section~\ref{sec:data_samples} contains details of the measured and simulated samples. Details of the extraction of the LJP density from the measurement are detailed in Section~\ref{sec:lund_plane}. The strategy for the corrections to stable-particle level is discussed in Section~\ref{sec:corrections}. The systematic uncertainties in the corrected distributions are described in Section~\ref{sec:syst_uncertainties}. The results on the corrected LJP density and their comparison with theoretical calculations are covered in Section~\ref{sec:results}. A summary of the measurement is presented in Section~\ref{sec:summary}.

Tabulated results are provided in the \HeHEPData record for this analysis \cite{hepdata}.

\section{The CMS detector and event reconstruction}\label{sec:cms_detector}

The central feature of the CMS apparatus is a superconducting solenoid of 6\unit{m} internal diameter, providing a magnetic field of 3.8\unit{T}. Within the magnetic volume are a silicon pixel and strip tracker, a lead tungstate crystal electromagnetic calorimeter (ECAL), and a brass and scintillator hadron calorimeter (HCAL), each composed of a barrel and two endcap sections. Forward calorimeters extend the pseudorapidity coverage provided by the barrel and endcap detectors. Muons are measured in gas-ionization detectors embedded in the steel flux-return yoke outside the solenoid. A more detailed description of the CMS detector, together with a definition of the coordinate system used and the relevant kinematic variables, is presented in Ref.~\cite{Chatrchyan:2008zzk}.

The silicon tracker used in 2016 measured charged particles within the range $\abs{\eta} < 2.5$. For nonisolated particles of $1 < \pt < 10\GeV$ and $\abs{\eta} < 1.4$, the track resolutions were typically 1.5\% in \pt and 25--90 (45--150)\mum in the transverse (longitudinal) impact parameter~\cite{CMS:2014pgm}. At the start of 2017, a new pixel detector was installed~\cite{Phase1Pixel}; the upgraded tracker measured particles up to $\abs{\eta} = 3.0$ with typical resolutions of 1.5\% in \pt and 20--75\mum in the transverse impact parameter~\cite{DP-2020-049} for nonisolated particles of $1 < \pt < 10\GeV$. According to simulation studies~\cite{DP-2017-015}, similar improvements are expected in the longitudinal direction.

The primary vertex (PV) is taken to be the vertex corresponding to the hardest scattering in the event, evaluated using tracking information alone, as described in Section 9.4.1 of Ref.~\cite{CMS-TDR-15-02}. The physics objects are jets, clustered using the anti-\kt jet finding algorithm~\cite{Cacciari:2008gp,Cacciari:2011ma} with the tracks assigned to a candidate vertex as inputs, and the associated missing transverse momentum, which is the negative vector \pt sum of those jets.

The particle-flow algorithm~\cite{CMS-PRF-14-001} aims to reconstruct and identify each individual particle in an event, with an optimized combination of information from the various elements of the CMS detector. The energy of photons is obtained from the ECAL measurement. The energy of electrons is determined from a combination of the electron momentum at the primary interaction vertex as determined by the tracker, the energy of the corresponding ECAL cluster, and the energy sum of all bremsstrahlung photons spatially compatible with originating from the electron track. The energy of muons is obtained from the curvature of the corresponding track. The energy of charged hadrons is determined from a combination of their momentum measured in the tracker and the matching ECAL and HCAL energy deposits, corrected for the response function of the calorimeters to hadronic showers. Finally, the energy of neutral hadrons is obtained from the corrected ECAL and HCAL energies.

For each event, hadronic jets are clustered from these reconstructed particles using the anti-\kt algorithm with $R = 0.4$ and 0.8, referred to as AK4 and AK8 jets in the following. Jet momentum is determined as the vectorial sum of all particle momenta in the jet. It is found from simulation to be, on average, within 5--10\% of the true momentum over the whole \pt spectrum and detector acceptance. Additional $\Pp\Pp$ interactions within the same or nearby bunch crossings (pileup) can contribute additional tracks and calorimetric energy depositions to the jet momentum. To mitigate this effect, charged particles identified to be originating from pileup vertices are discarded and an offset correction is applied to correct for remaining contributions for both AK4 and AK8 jets, as described in Ref.~\cite{CMS-PRF-14-001}. Jet energy corrections are derived from simulation to bring the measured response of jets to that of particle-level jets on average. In situ measurements of the momentum balance in dijet, photon$+$jet, $\PZ+$jet, and multijet events are used to account for any residual differences in the jet energy scale between data and simulation~\cite{CMS:2016lmd, DP-2021-033}. The populations of AK4 and AK8 jets are calibrated separately using the techniques described in Refs.~\cite{CMS:2016lmd, DP-2021-033}. The jet energy resolution amounts typically to 5\% at 1\TeV~\cite{CMS:2016lmd, DP-2021-033}. Additional selection criteria are applied to each jet to remove jets potentially dominated by anomalous contributions from various subdetector components or reconstruction failures~\cite{CMS:2016lmd, DP-2021-033}. For the charged-particle tracks used for the jet substructure extraction, further selection requirements are applied for the mitigation of residual pileup particles and to reject badly reconstructed objects, such as artifacts from detector noise~\cite{Sirunyan:2020foa}. If there are charged-particle tracks in the jet that do not belong to any vertex, they are used for the jet substructure observable only if their distance of closest approach to the primary vertex along the beam axis is smaller than 0.3\cm~\cite{Sirunyan:2020foa}.

Events of interest are selected using a two-tiered trigger system. The first level (L1), composed of custom hardware processors, uses information from the calorimeters and muon detectors to select events at a rate of around 100\unit{kHz} within a fixed latency of 4\mus~\cite{CMS:2020cmk}. The second level, known as the high-level trigger (HLT), consists of a farm of processors running a version of the full event reconstruction software optimized for fast processing, and reduces the event rate to around 1\unit{kHz} before data storage~\cite{CMS:2016ngn}.

\section{Data and simulated samples}\label{sec:data_samples}

The analysis uses $\Pp\Pp$ collision data collected by the CMS experiment in 2016--2018 at a center-of-mass energy of $\sqrt{s}=13\TeV$, corresponding to an integrated luminosity of 138\fbinv~\cite{CMS-LUM-17-003,CMS-PAS-LUM-17-004,CMS-PAS-LUM-18-002}. Events with high-\pt jets are collected with triggers requiring at least one AK4 or AK8 jet. The minimum jet \pt of 700\GeV considered in this analysis is well beyond the turn-on region of the trigger efficiency curve. The residual inefficiency is smaller than 1\% and is negligible in comparison with the statistical uncertainties of the measurement. No corrections are applied to account for residual trigger inefficiencies.

The main processes of interest in this analysis are events composed of jets produced through the strong interaction. Their simulation is performed with two main combinations of MC event generators to correct for the detector response and to derive systematic uncertainties for the measurement. A sample of jet events is generated at leading order (LO) with \PYTHIA{8.240}~\cite{Sjostrand:2014zea}, which implements a \pt-ordered shower and where the hadronization of quarks and gluons into stable hadrons is described by the Lund string model~\cite{Andersson:1983ia,Sjostrand:1984iu}. The \PYTHIA{8} parameters for the UE are set according to the CP5 tune~\cite{CMS:2019csb}. A second sample is generated at LO with \HERWIG{7.1.4}~\cite{Ball:2014uwa} with the CH3 tune~\cite{Sirunyan:2020pqv} to assess systematic uncertainties related to the modeling of the parton shower and hadronization. In \HERWIG{7}, the parton shower follows angular-ordered radiation~\cite{Gieseke:2003rz}, and the hadronization is described by the cluster fragmentation model~\cite{Webber:1983if}. The next-to-NLO (NNLO) NNPDF 3.1~\cite{Ball:2014uwa} parton distribution functions (PDFs) with $\alpS(m_{\PZ}) = 0.118$ are used for both MC generators, where $m_{\PZ}$ is the \PZ boson mass.

We verified that the contribution of processes beyond those produced purely by the strong interaction is negligible within the precision of the measurement. These processes beyond pure QCD interactions include vector boson production in association with jets and top quark production. Individually, the jets from these processes have distinct signatures on the primary LJP, particularly for boosted decay topologies \cite{Dreyer:2018nbf}, but these signatures are negligible when aggregated with the much larger contribution of QCD multijet processes, according to particle-level \PYTHIA{8} CP5 simulation studies. The \PYTHIA{8} CP5 and \HERWIG{7} CH3 samples used for the corrections in the analysis rely on QCD multijet processes for the event generation.

The \PYTHIA{8} CP5 and \HERWIG{7} CH3 generated samples are passed through a detailed simulation of the CMS detector using \GEANTfour~\cite{Agostinelli:2002hh}. To simulate the effect of pileup, multiple inelastic events are generated using \PYTHIA{8} and are superimposed on each primary interaction event. The MC simulated events are reweighed to reproduce the observed pileup activity in data.

Additional particle-level predictions are compared with the corrected distributions. Predictions are computed at LO with \PYTHIA{8.303} with CP2, CUEP8M1, and Monash tunes~\cite{Skands:2014pea, Khachatryan:2015pea, CMS:2019csb}. The Monash tune ~\cite{Skands:2014pea} is used as a baseline for the \PYTHIA{8} tunes developed by the CMS Collaboration, which was tuned mostly to \EE data at LEP and SLD, to $\Pp\PAp$ data at the SPS and Tevatron, and to $\Pp\Pp$ data from Run 1 at the LHC. The comparison with the predictions from the CUEP8M1 tune~\cite{Khachatryan:2015pea}, which was widely used in $\sqrt{s} = 7$ and 8\TeV analyses by CMS, illustrates the effects of including 13\TeV data in the development of the CP tunes of \PYTHIA{8}. In the context of this analysis, the main difference of the CP5 tune with respect to the other tunes of \PYTHIA{8} is the value of \alpS for final-state radiation (FSR), \alpSFSR, which is $\alpSFSR = 0.118, 0.130, 0.1365$, and 0.1365 for the CP5, CP2, CUEP8M1, and Monash tunes. Additional \PYTHIA{8.303} predictions calculated using the \DIRE~\cite{Hoche:2015sya} and \VINCIA~\cite{Giele:2007di} parton showers are presented. The \DIRE shower implements a \pt-ordered color dipole shower, where the radiator-spectator particle pairs evolve simultaneously, and it includes higher-order corrections, such as triple-collinear or double-soft parton emissions. The version of \VINCIA in \PYTHIA{8.303} uses the antenna sector shower formalism ~\cite{Gehrmann-DeRidder:2005btv, Giele:2007di,Brooks:2020upa}. The branching kernels, known as antenna functions, treat coherent sums of parton pairs without requiring a separation into radiators and spectators. Both \VINCIA and \DIRE have their respective set of tuning parameters in \PYTHIA{8}. We also present predictions calculated with \HERWIG{7.2.0}~\cite{Bellm:2015jjp,Bellm:2019zci} using a dipole parton shower, which is an alternative to the standard angular-ordered shower of \HERWIG{7}. The dipole parton shower available in \HERWIG{7} uses the Catani--Seymour dipole factorization formalism~\cite{Catani:1996vz}. In addition to \HERWIG{7} and \PYTHIA{8}, we also present predictions generated with \SHERPA{2.2.10} ~\cite{Bothmann:2019yzt, Krauss:2001iv,Gleisberg:2008fv}. The parton shower in \SHERPA{2} is also based on the Catani--Seymour dipole factorization approach~\cite{Schumann:2007mg}, and hadrons are formed by a modified cluster hadronization model~\cite{Winter:2003tt}.

We also consider particle-level predictions of \HERWIG{7.2.0} with different choices of the ordering scale variable in its angular-ordered shower~\cite{Bellm:2019zci,Bewick:2019rbu}. This choice determines which quantity is preserved for subsequent emissions, which affects both the momentum recoil assignment in the parton shower as well as its logarithmic accuracy. Depending on the recoil scheme, they can reach NLL accuracy for certain classes of global observables, such as event shapes and jet-resolution variables~\cite{Banfi:2004yd}. In \HERWIG{7}, there are four different options for the recoil scheme. The transverse momentum preserving scheme (\pt scheme) ensures the independence of successive soft-collinear emissions that are well separated in rapidity, but it can produce an excessive amount of hard radiation in phase space regions with no logarithmic enhancements. The virtuality-preserving scheme ($q^2$ scheme) avoids overpopulating such region of phase space, but breaks the independence of successive emissions. The dot-product-preserving scheme ($q_1\cdot q_2$ scheme) retains features of both \pt and $q^2$ schemes, although to some extent the $q_1\cdot q_2$ scheme can overpopulate the nonlogarithmically-enhanced phase space region. Thus, a recoil scheme with a phase space veto that suppresses events with large-virtuality partons was introduced ($q_1\cdot q_2+\text{veto}$ scheme). For the \HERWIG{7} CH3 sample used for the corrections to particle level, the $q^2$ scheme is used, which was the default setting of \HERWIG{7.1.4} when the CH3 tune was developed.

The \PYTHIA{8} and \SHERPA{2} generators apply a model where multiple-parton interactions (MPI) are interleaved with the parton shower~\cite{Sjostrand:1987su}, whereas the \HERWIG{7} generator models the spatial overlap between the colliding protons through a Fourier transform of the electromagnetic form factor of the proton, which plays the role of an effective inverse proton radius~\cite{Butterworth:1996zw,Borozan:2002fk,Bahr:2008dy,Gieseke:2012ft}. The MPI parameters of the generators are tuned to measurements in $\Pp\Pp$ collisions at the LHC~\cite{Khachatryan:2015pea}. The NNLO NNPDF 3.1 PDFs with $\alpS(m_{\PZ}) = 0.118$ are used for all MC event generator predictions.

\section{Event selection and jet substructure extraction}\label{sec:lund_plane}

Jets with $\pt > 700\GeV$ and $\abs{y}<1.7$ are selected for the measurement. As motivated in Refs.~\cite{Dreyer:2018nbf,Lifson:2020gua}, a high-\pt jet selection yields a larger phase space for hard radiation inside the jet in a region where nonperturbative effects are small. The maximum \kt at a particular \DeltaR is given by $\kt^\text{max} = \frac{1}{2} \pt^\text{jet} \DeltaR$, which corresponds to a diagonal line in the LJP. Due to the inclusive jet selection, all emissions share the same minimum kinematical limit. Both AK4 and AK8 jets are contained within the tracker acceptance with the jet rapidity selection.

\begin{figure}[!htb]
  \centering
  \includegraphics[width=0.49\textwidth]{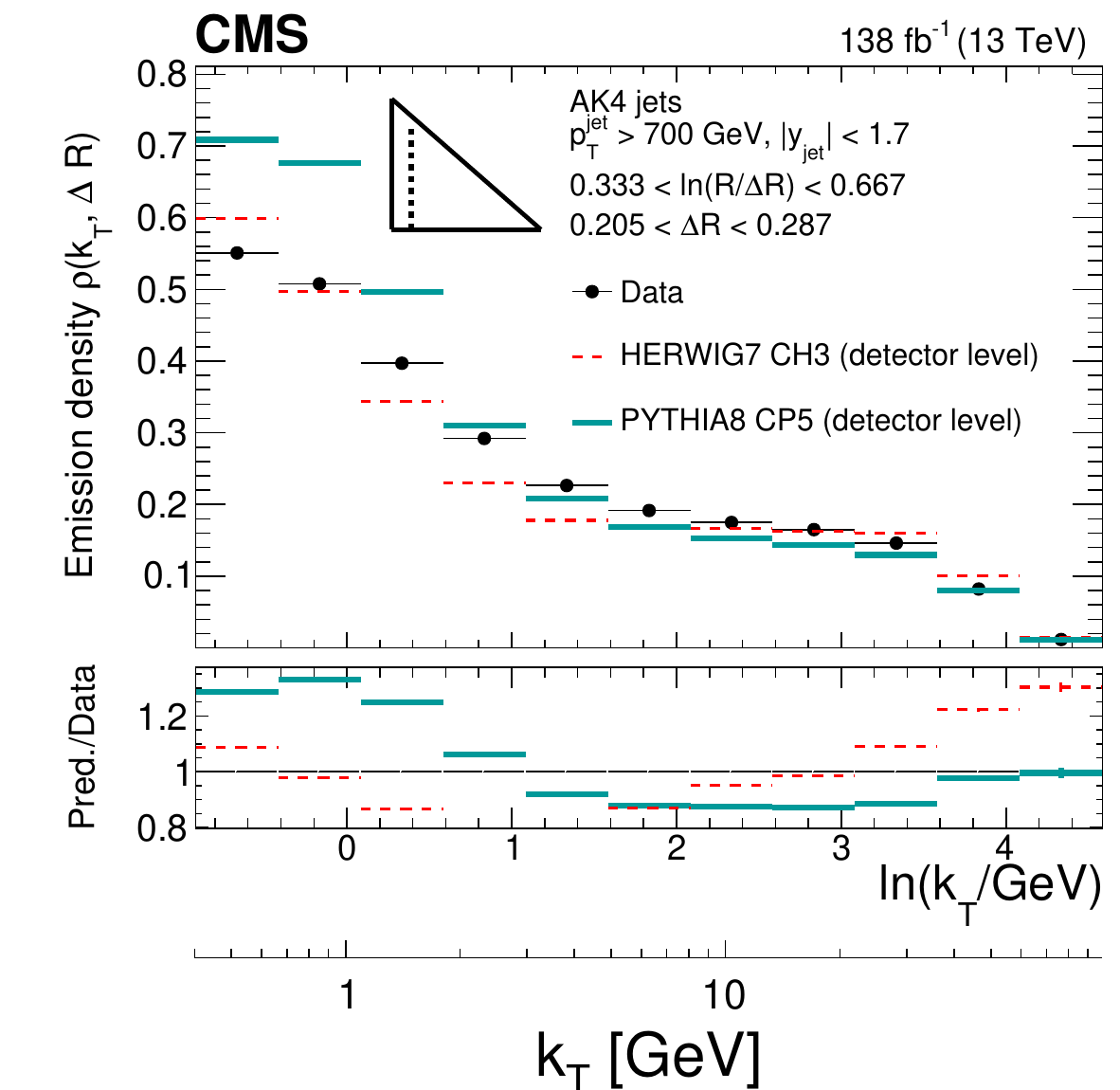}
  \includegraphics[width=0.49\textwidth]{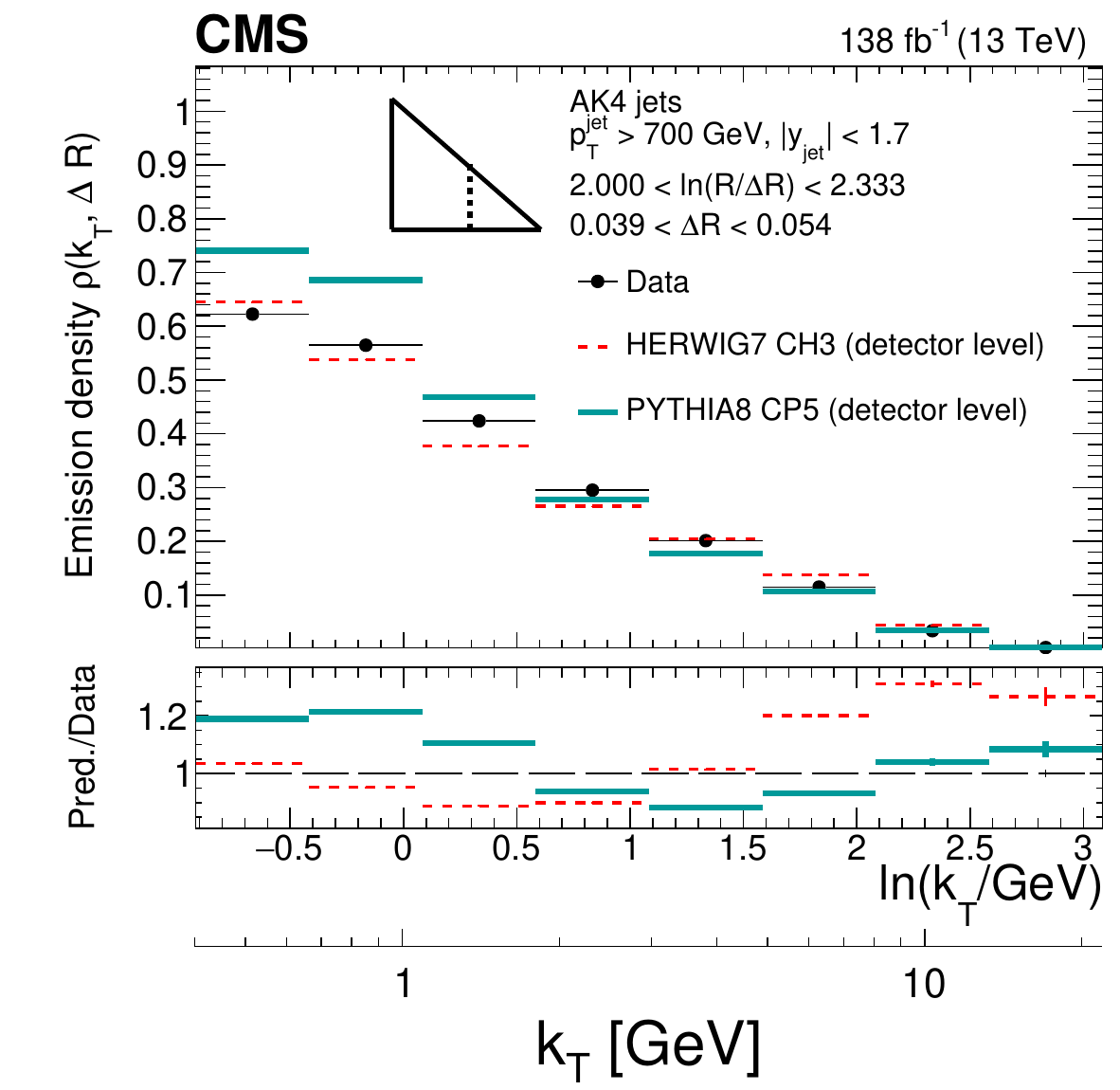}
  \includegraphics[width=0.49\textwidth]{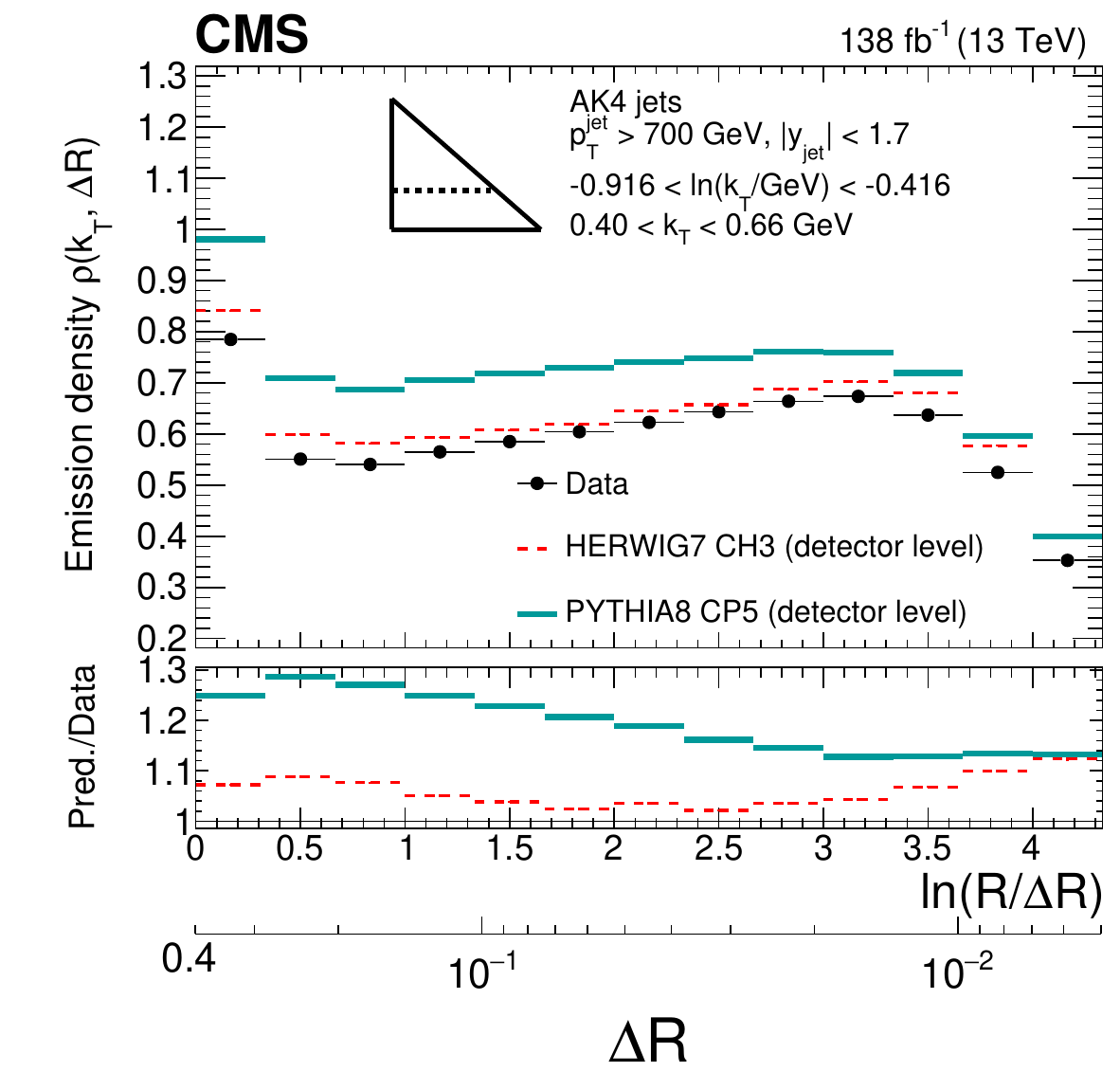}
  \includegraphics[width=0.49\textwidth]{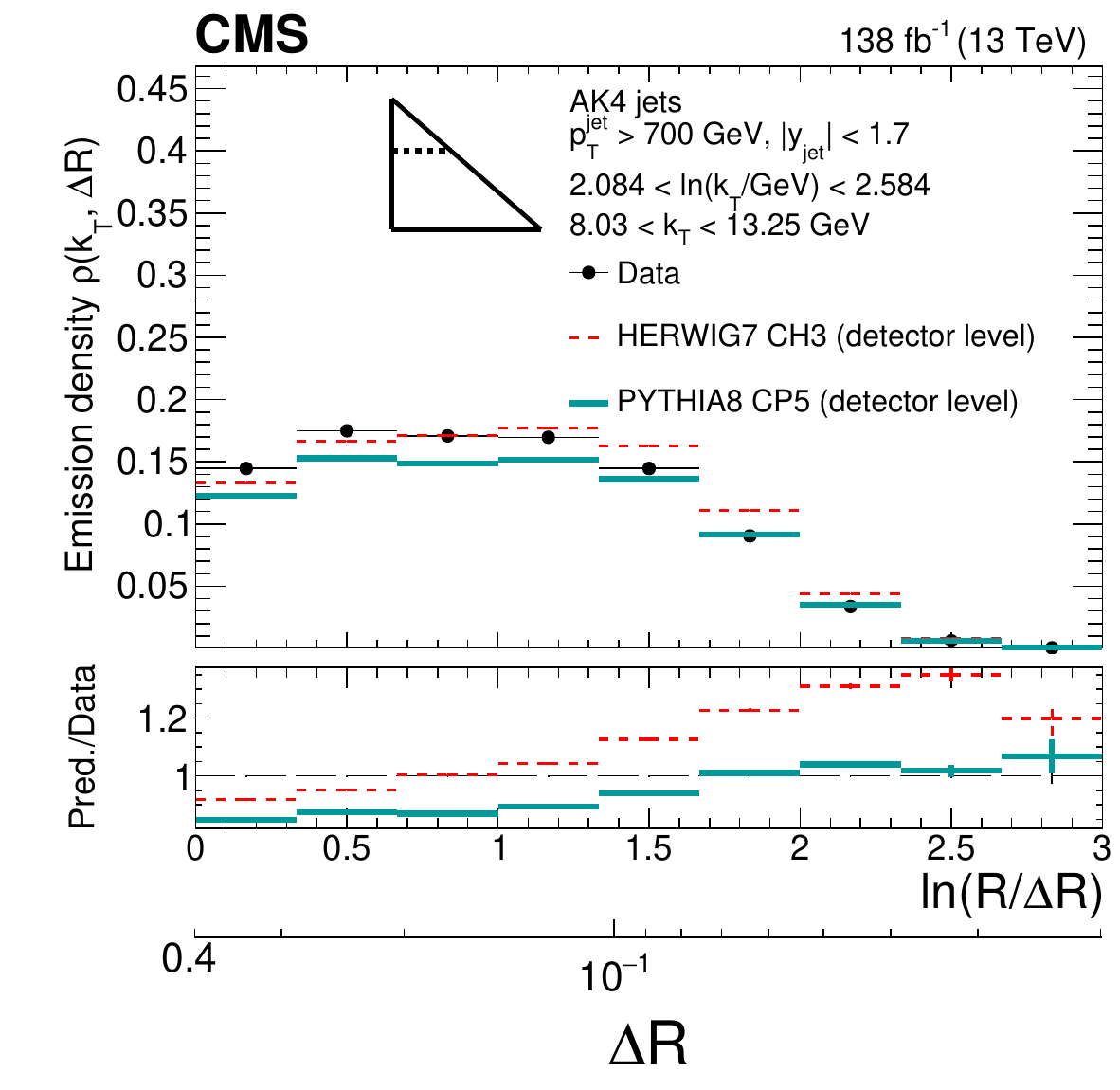}
  \caption{Detector-level distributions of measured and MC-simulated events generated with \PYTHIA{8} CP5 and \HERWIG{7} CH3 for four different slices of the charged-particle LJP, as indicated by the triangular diagrams in the plots. The lower panels in the plots show the ratio of the predictions with respect to the data. Only statistical uncertainties are included here. The comparison shows that neither \HERWIG{7} CH3 nor \PYTHIA{8} CP5 are able to describe the data well in various regions of the LJP. The vertical bars represent the statistical uncertainties, which are smaller than the markers for most of the bins.}
  \label{fig:substructureDataMcSpectrumsDetectorLevel}
\end{figure}

We follow the prescription described in Section~\ref{sec:introduction}. The charged-particle constituents of the original anti-\kt jet are reclustered using the CA algorithm to construct the primary LJP. We use the anti-\kt clustering algorithm for the initial clustering of the jet using neutral and charged particles, since these are the jets for which the jet energy scale and resolution are determined in data collected in high pileup conditions at the LHC. The use of another algorithm for the initial clustering affects mostly the $\DeltaR \approx R$ region, whereas it leaves the collinear region mostly unaffected~\cite{Dreyer:2018nbf,Lifson:2020gua}. Although the original anti-\kt jet is clustered using neutral and charged particle-flow candidates, the primary LJP is calculated using only its charged-particle constituents. This allows us to resolve hard collinear emissions and mitigate pileup better. Due to the approximate isospin symmetry of the strong force, the radiation pattern of the jet does not depend on the electric charge of the final-state particles. Although the charged-particle jet substructure is not infrared and collinear safe, this choice does not affect the comparison to theoretical calculations of the primary LJP density, since it can be incorporated as part of the nonperturbative corrections~\cite{Lifson:2020gua}. The measured charged-particle constituents are required to have $\pt > 1\GeV$ to further suppress the contributions of residual pileup particles and to avoid the decrease in track reconstruction efficiency below 1\GeV. The measurement is reported for splitting angles as small as $\DeltaR \approx 0.005$, which is a \DeltaR separation where pixel cluster merging effects can efficiently be mitigated with the dedicated cluster splitting algorithms of CMS~\cite{CMS-DP-2019-007}. The largest splitting angle reported in the measurement is $\DeltaR = 0.8$, which is the distance parameter of AK8 jets. The \kt range reported in the measurement spans 0.4 to 720\GeV. The largest value of \kt is accessed only with $R = 0.8$ jets. Figure~\ref{fig:substructureDataMcSpectrumsDetectorLevel} shows different slices of the primary LJP density measured in data. The detector-level predictions of \HERWIG{7} CH3 and \PYTHIA{8} CP5 are shown in the same plots. Throughout the paper, four different regions of the LJP are presented, each probing different aspects of the radiation pattern of the jet. The vertical slice at large \DeltaR shows the density of emissions in a wide range of \kt values, and is mostly sensitive to the parton cascade evolution in the jet. A vertical slice at small \DeltaR shows the behavior of the density of emissions for collinear emissions. A horizontal slice at low \kt shows a region of the LJP dominated by hadronization effects and the underlying event activity as a function of \DeltaR, whereas a horizontal slice at high \kt shows the density of emissions for hard emissions and is less sensitive to hadronization effects.

\clearpage

\section{Unfolding}
\label{sec:corrections}

\begin{figure}
  \centering
  \includegraphics[width = .54\textwidth]{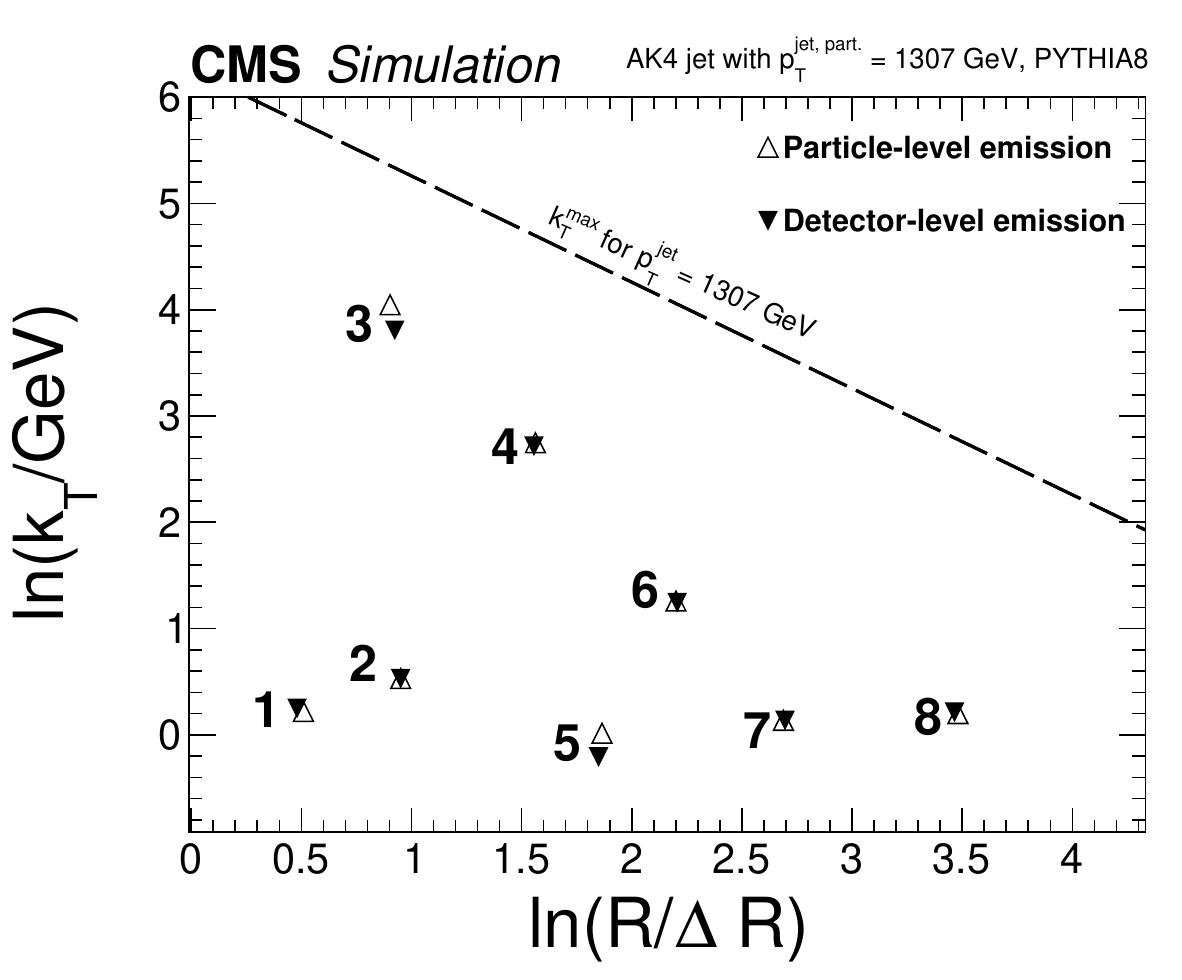}
  \includegraphics[width = .454\textwidth]{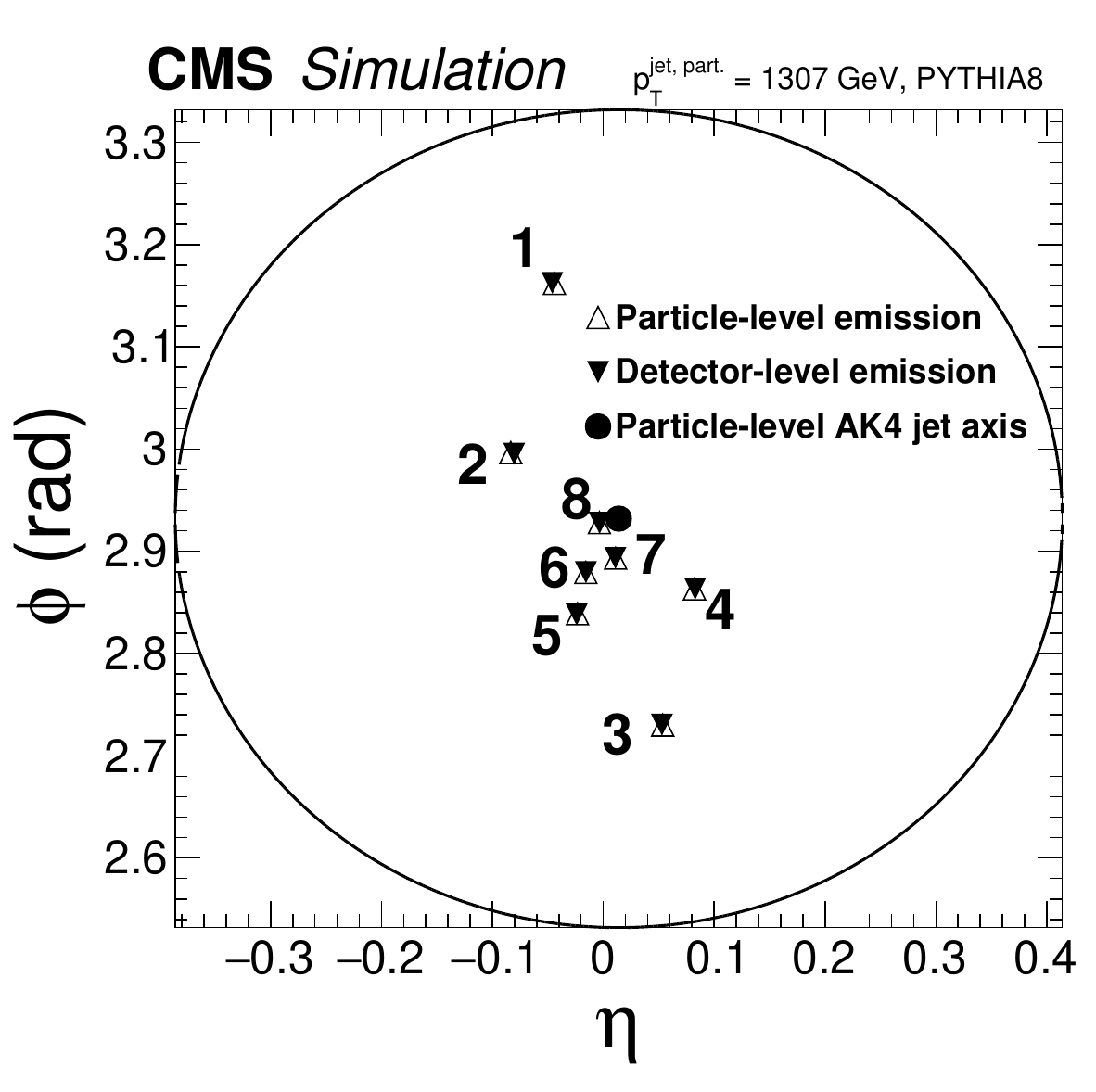}
  \caption{Event displays of a simulated AK4 jet at detector level (solid triangles) and particle level (open triangles). The right-hand side plot represents the $\eta$ and $\phi$ coordinates of the emissions in the CMS coordinate system to illustrate the geometrical matching used for the corrections in the measurement. The center of the particle-level anti-\kt jet is represented by the solid circular marker. The circular line with radius $R = 0.4$ serves as a proxy for the anti-\kt distance parameter used to cluster the AK4 jet. The Lund plane on the left plot is associated with the same jet, and is filled with the primary emissions from the CA declustering from left to right (from large to small angles). The numbers in both plots represent the order of the emission of the primary CA tree declustering sequence.}
  \label{fig:matching_sketches}
\end{figure}

The measured detector-level distributions are corrected to particle level using corrections derived from simulation. The particle level is defined by the set of particles that have a lifetime longer than $10^{-8}$\unit{s} as given by the MC event generator. Neutral and charged particles are used for the anti-\kt jets at particle level, jet clustering at the particle level, as is done for the detector-level anti-\kt, which is taken into account for the correction of the full jet \pt. The jet substructure is corrected to the charged stable-particle level, where the CA reclustering is done using only the charged particles of the anti-\kt jet, as is done at detector level. The corrections to particle level are applied sequentially in three steps. First, a set of bin-by-bin corrections is applied to the detector-level distribution to account for the background (purity corrections). In this measurement, the background consists of detector-level emissions that are not paired with particle-level emissions. Then, to correct for bin-to-bin migrations due to detector effects, a multidimensional unfolding of the LJP emissions is applied. After the correction for bin-to-bin migration effects, a set of bin-by-bin corrections to account for particle-level emissions that are not reconstructed at detector level (efficiency corrections) is applied. At particle level, jets are clustered without a \pt requirement on the particles used for the jet clustering. The particle- and detector-level distributions have the same binning in the fiducial region. The particle-level distributions have additional underflow and overflow bins to account for phase space migrations, which are removed after the full chain of corrections.

To derive the corrections, a mapping between the detector- and particle-level emissions is necessary. The response matrix is determined from simulation for geometrically matched emissions, which is described in the following. To match the detector- and particle-level emissions, for a given detector-level emission, we iterate through the list of particle-level emissions and calculate their distance relative to the detector-level emission in $\eta$--$\phi$. The particle-level emission that is closest to the detector-level emission is the candidate for the particle-level matched emission. Then, a similar iteration is done for a given particle-level emission, and iterating through the detector-level emissions. Only particle-level and detector-level emission pairs with a $\DeltaR_\text{match}$ distance of less than 0.1 units are included for the matching procedure. We only consider a pair of emissions to be matched if the same pair of emissions is found when iterating through the list of detector- and particle-level list of emissions. As a result, no detector- (particle-)level emission is assigned to two different particle- (detector-)level emissions. The corrections are derived using \PYTHIA{8} CP5 and \HERWIG{7} CH3 simulated events. Neither of the detector-level predictions of \PYTHIA{8} CP5 and \HERWIG{7} CH3 are able to describe the data in the entire LJP. The nominal set of corrections is derived using the sample of \PYTHIA{8} CP5 simulated events, since we can use its parton shower event weights to propagate uncertainties related to the renormalization scale choice for FSR and initial-state radiation (ISR), as described in Section \ref{sec:syst_uncertainties}. The sample of \HERWIG{7} CH3 simulated events is used to estimate biases in the unfolding corrections and to compute the systematic uncertainties associated with the parton shower and hadronization model used for the corrections, as described in Section~\ref{sec:syst_uncertainties}.

The purity corrections have values smaller than 5\% in most of the LJP, except in the region of wide-angle, soft emissions, where it has values of the order of 20\%. This region is dominated by UE activity particles and residual pileup particles. The efficiencies are above 90\% throughout most of the LJP. The efficiency drops to 70\% in the large \DeltaR and low \kt region and decreases with decreasing \DeltaR down to 40--50\%. The phase space region of low \kt and large \DeltaR is affected by the charged-particle $\pt > 1\GeV$ requirement that is applied at detector level, resulting in a higher number of splittings at particle level than at detector level, which leads to a reduction in the reconstruction efficiency. In the small \DeltaR region, particle-level subjets can be lost in the reconstruction due to pixel cluster merging effects that are not effectively mitigated by the cluster splitting algorithm~\cite{CMS-DP-2019-007}. Pixel cluster merging leads to an inefficiency for collinear emissions, which increase with smaller \DeltaR. This is one component of the overall track reconstruction efficiency, and it is partially mitigated by restricting the analysis to splittings with $\DeltaR > 0.005$. The $\pt > 1\GeV$ requirement at detector level for the charged-particle constituents of the jet reduces the number of constituents of the subjets, which also contributes to the reduction of the reconstruction efficiency at small \kt and small \DeltaR. The purity and efficiency corrections calculated with the \HERWIG{7} CH3 and \PYTHIA{8} CP5 samples are consistent with each other within 1 to 7\% throughout most of the LJP.

An example of the correspondence of the detector- and particle-level emissions in a simulated jet is shown in Fig.~\ref{fig:matching_sketches}. For most simulated jets, there is a close correspondence in the $\eta$--$\phi$ coordinate system, with smearing effects and particle losses affecting the reconstruction of \kt in the LJP, and angular resolution smearing at small \DeltaR. The correspondence between the respective primary LJP emissions might be lost due to detector effects. This can occur when a different branch of the jet tree of primary emissions is followed at detector level relative to the one followed at particle level. The presence of these mismatches is mitigated with the aforementioned requirement of using unique geometrically matched emissions for the corrections, and they are accounted for using the bin-by-bin purity and efficiency corrections. The residual amount of mismatches represents less than a few percent of the total number of matched emissions in simulation. The residual mismatches include cases where the particle-level emissions are merged or fragmented at detector level, or cases where the harder (softer) subjet becomes the softer (harder) subjet due to momentum smearing effects, which becomes harder to mitigate in the collinear region. These mismatches are observed in both \HERWIG{7} CH3 and \PYTHIA{8} CP5 simulated events.

A multidimensional unfolding of the number of emissions \nEmissions is performed to correct for the bin-to-bin migrations caused by detector effects. The dimensions that are considered in the correction are the anti-\kt jet \pt, and the \kt and \DeltaR of the emissions. The number of jets \nJets, which is used to calculate the per-jet average LJP density, is also corrected to account for migration effects in the anti-\kt jet \pt. The response matrix used for the unfolding is calculated using events simulated with \PYTHIA{8} CP5. To assess if the deconvolution problem is sensitive to small perturbations in the input distribution, we calculate the condition number of the probability matrix, which is defined as the ratio of its largest to smallest nonzero eigenvalues. The probability matrix is obtained by normalizing the response matrix such that the probability of reconstruction in the different detector-level bins adds up to 1 for each particle-level bin. A condition number of order unity suggests that the probability matrix is well-conditioned. The condition number of the probability matrix used in the unfolding of the LJP of emissions is much larger than unity, suggesting that the deconvolution problem is ill-conditioned, and that regularized unfolding is required. For the present measurement, we perform the regularized matrix inversion using iterative D'Agostini unfolding with early stopping \cite{DAgostini:1994fjx}. We use its implementation in the \textsc{RooUnfold} package \cite{RooUnfold}. Regularized unfolding introduces a bias towards an input particle-level spectrum (\PYTHIA{8} CP5 or \HERWIG{7} CH3 in this analysis) in order to control the sensitivity to small perturbations in the input distribution. The particle-level spectrum that is used in regularized unfolding is referred to as the ``prior distribution.'' There is model dependence in the response matrix as well; the detector response for subjets simulated with \HERWIG{7} CH3 is different from the response of subjets simulated with \PYTHIA{8} CP5. These two components of model dependence are included in the systematic uncertainties described in Section~\ref{sec:syst_uncertainties}.

Since there are multiple entries in the LJP histogram for a given jet, the detector-level bins are statistically correlated. These statistical correlations manifest as off-diagonal elements in the covariance matrix built from statistical uncertainties. We find statistical bin-to-bin correlations of 5--10\%. The hard, wide-angle emissions have weak correlations with the rest of the LJP. Emissions at low \kt and large \DeltaR have the largest correlations with the rest of the LJP. In this region of phase space, the contribution from the UE and residual pileup particles is important. The covariance matrix of statistical uncertainties is used as an input for the unfolding procedure.

The number of iterations, which plays the role of the regularization parameter in iterative D'Agostini unfolding, is optimized based on $\chi^2$ goodness-of-fits tests at detector level. At each iteration, the detector-level distribution associated with the unfolded distribution is calculated by multiplying the unfolded distribution by the migration matrix. We call this ``forward folding,'' and the detector-level distribution obtained after matrix multiplication as the ``forward-folded distribution.'' To quantify the compatibility between the forward-folded distribution and the input measured distribution, the $\chi^2$ of the forward-folded distribution and the measured detector-level distribution is calculated at a given iteration. In D'Agostini unfolding, the agreement between the input distribution and the forward-folded distributions improves monotonically at each iteration. To avoid overfitting the unfolded distributions to the statistical fluctuations present in the measured distributions, the algorithm is stopped at the iteration at which the corresponding $p$-value is above 0.05. The covariance matrix of statistical uncertainties measured at detector level is used for these goodness-of-fit tests at each iteration. Twelve iterations and eight iterations are used for AK4 jets and AK8 jets, respectively.

\begin{figure}[!htb]
  \centering
  \includegraphics[width=0.49\textwidth]{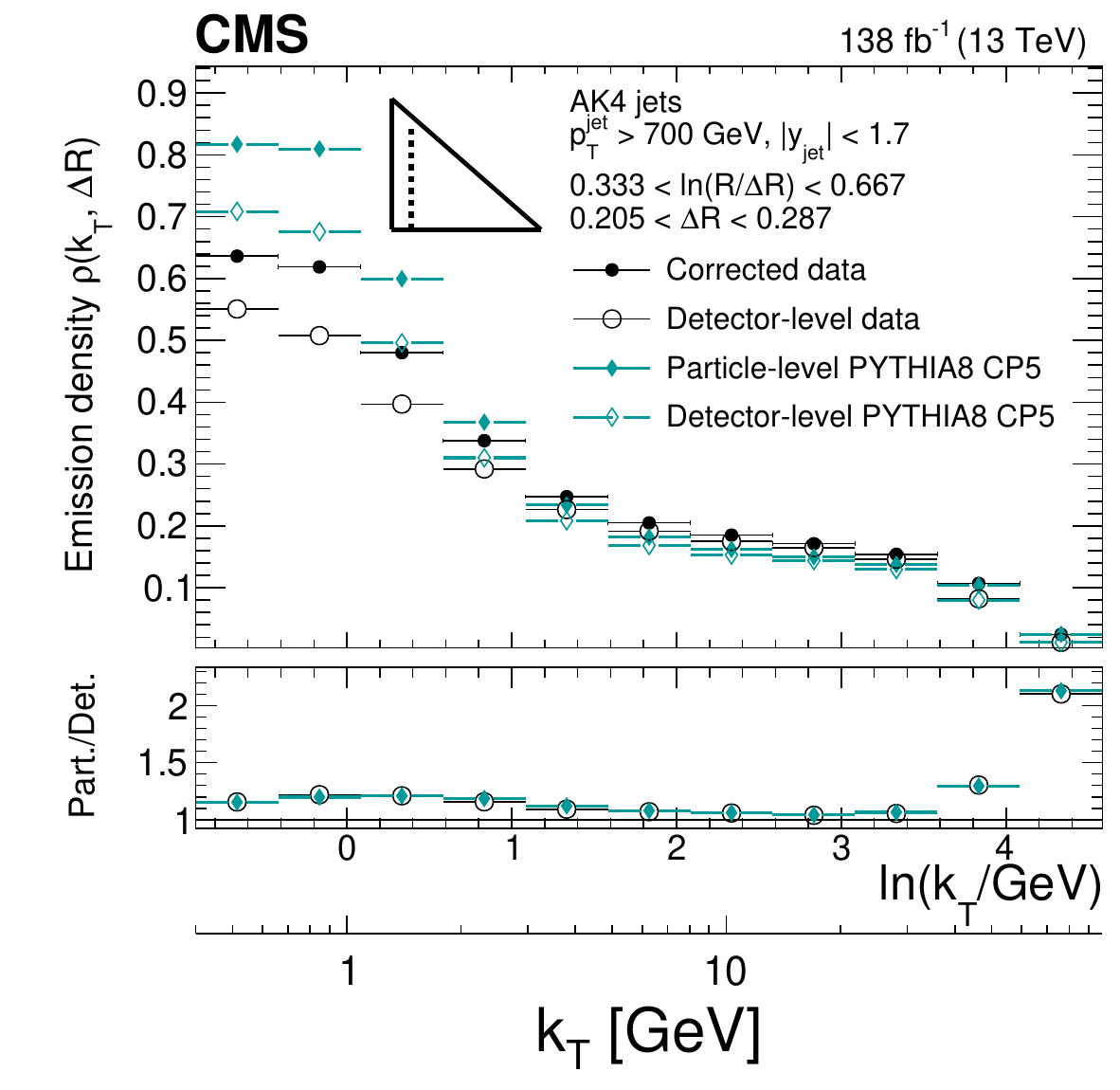}
  \includegraphics[width=0.49\textwidth]{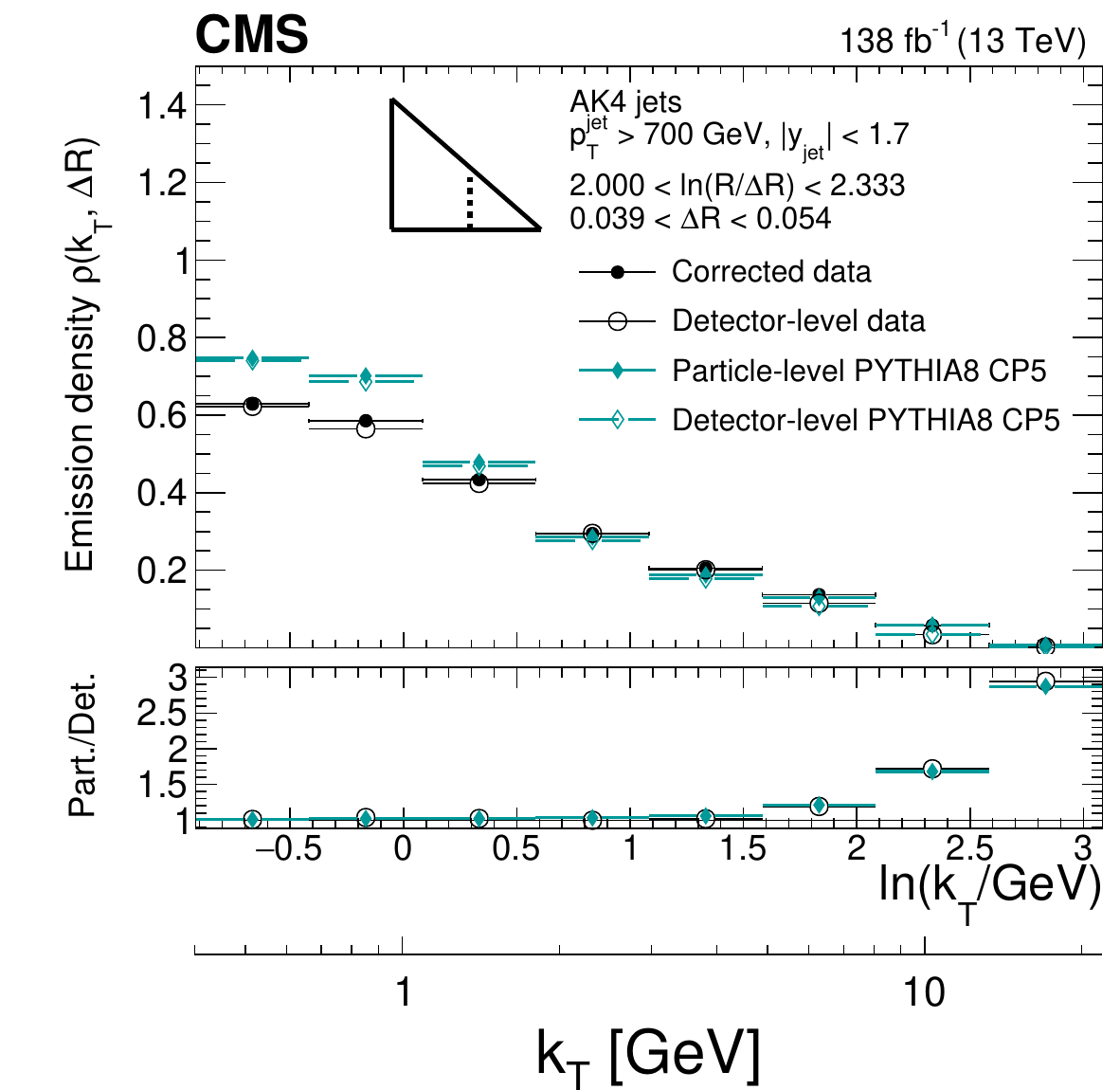}
  \includegraphics[width=0.49\textwidth]{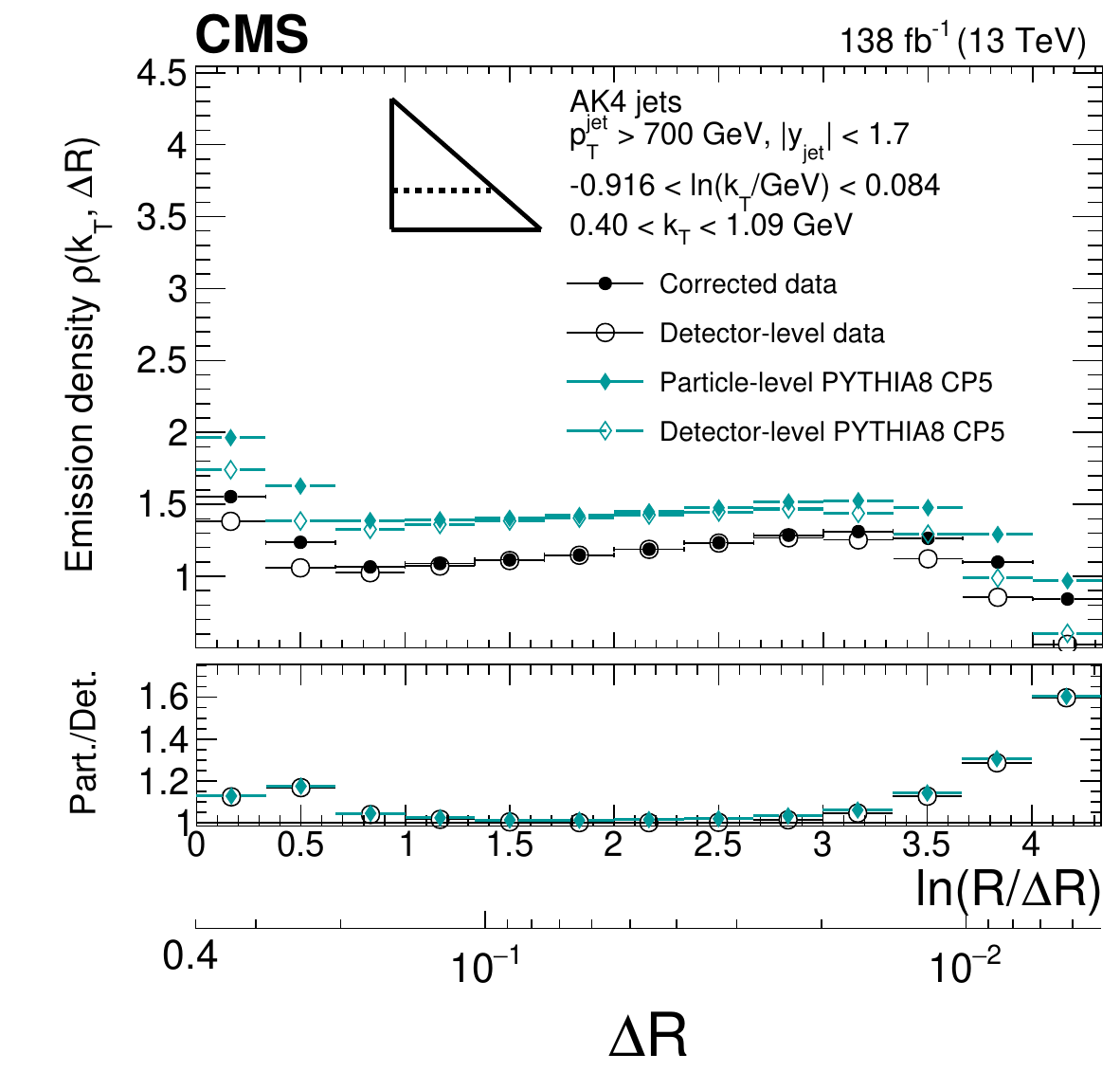}
  \includegraphics[width=0.49\textwidth]{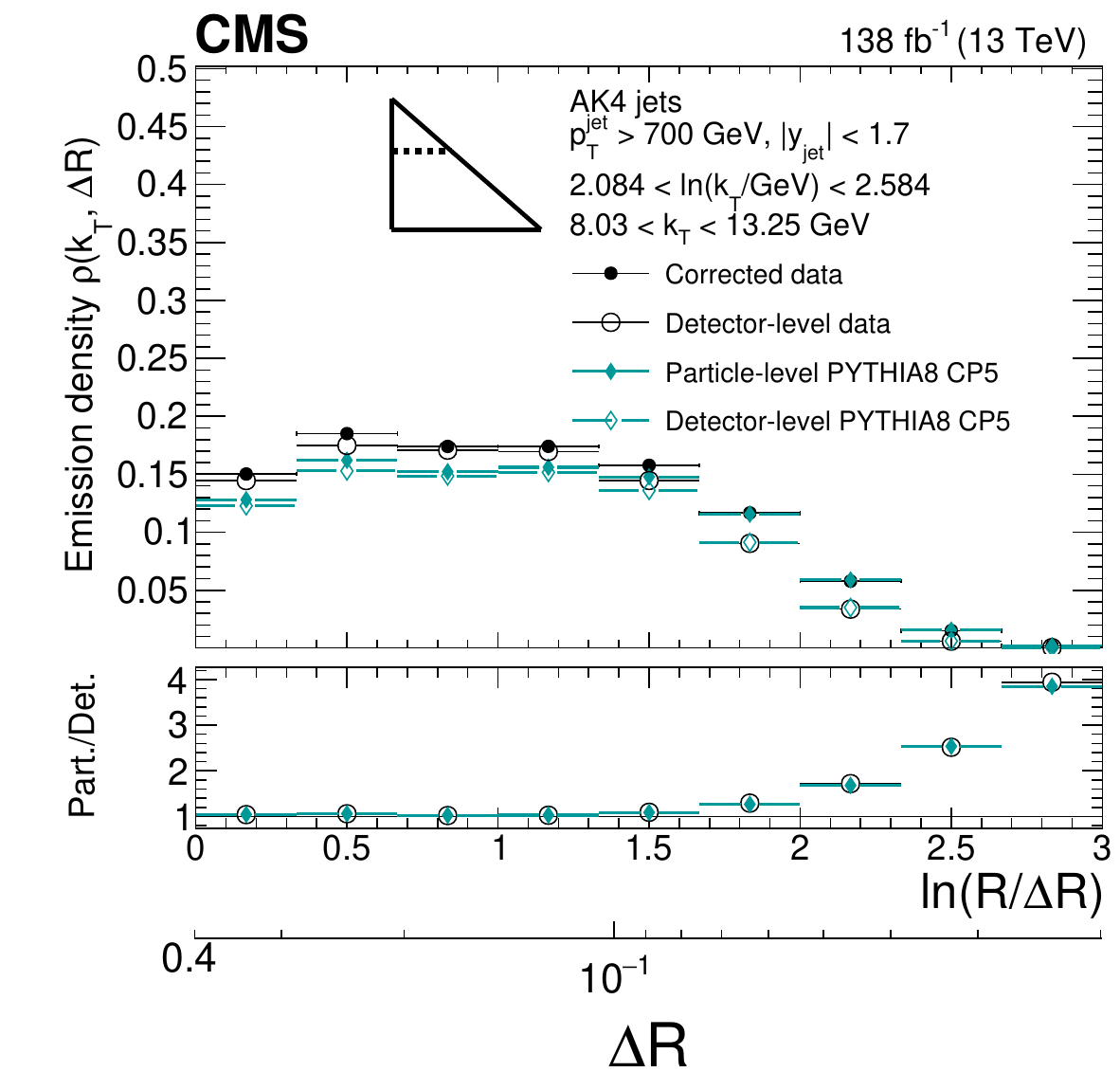}
  \caption{Detector-level (open symbols) and particle-level (closed symbols) distributions for the data and MC simulated events of \PYTHIA{8} CP5. Only statistical uncertainties are included in these plots, which are smaller than the markers for most of the bins. The lower panels in the plot show the ratio of the particle-level to the respective detector-level distributions, which is used as a metric for the effective modifications of the charged-particle LJP density because of the detector effects. The size of the corrections can be inferred from the ratio of the particle-level to the detector-level distributions, which are larger closer to the kinematical edge of the LJP.}
  \label{fig:substructureDataMcSpectrums}
\end{figure}

To illustrate the effective modifications of the particle-level LJP density because of the detector effects, which are corrected with the unfolding procedure, Figure~\ref{fig:substructureDataMcSpectrums} shows the distributions of four different slices of the LJP density measured in data using AK4 jets at both the detector and particle (unfolded) levels. The \PYTHIA{8} CP5 predictions are used as a reference in Fig.~\ref{fig:substructureDataMcSpectrums} at both the detector and particle levels. Due to the finite tracking efficiency and the $\pt > 1 \GeV$ requirement for charged particles at the detector level, the detector-level \kt of a subjet is, on average, smaller than its corresponding particle-level \kt. This effect tends to increase with \kt, because subjets with higher \kt have larger charged-particle multiplicities. Together with the fast drop of the LJP density at the kinematical edge of the LJP, this leads to large a depletion of emissions in that region, which explains the difference between the comparison detector- and particle-level LJP densities in that region of the LJP, as shown in Fig.~\ref{fig:substructureDataMcSpectrums}. At low \kt and large \DeltaR, the modification of the LJP density at detector level relative to particle level is due to the presence of residual pileup particles that populate that region of phase space as well as the $\pt > 1\GeV$ threshold used at detector level.

\section{Systematic uncertainties}\label{sec:syst_uncertainties}

Theoretical and experimental uncertainties are propagated by repeating the unfolding procedure with variations of the response matrix, prior distribution, purity, and efficiency corrections. The following systematic uncertainties are considered:

\textit{Shower and hadronization modeling uncertainty:} In D'Agostini iterative unfolding, there are two main sources of model dependence: one of them is the response matrix, which describes the bin-to-bin migrations from particle-level bins to detector-level bins; and the second one is the prior distribution, which is used for regularization (the latter can be identified with the simulated particle-level spectrum). To evaluate the uncertainties from these two sources, the unfolding procedure is repeated with uncorrelated variations of the prior distribution used for regularization and the response matrix, one variation at a time. For the prior bias variation, the distribution of \HERWIG{7} CH3 is assumed instead of the one of \PYTHIA{8} CP5, while keeping the response matrix calculated with \PYTHIA{8} CP5 events. For the response matrix variation, the \HERWIG{7} CH3 response matrix is used, while keeping the \PYTHIA{8} CP5 spectrum for the prior distribution. The difference of each variation with respect to the nominal results, which are calculated with \PYTHIA{8} CP5 for both the prior distribution and response matrix, is used to calculate the corresponding systematic uncertainty. Since neither \PYTHIA{8} CP5 nor \HERWIG{7} CH3 describe the data well everywhere in the primary LJP, the two components of this systematic uncertainty are symmetrized. The resulting response matrix and prior bias uncertainties are treated as uncorrelated between each other. The prior bias and response matrix components have similar contributions in the bulk of the LJP (the region away from the kinematical edge) at large $\DeltaR \sim R$, whereas the component associated with the response matrix is dominant at small \DeltaR. The total systematic uncertainty is of the order of 2--7\% across the bulk of the distribution, increasing to about 10\% close to the kinematical edge of the LJP.

 \textit{Parton shower scale uncertainties:} To estimate the theoretical uncertainty due to missing higher-order corrections in the perturbative calculation of the parton shower of \PYTHIA{8} CP5, we consider variations of the renormalization scale for FSR and ISR, denoted symbolically with ($\mu^{\mathrm{FSR}}, \mu^{\mathrm{ISR}}$). We repeat the unfolding procedure for six different variations of the renormalization scale by factors of ($\frac{1}{2}$, $\frac{1}{2}$), ($\frac{1}{2}$,1), (1,$\frac{1}{2}$), (2,2), (2,1), and (1,2)~\cite{Cacciari:2003fi, Catani:2003zt}. The corresponding uncertainties of the unfolded distributions are of the order of 0.5--3.0\%. Generally, the variations by factors of 2 and 1/2 have symmetric effects on the LJP density and in the associated systematic uncertainties. The effect of the variations of the ISR renormalization scale are smaller than the ones from FSR. The ($\frac{1}{2}$,1) and (2,1) variations are referred to as FSR down and up, respectively, and likewise for ISR down and up.

\textit{Tracking efficiency uncertainties:} For the measurement of the primary LJP density using charged-particle tracks, the most crucial data-to-simulation difference is given by the tracking efficiency uncertainty. The effect of losing one or more particles with \pt of a few \GeV has an important impact on the \kt detector response, particularly for subjets close to the kinematical edge of the LJP, where the LJP density is steeply falling and where subjets tend to have larger particle multiplicities. To account for the mismodeling of track reconstruction in the simulation in high-density environments, such as in the vicinity of the jet core~\cite{CMS:2016lmd}, we propagate a track reconstruction efficiency uncertainty through the unfolding procedure. This is done by randomly removing 3\% of the tracks in simulation. This approach is a conservative way of covering data-to-simulation differences for tracking in jets, as found in the context of the jet energy calibration of high-\pt jets for 2016--2018 data analyses. The dependence of the tracking efficiency uncertainty in the track density is not known; the 3\% uncertainty value is expected to cover such residual dependence. This value is larger than the tracking efficiency uncertainty for isolated tracks, which ranges from 2.2--2.7\% for 2016--2018 data~\cite{DP-2022-012}. The unfolding is repeated with this variation, and the symmetrized difference of the unfolded distributions with this variation relative to the nominal unfolded distributions is taken as a systematic uncertainty. Since there was a new pixel detector installed in 2017~\cite{Phase1Pixel}, the tracking efficiency uncertainty of 2016 is decorrelated from the one of 2017--2018. The symmetrized difference of the unfolding with these variations relative to the nominal unfolded distributions is taken as a systematic uncertainty. The uncertainty is about 1--2\% in the bulk of the primary LJP and increases up to about 15--25\% in the kinematical edge of the LJP. This is because the density of emissions falls rapidly at the edge of the LJP and because the average number of subjet constituents is larger in that region; emissions close to the edge of the LJP behave more jet-like than those in the bulk of the LJP. On average, the detector-level \kt is lower than the corresponding particle-level \kt. The track reconstruction efficiency affects such correspondence between the detector-level \kt and the particle-level \kt for a given pair of matched emissions.

\textit{Response matrix statistical uncertainties: } The statistical uncertainties of the simulated sample that is used to derive the migration matrix are propagated through the unfolding procedure, which results in a contribution to the covariance matrix.

\textit{Pileup modeling:} We vary the value of the inelastic cross section used to generate the number of pileup interactions in the simulation. The distribution of pileup interactions in the simulation is reweighed with $\pm4.6$\% variations of the $\Pp\Pp$ inelastic cross section at 13\TeV relative to the nominal value of 69.2\unit{mb}. The resulting systematic uncertainty is less than 1\% in most of the LJP and increases to about 2\% at the kinematical edge of the LJP.

\textit{Jet energy scale and resolution:} The jet energy scale uncertainty is propagated through the unfolding by shifting the jet \pt at detector level in the simulation according to the $\eta$-\pt dependent jet energy scale uncertainties~\cite{CMS:2016lmd}. This uncertainty has an effect through the $\pt > 700\GeV$ selection requirement. The uncertainty is smaller than 1\% throughout the unfolded LJP density. We also consider the uncertainties in the jet energy resolution measurement~\cite{CMS:2016lmd}. The jet \pt is further smeared at detector level in simulation in order to better reproduce the jet energy resolution measured in data~\cite{CMS:2016lmd}. Such a smearing procedure comes with an associated systematic uncertainty, which is propagated through the unfolding procedure. The respective uncertainty is at the per-mille level in the corrected primary LJP density.

The various sources of uncertainty are considered to be independent and their effects are added in quadrature in a given bin. The total experimental uncertainties range from 2--7\% in the bulk of the LJP. The systematic uncertainties can go up to about 25\% at the edge of the LJP, which is the region where the smearing effects are stronger and where the LJP density drops faster. The uncertainties of each source are considered as bin-to-bin fully correlated, with the exception of the systematic uncertainty related to the response matrix statistical uncertainties.

\begin{figure}
  \centering
  \includegraphics[width = 1.\textwidth]{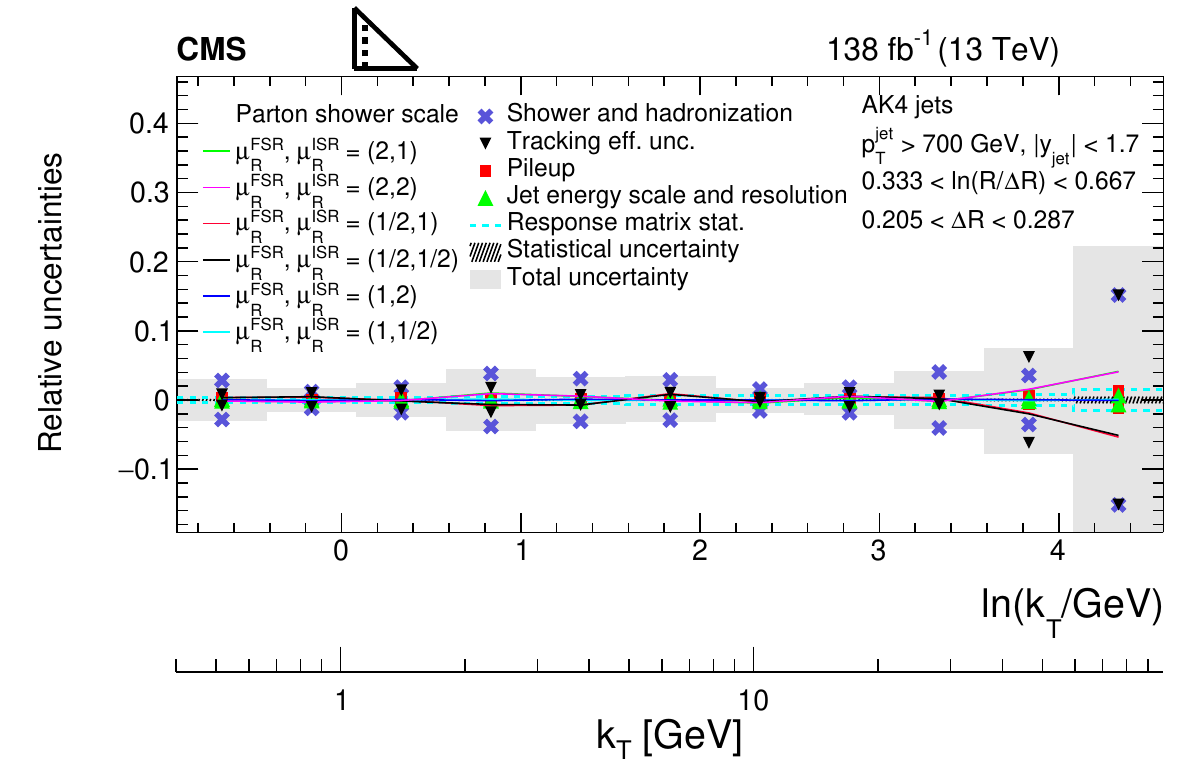}
  \includegraphics[width = 1.\textwidth]{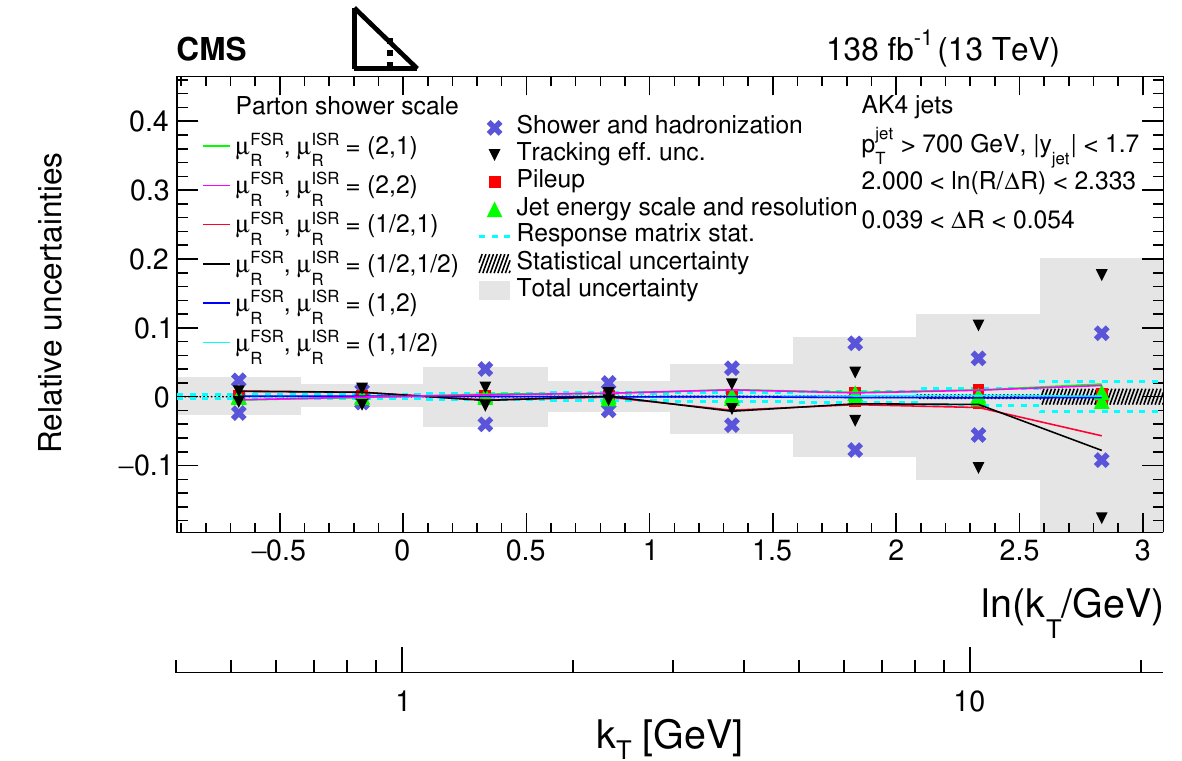}
  \caption{Different components of the systematic uncertainties for AK4 jets for two different vertical slices of the charged-particle LJP density. The upper plot is for large angles $0.205 < \DeltaR < 0.287$, and the lower plot is for small angles $0.039 < \DeltaR < 0.054$. The total experimental uncertainty is represented by the filled area. The statistical uncertainties in the data are represented by the hashed band.}
  \label{fig:sysuncert_ak4}
\end{figure}

\begin{figure}
  \centering
  \includegraphics[width = 1.\textwidth]{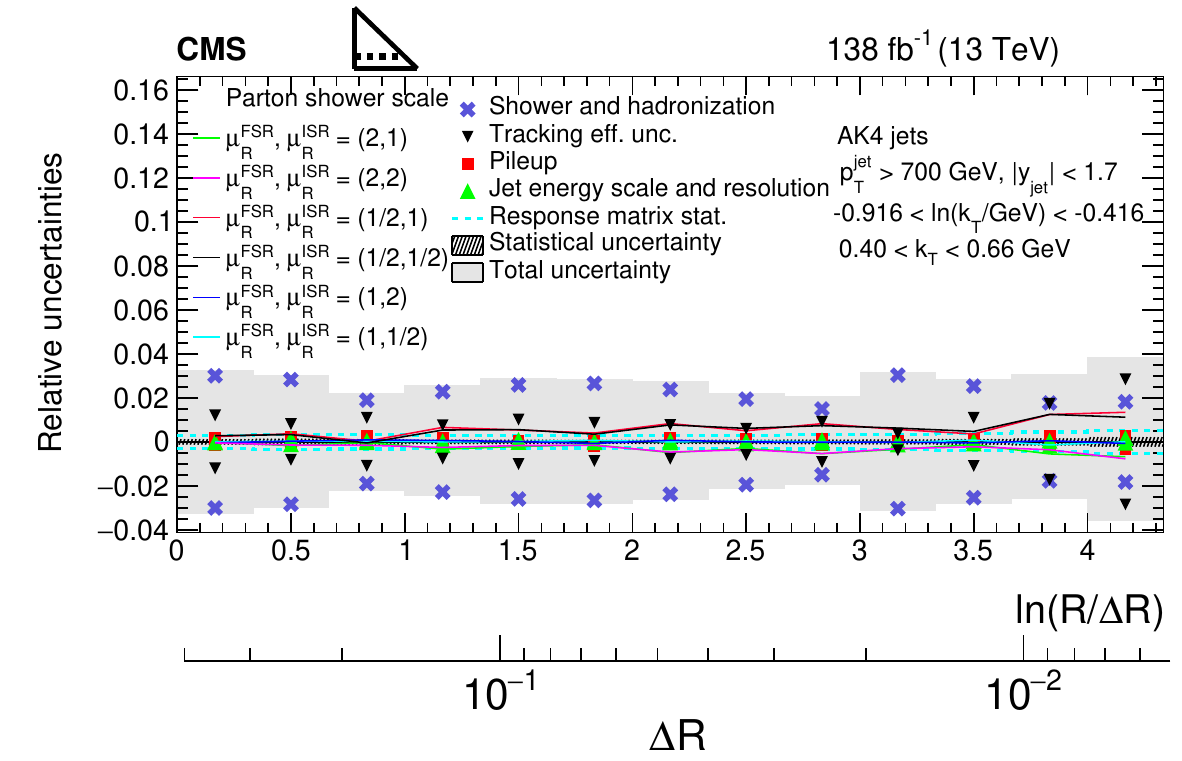}
  \includegraphics[width = 1.\textwidth]{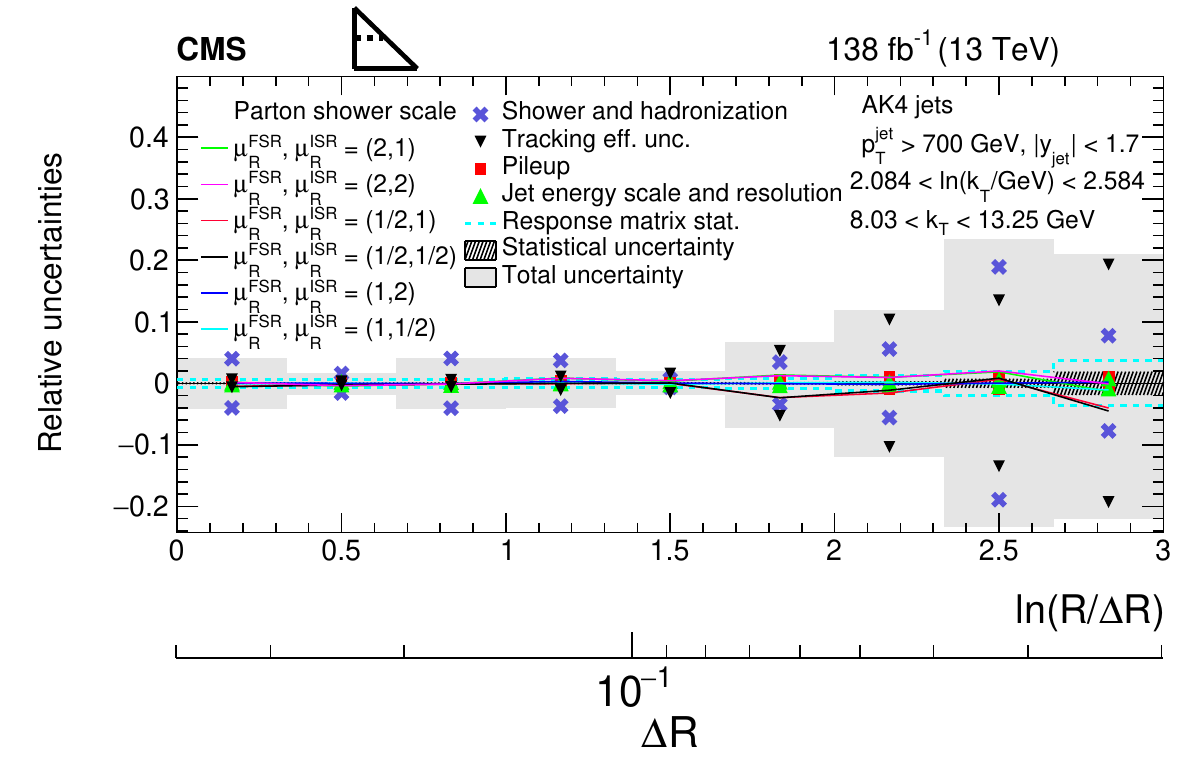}
  \caption{Different components of the systematic uncertainties for AK4 jets for different horizontal slices of the charged-particle LJP density. The upper plot is for low \kt of $1.09 < \kt < 1.79\GeV$, and the lower plot is for higher \kt of $8.03 < \kt < 13.25\GeV$. The total experimental uncertainty is represented by the filled area. The statistical uncertainties in the data are represented by the hashed band.}
  \label{fig:sysuncert_ak4_deltaR}
\end{figure}

Figures~\ref{fig:sysuncert_ak4} and \ref{fig:sysuncert_ak4_deltaR} show the different contributions of the systematic uncertainties in four different slices of the LJP for AK4 jets. The parton shower and hadronization uncertainties are the leading systematic uncertainties in the bulk of the LJP, which have typical values of 2--7\% throughout the LJP, reaching up to 20\% in certain bins close to the kinematical edge of the LJP. This systematic uncertainty reflects the difference in the detector response at the subjet level between \HERWIG{7} CH3 and \PYTHIA{8} CP5. The tracking efficiency uncertainty becomes more prominent as one approaches the LJP kinematical edge, reaching values of up to 25\%. Other contributions to the experimental uncertainties, which are due to the modeling of event-level properties in simulation or jet-level corrections and reconstruction, are less significant and typically below 1\% in the bulk of the LJP. Luminosity calibration uncertainties and PDF set uncertainties have a negligible effect for this measurement because of cancellations in the per-jet normalization of the average density of emissions.

\clearpage

\clearpage

\clearpage

\section{Results}\label{sec:results}

\begin{figure}
  \centering
  \includegraphics[width = 0.9\textwidth]{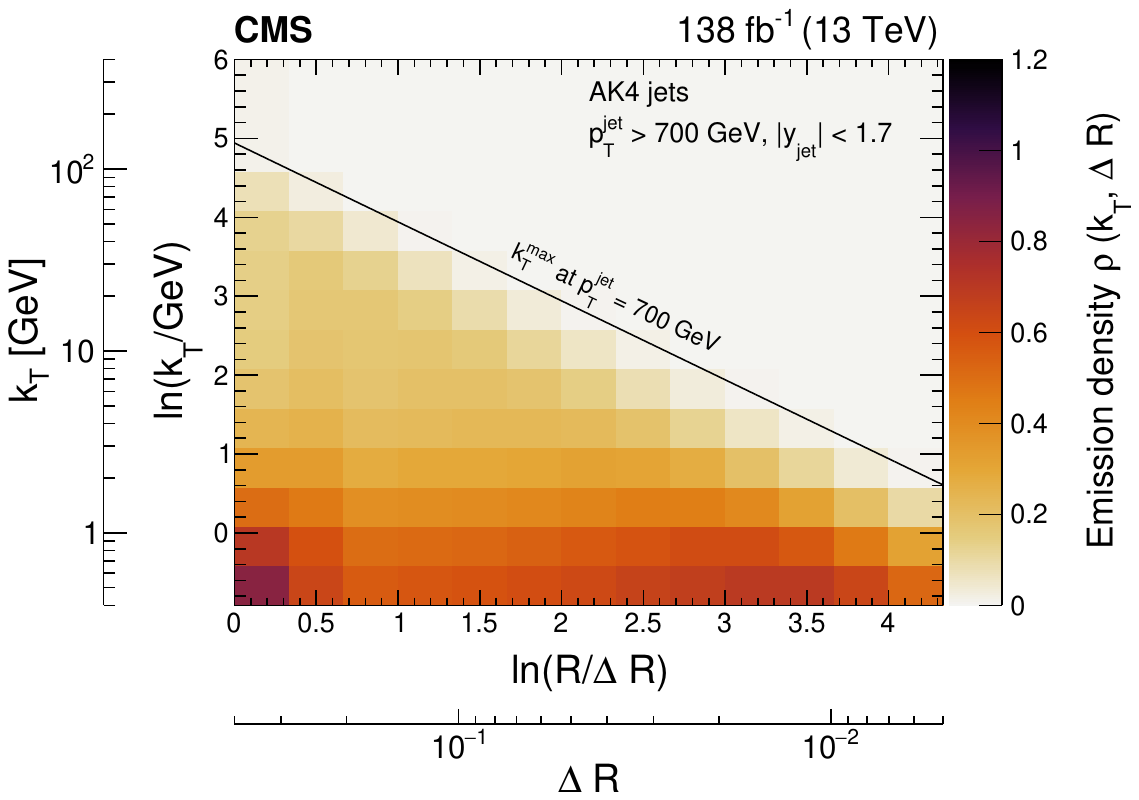}
  \includegraphics[width = 0.9\textwidth]{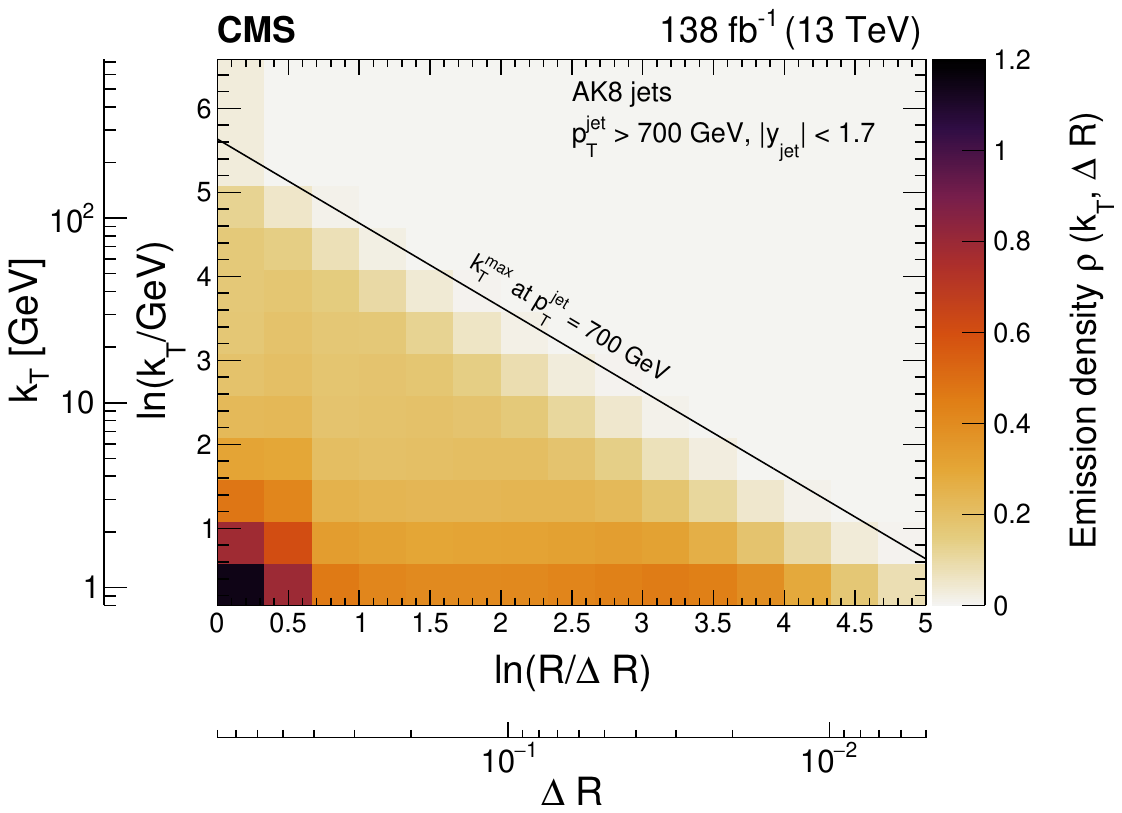}
  \caption{Two-dimensional distributions of the charged-particle primary LJP densities corrected to particle level for AK4 jets (upper plot) and AK8 jets (lower plot). The diagonal line in both plots represents the kinematical limit of the emissions for a jet with $\pt^\text{jet} = 700\GeV$. \label{fig:fullycorrected_2dplane}}
\end{figure}

\begin{figure}
  \centering
  \includegraphics[width=0.49\textwidth]{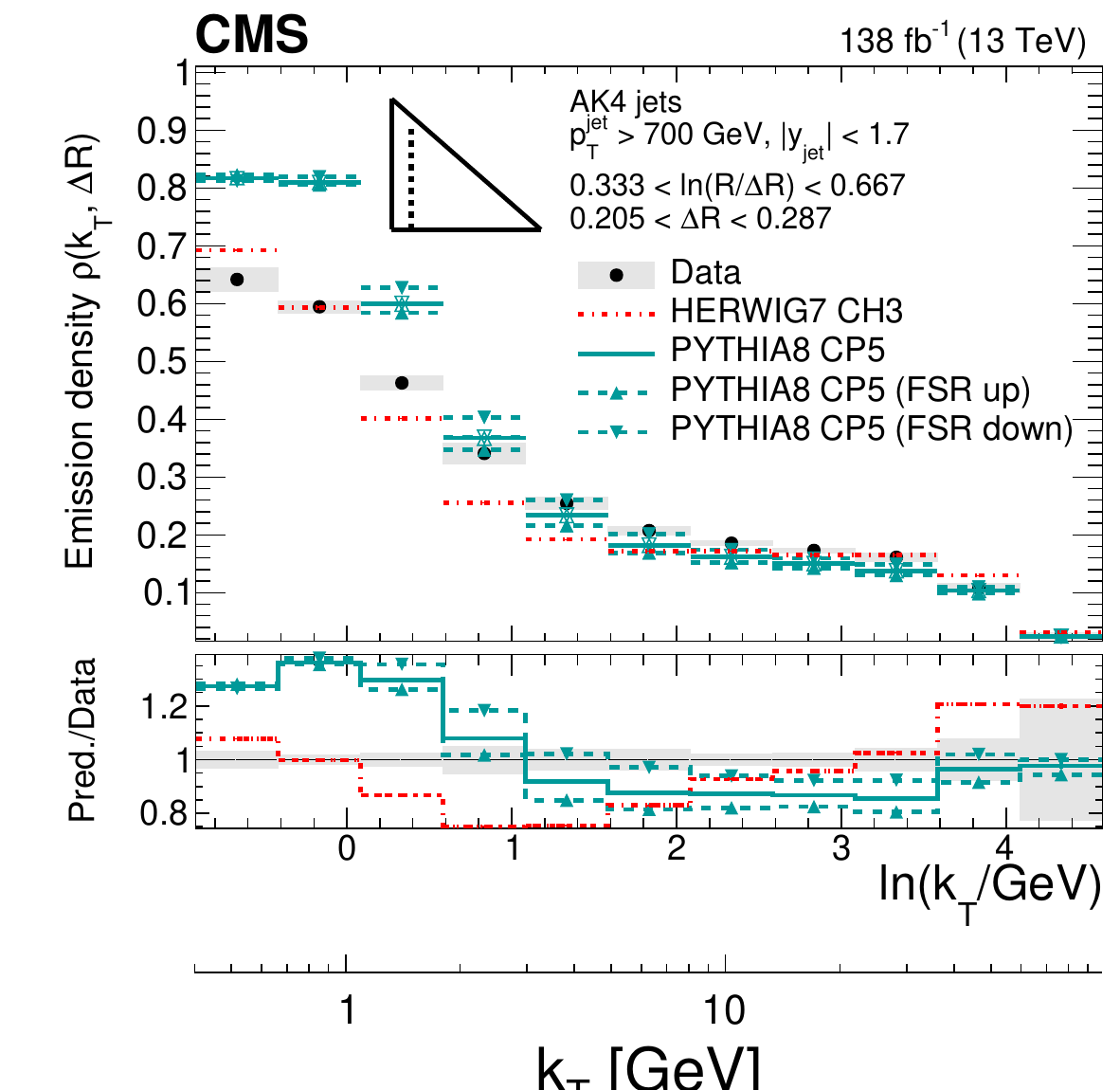}
  \includegraphics[width=0.49\textwidth]{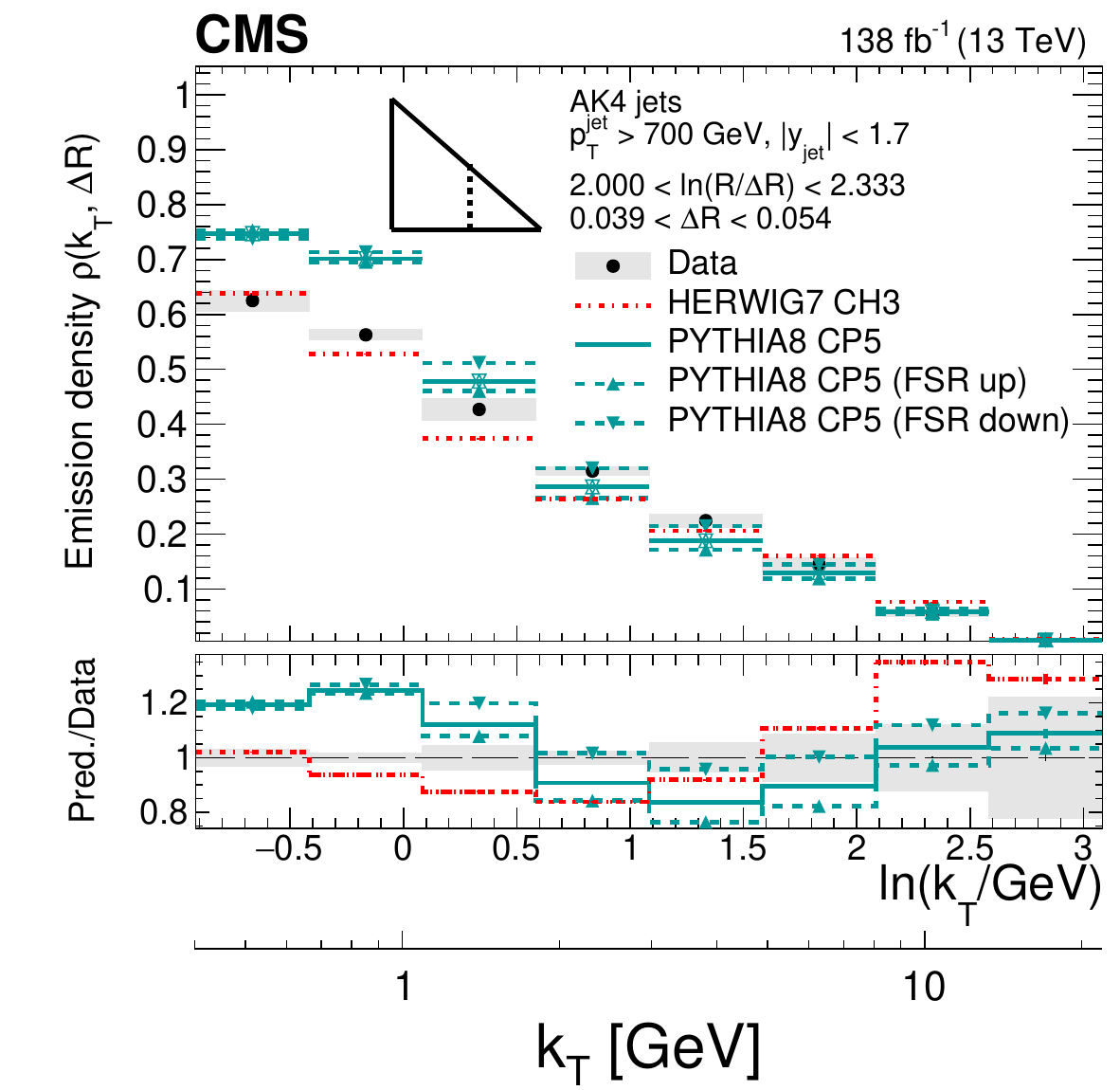}
  \includegraphics[width=0.49\textwidth]{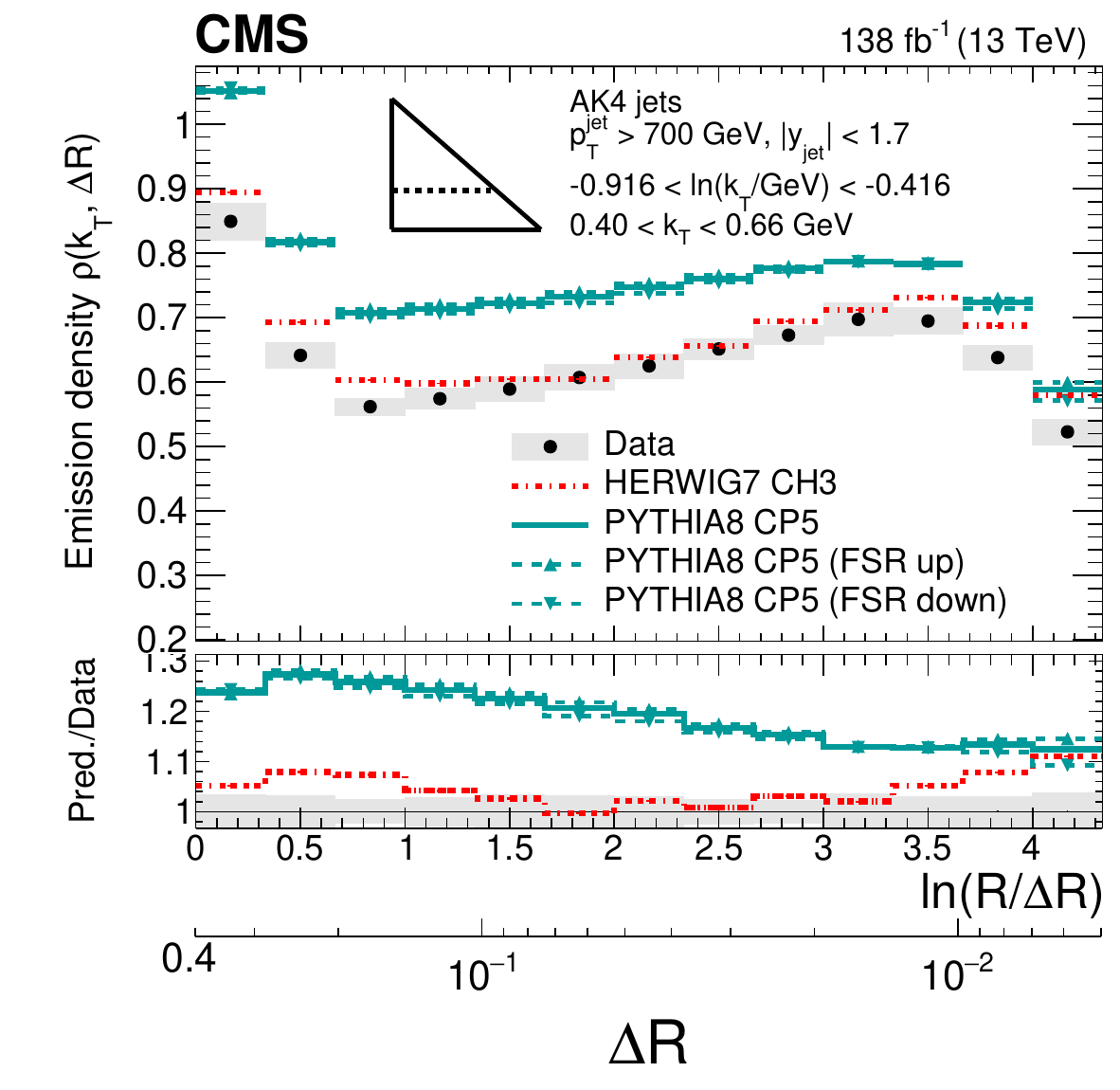}
  \includegraphics[width=0.49\textwidth]{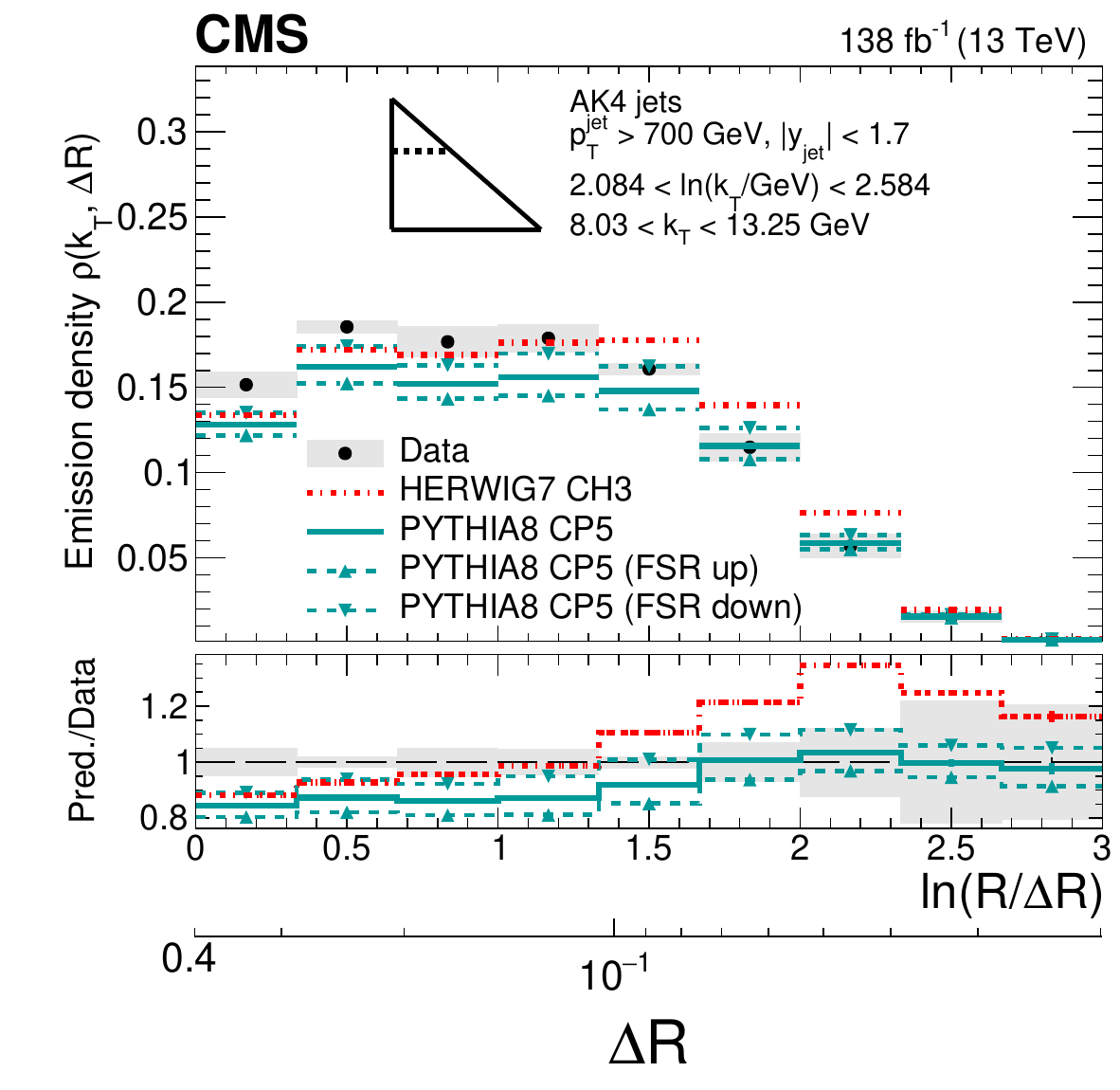}
  \caption{Four slices of the charged-particle primary LJP density of AK4 jets compared with predictions by \PYTHIA{8} CP5 and \HERWIG{7} CH3. Variations of the ISR and FSR scales for \PYTHIA{8} CP5 predictions are shown as well. The band represents the total experimental uncertainty. The upper two plots correspond to vertical slices for fixed \lnDeltaR (large angles on upper-left, small angles on upper-right). The lower two plots correspond to two different horizontal slices for fixed \lnkT: the lower-left plot contains low-\kt splittings, whereas the lower-right plot contains high-\kt splittings, which populate mostly wide-angle radiation. \label{fig:LPslicesak4}}
\end{figure}

\begin{figure}
  \centering
  \includegraphics[width=0.49\textwidth]{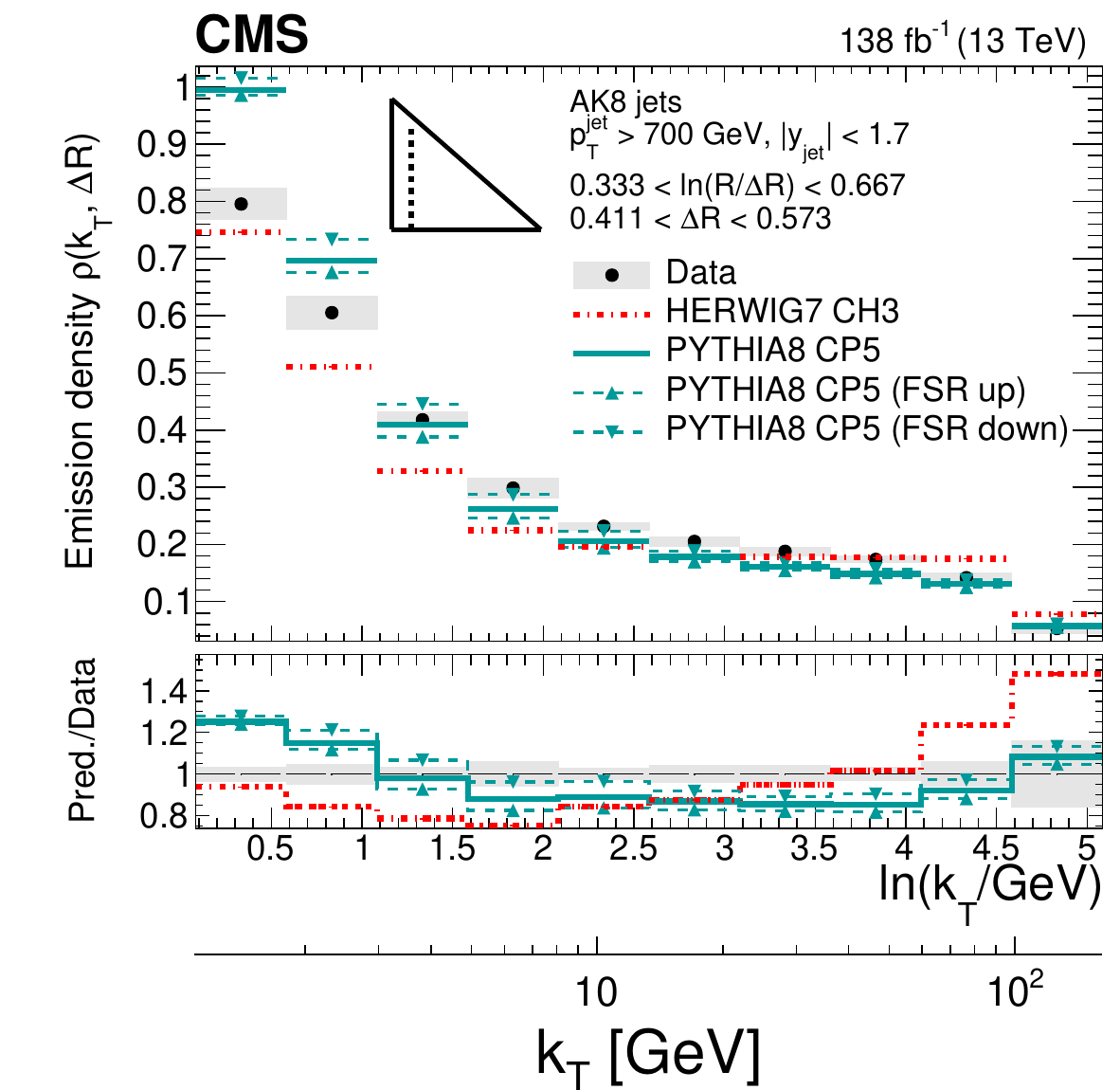}
  \includegraphics[width=0.49\textwidth]{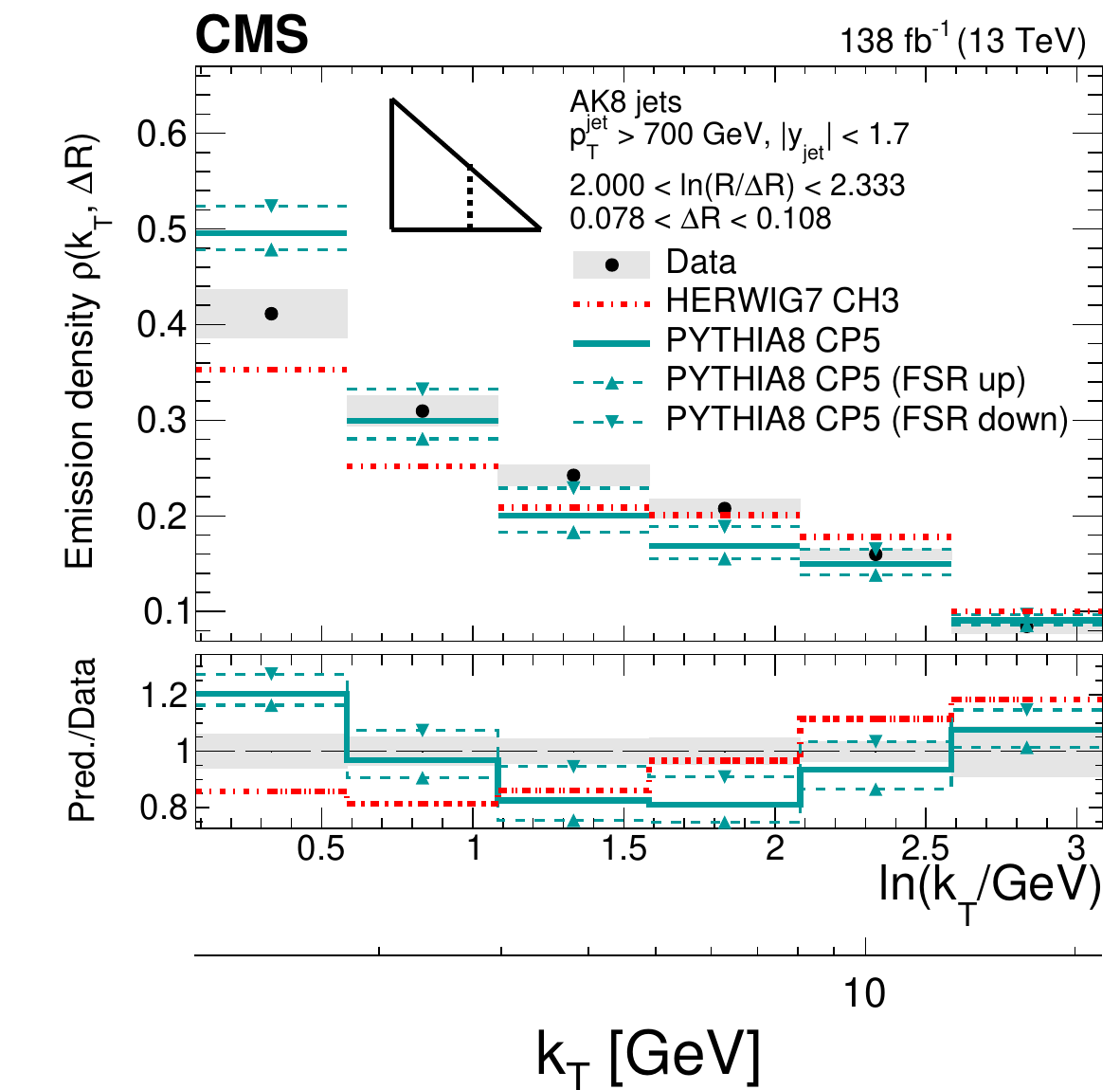}
  \includegraphics[width=0.49\textwidth]{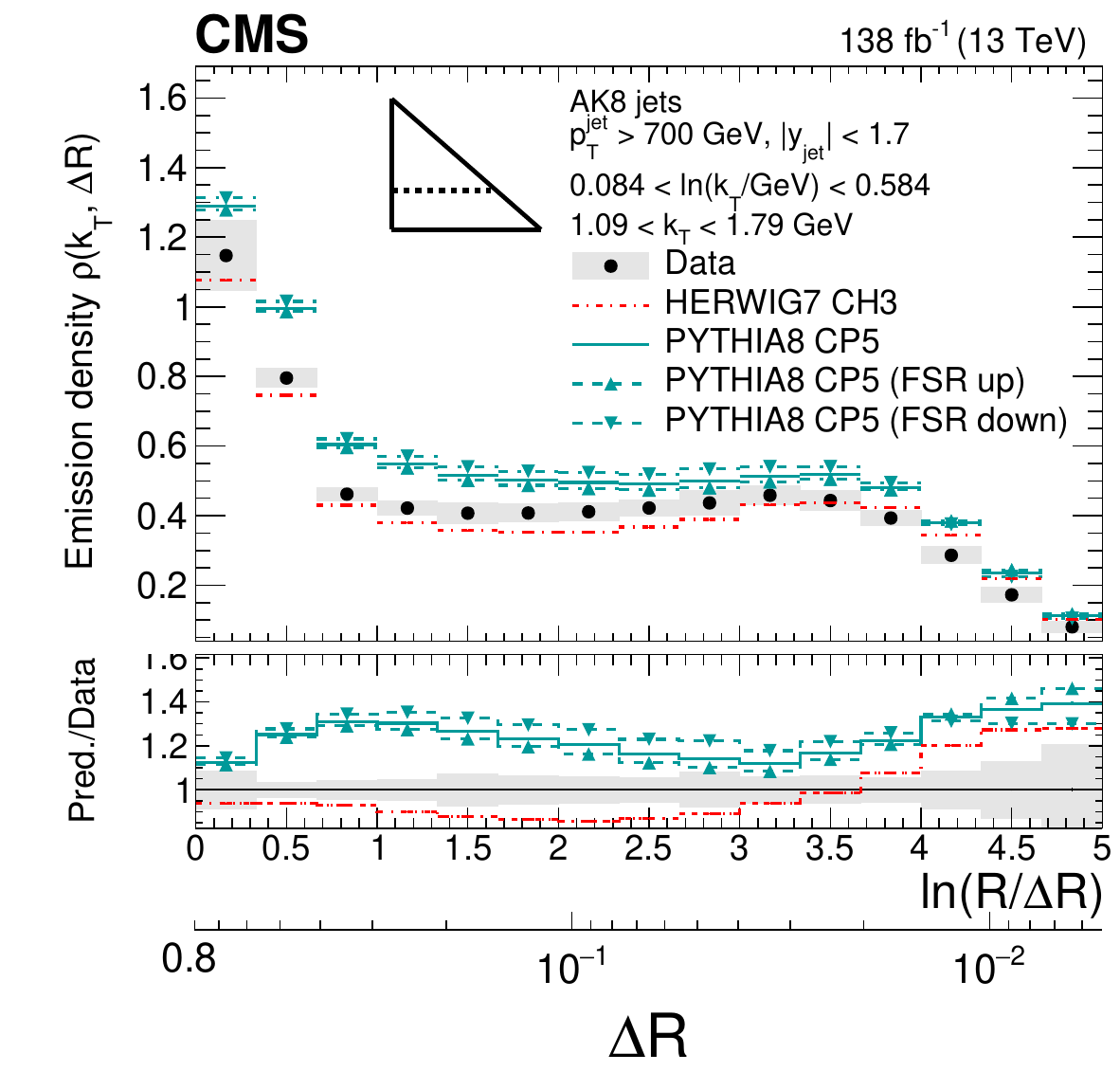}
  \includegraphics[width=0.49\textwidth]{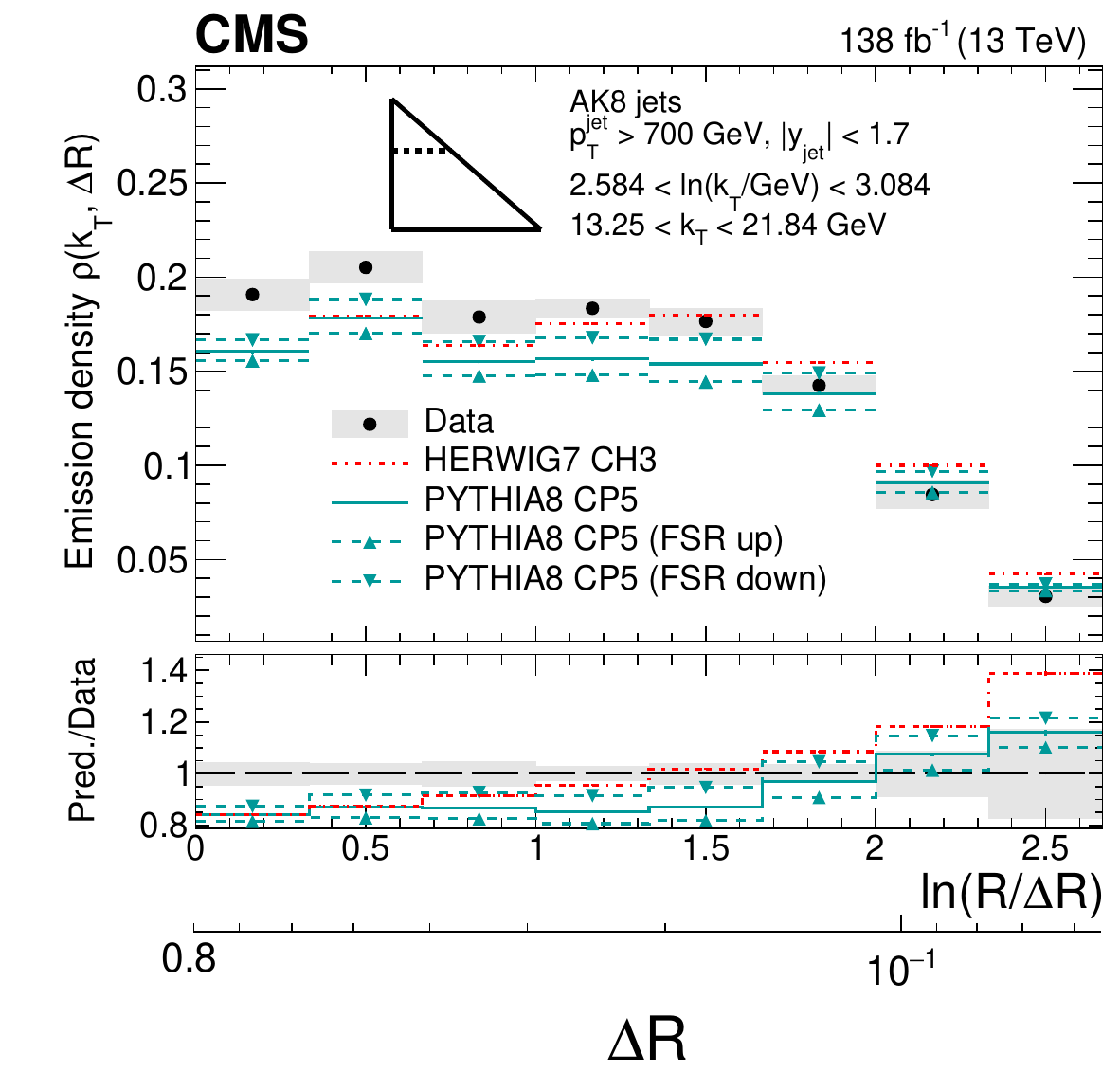}
  \caption{Four slices of the charged-particle primary LJP density of AK8 jets compared with predictions by \PYTHIA{8} CP5 and \HERWIG{7} CH3. Variations of the ISR and FSR scales for \PYTHIA{8} CP5 predictions are shown as well. The band represents the total experimental uncertainty. The upper two plots correspond to vertical slices for fixed \lnDeltaR (large angles on upper-left, small angles on upper-right). The lower two plots correspond to two different horizontal slices for fixed \lnkT: the lower-left plot corresponds to low-\kt splittings and spans the full range in \lnDeltaR, whereas the lower-right plot corresponds to high-\kt splittings, which populate mostly wide-angle radiation.}
  \label{fig:LPslicesak8}
\end{figure}

\begin{figure}
  \centering
  \includegraphics[width=0.49\textwidth]{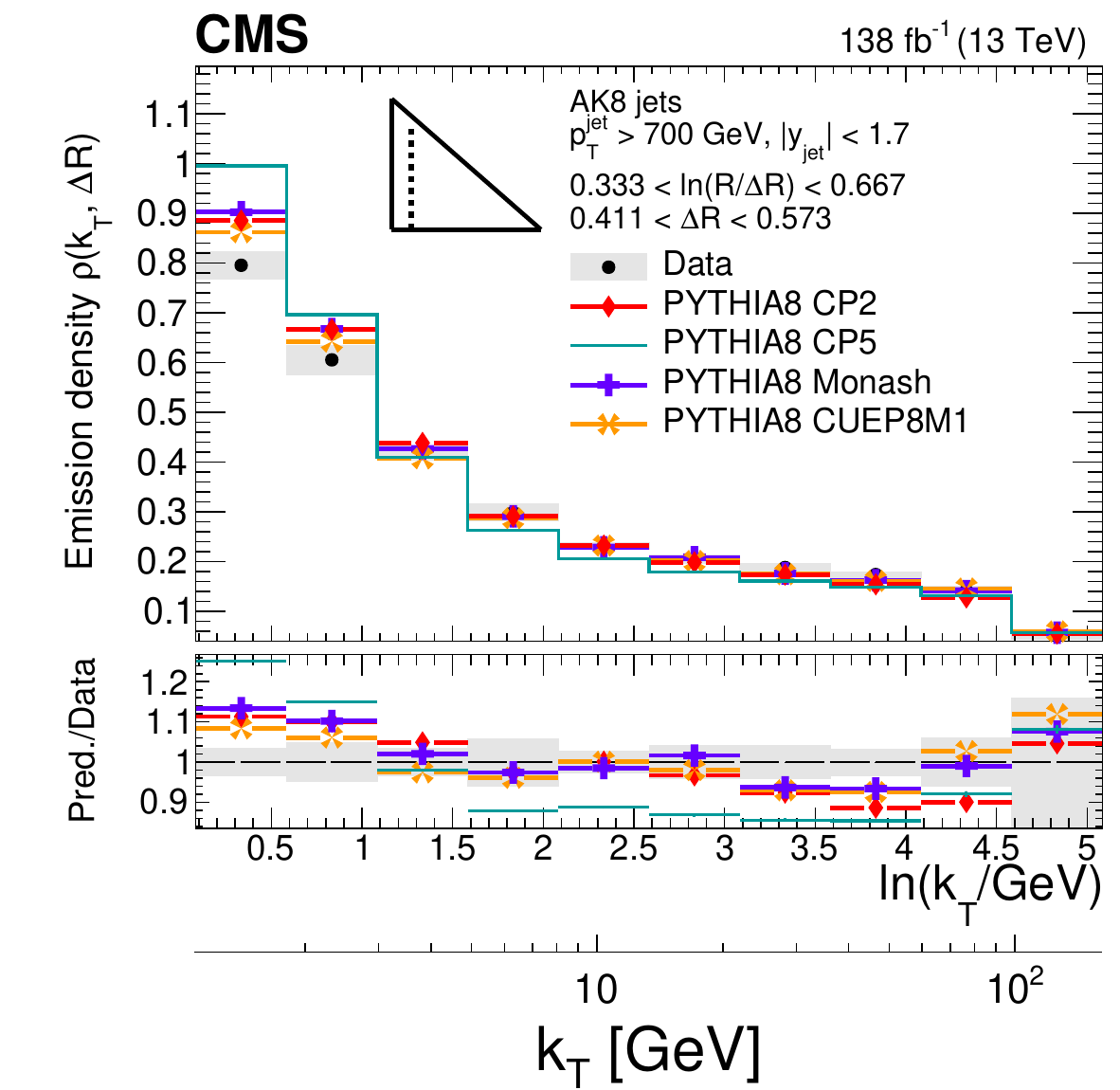}
  \includegraphics[width=0.49\textwidth]{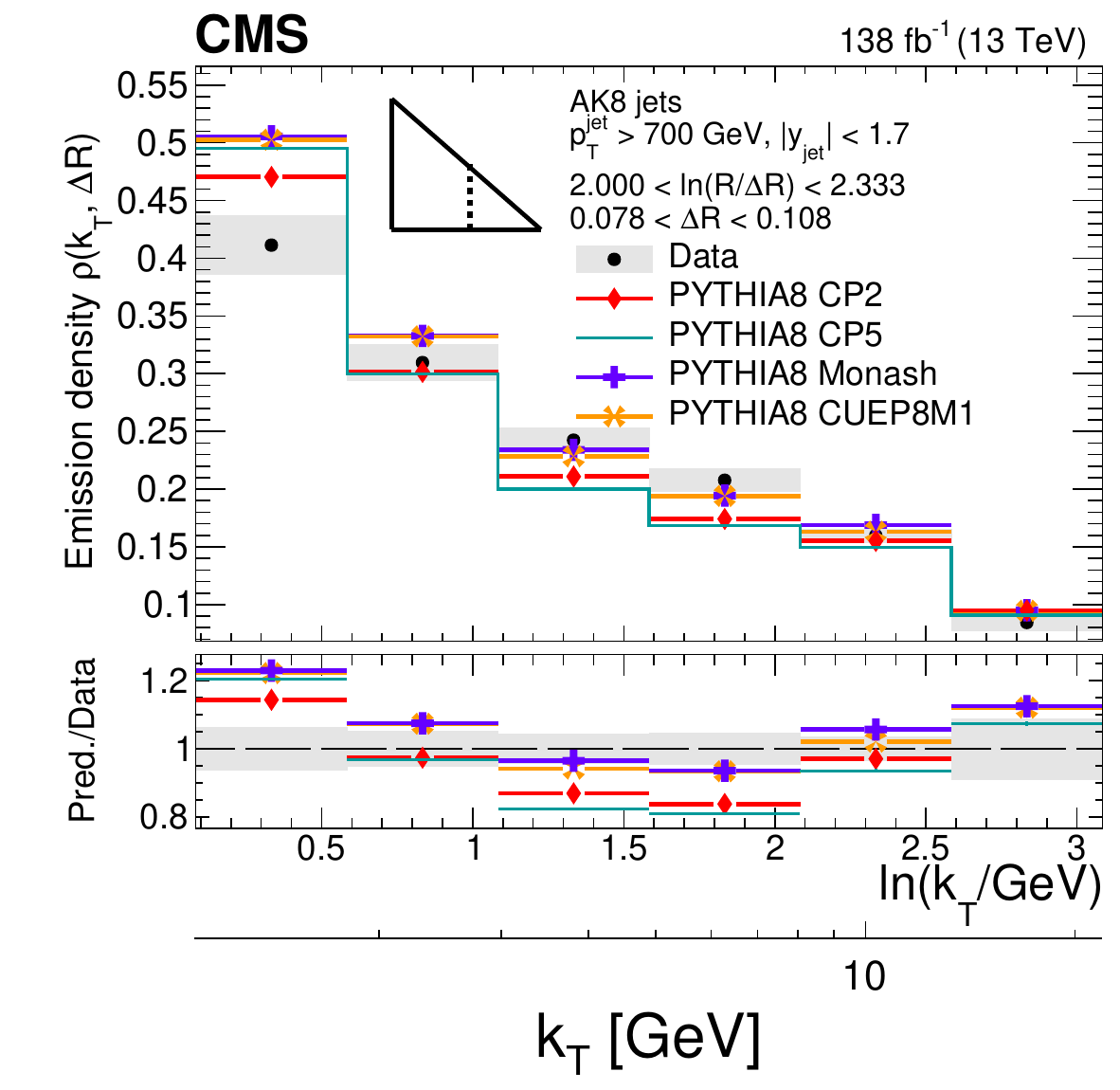}
  \includegraphics[width=0.49\textwidth]{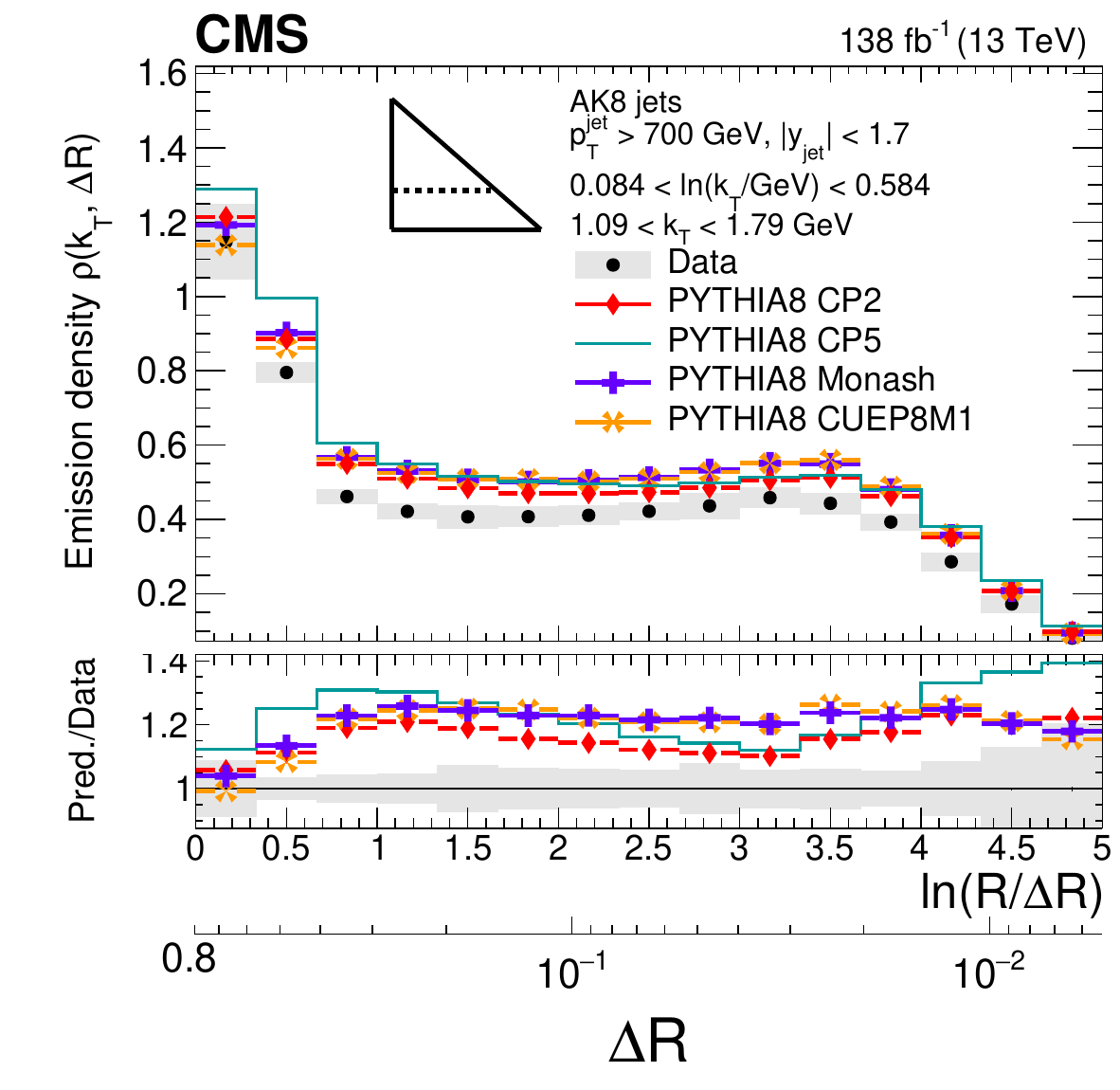}
  \includegraphics[width=0.49\textwidth]{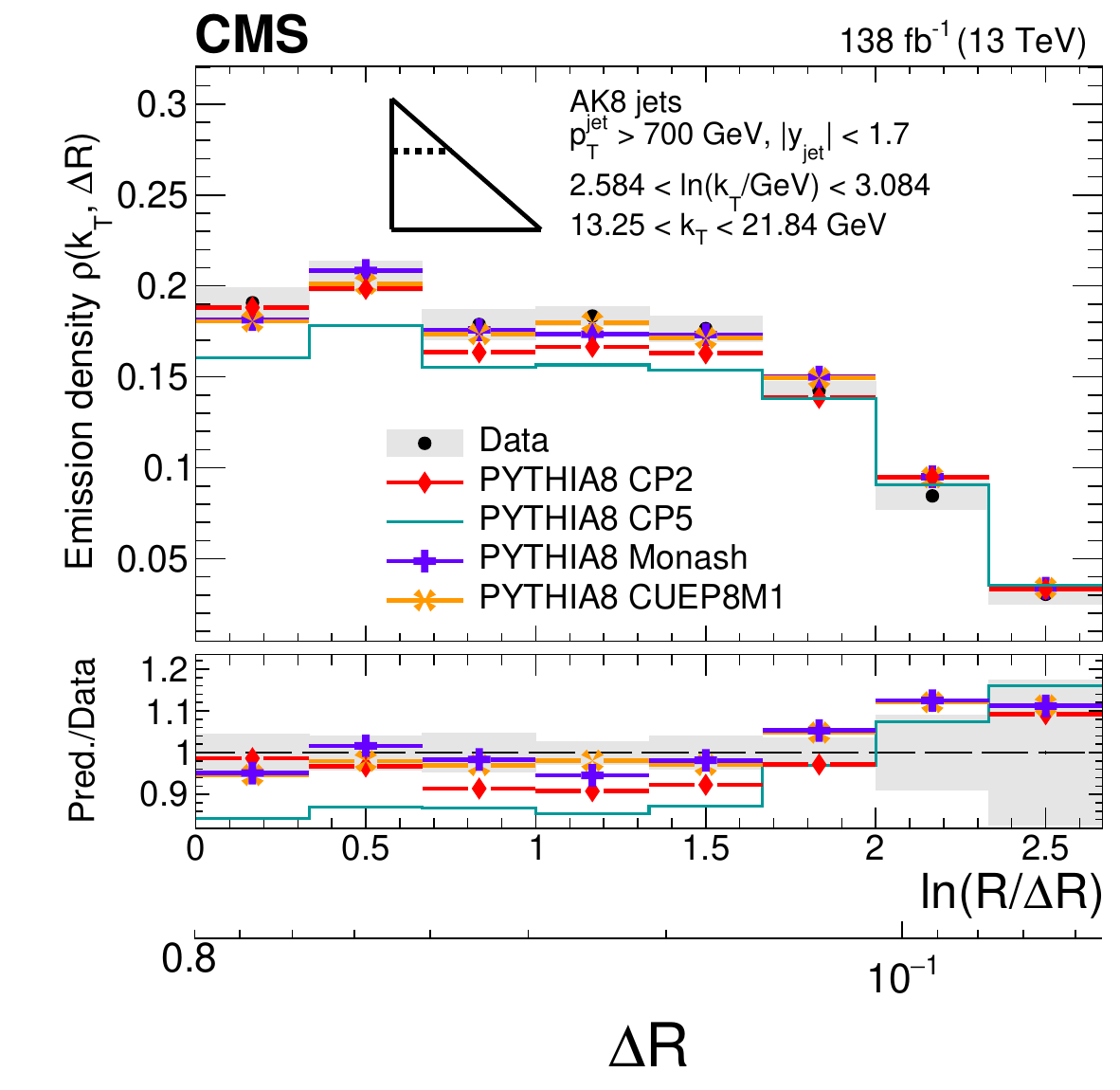}
  \caption{Four different slices of the charged-particle primary LJP density of AK8 jets compared with predictions generated with \PYTHIA{8} using tunes CP2, CP5, Monash, and CUEP8M1. The most important difference between the tunes is the value of \alpSFSR, as described in the text. The band represents the total experimental uncertainty. The upper two plots correspond to vertical slices of the LJP for fixed \lnDeltaR (large angles on upper-left, small angles on upper-right). The lower two plots correspond to two different horizontal slices for fixed \lnkT: the lower-left plot corresponds to low-\kt splittings and spans the full range in \lnDeltaR, whereas the lower-right plot corresponds to high-\kt splittings, which populate mostly wide-angle radiation. Statistical uncertainties in data and MC-simulated events are represented by vertical bars, which are smaller than the markers in most of the bins.}
  \label{fig:LPslicesak8_pythia8tunes}
\end{figure}

\begin{figure}
  \centering
  \includegraphics[width=0.49\textwidth]{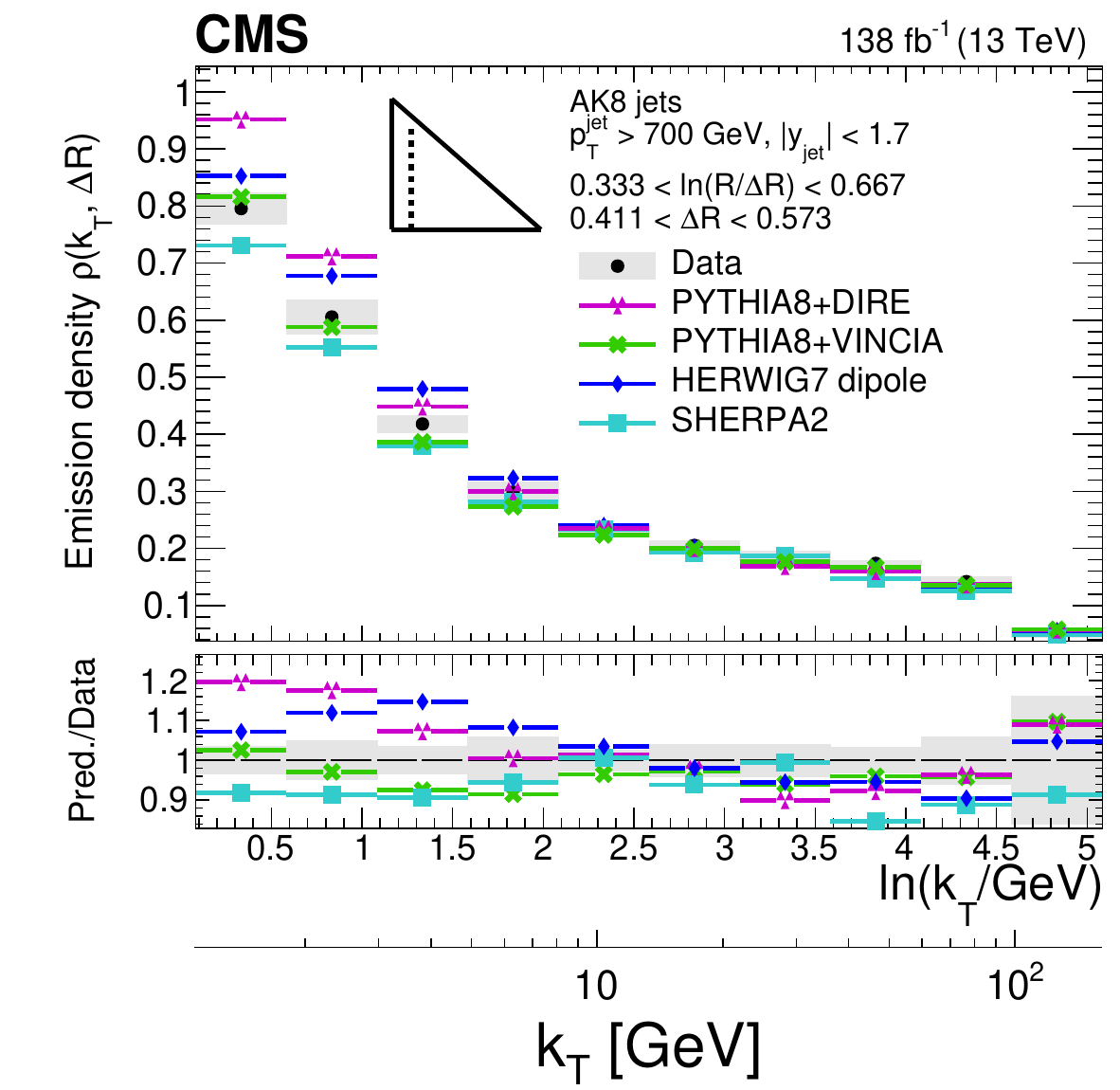}
  \includegraphics[width=0.49\textwidth]{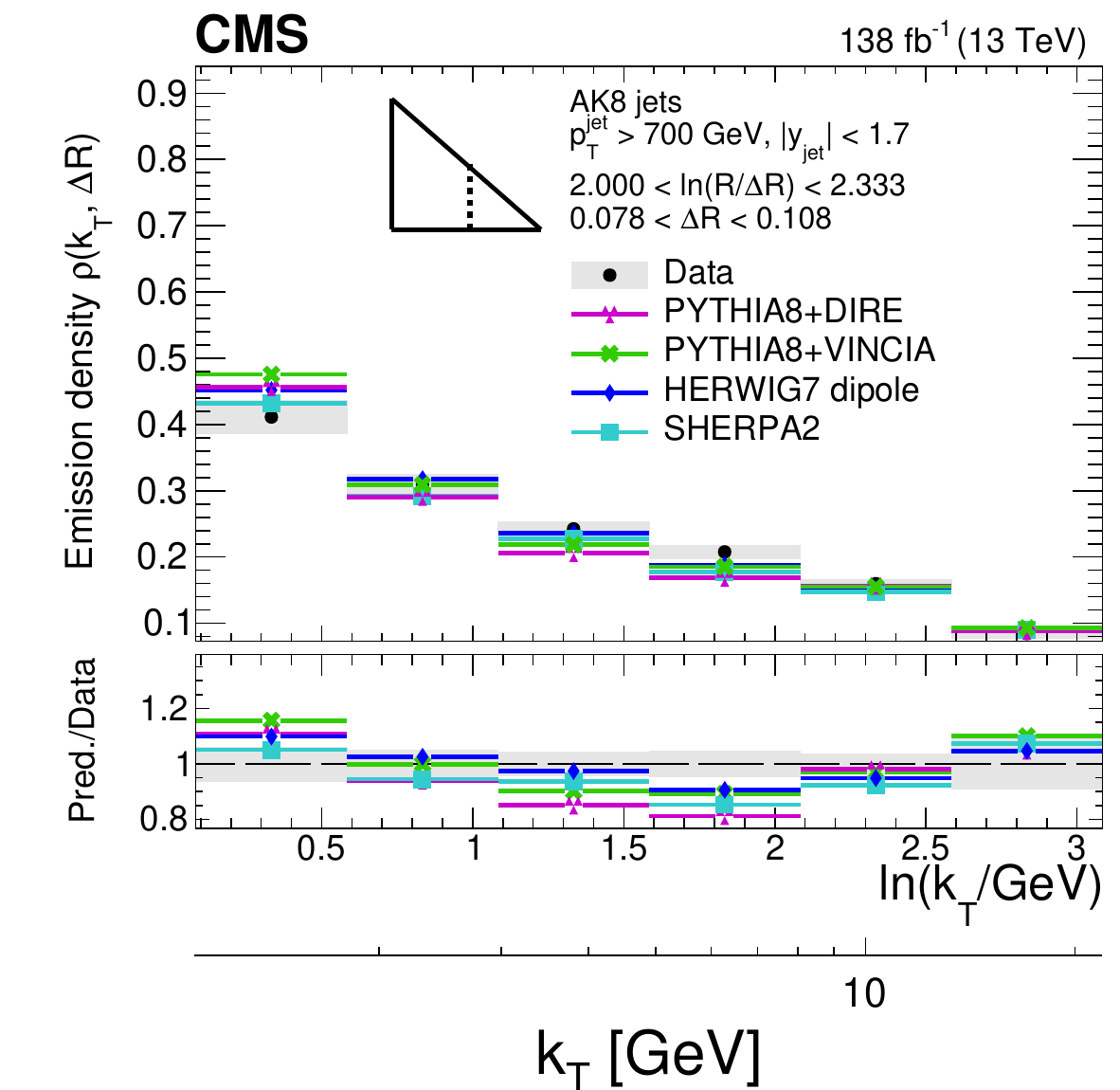}
  \includegraphics[width=0.49\textwidth]{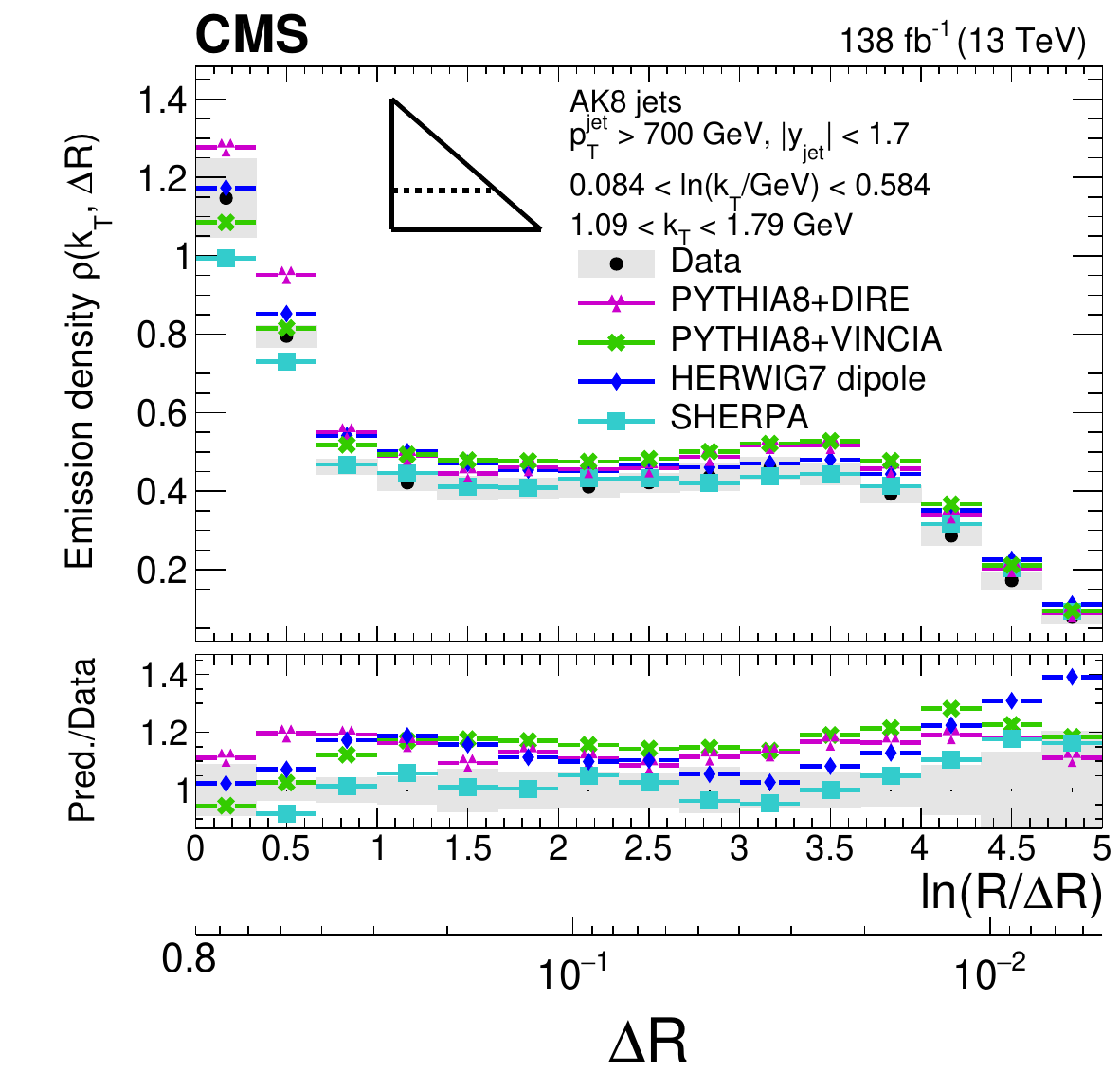}
  \includegraphics[width=0.49\textwidth]{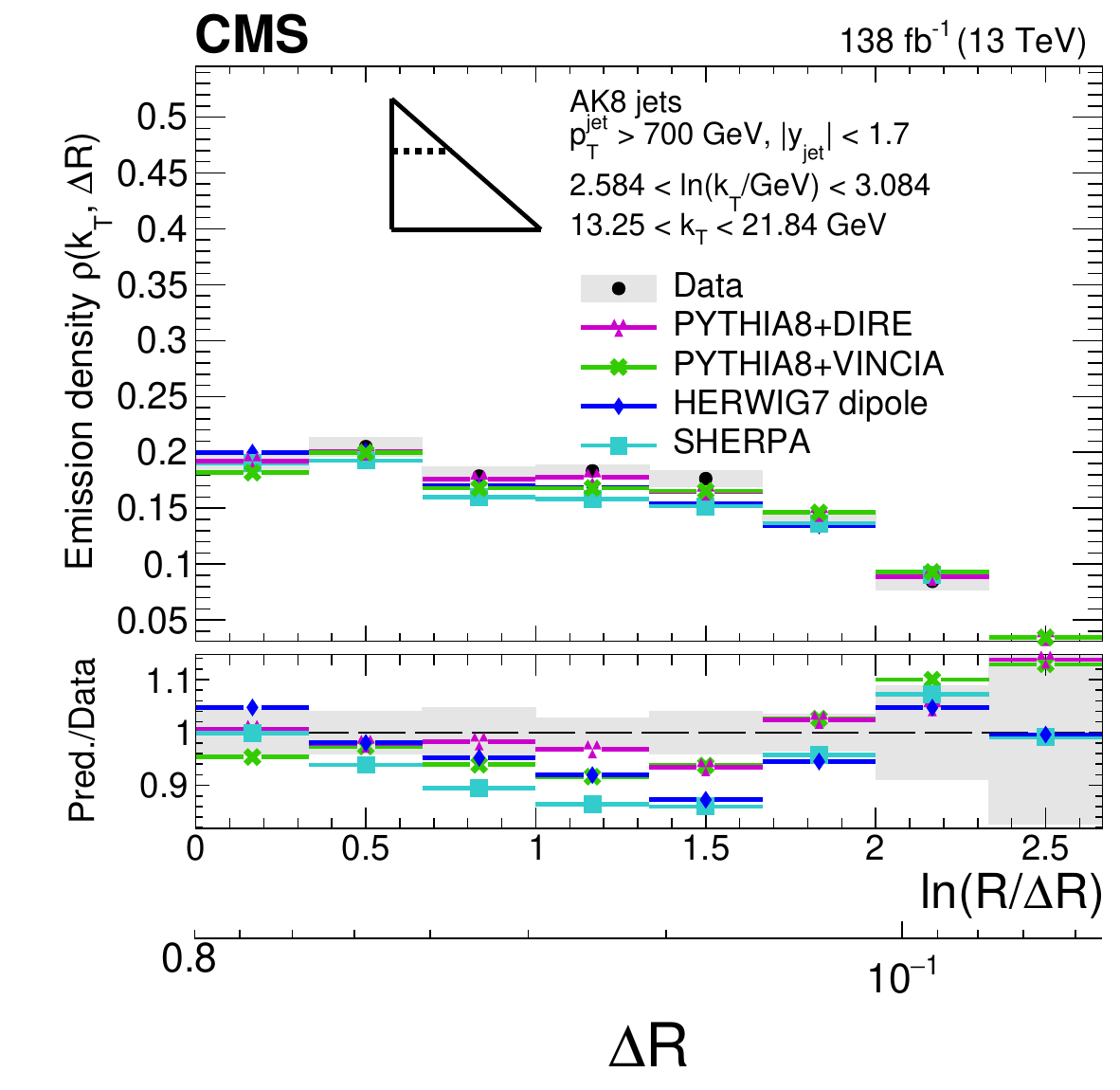}
  \caption{Four different slices of the charged-particle primary LJP density of AK8 jets compared with predictions by \PYTHIA{8}+\VINCIA, \PYTHIA{8}+\DIRE, \HERWIG{7} with dipole shower, and \SHERPA{2}. The band represents the total experimental uncertainty. The upper two plots correspond to vertical slices of the LJP for fixed \lnDeltaR (large angles on upper-left, small angles on upper-right). The lower two plots correspond to two different horizontal slices for fixed \lnkT: the lower-left plot corresponds to low-\kt splittings and spans the full range in \lnDeltaR, whereas the lower-right plot corresponds to high-\kt splittings, which populate mostly wide-angle radiation. Statistical uncertainties in data and MC-simulated events are represented by vertical bars, which are smaller than the markers in most of the bins.}
  \label{fig:LPslicesak8_dipole}
\end{figure}

The unfolded primary LJP densities are presented in Fig.~\ref{fig:fullycorrected_2dplane} for AK4 and AK8 jets. The primary LJP of AK8 jets give access to emissions at large \DeltaR and high \kt. In these two-dimensional plots, one can readily see the plateauing of the emission density at high \kt, as well as the fast growth of emissions at low \kt, as expected from the dependence of the density of emissions with \alpS. The angular region, where the density of emissions plateaus, is much broader for AK8 jets (starting at about $\DeltaR < 0.4$), since clustering effects are mitigated due to the larger distance parameter $R$~\cite{Lifson:2020gua}.

In Figs. \ref{fig:LPslicesak4} and \ref{fig:LPslicesak8}, we project the primary LJP density distribution onto the \lnkT axis in a window of large-angle splittings (upper-left plot) and in a window of small-angle splittings (upper-right plot). We also project the LJP onto the \lnDeltaR axis in a window of low-\kt splittings (lower-left plot) and in a window of high-\kt splittings (lower-right plot). The low-\kt splittings populate a wide range in \DeltaR, whereas the hard splittings populate mostly the wide-angle radiation region. The shapes of the distributions are similar for AK4 and AK8 jets except that the soft and large-angle splittings are more abundant for AK8, which is because of the larger contribution of the UE due to the larger \DeltaR interval for the first bin in \lnDeltaR.

The unfolded distributions are compared with a number of MC event generator predictions at particle level. The MC generator predictions differ in their implementation of color coherence in the parton showers as well as their logarithmic accuracy in different regions of the LJP. There are differences also in the modeling of the UE activity and hadronization effects. First, we discuss the comparison with the predictions based on events generated with \HERWIG{7} CH3 and \PYTHIA{8} CP5, including the FSR scale variations of the latter. Figure ~\ref{fig:LPslicesak4} (lower-left) shows that \PYTHIA{8} CP5 overestimates the number of emissions relative to data by approximately 15--20\%, whereas \HERWIG{7} CH3 is in better agreement with the data within 5--10\%. For \PYTHIA{8} CP5, we show also the predictions for different renormalization scale variations. The FSR up and down scale variations generate a theoretical uncertainty band of about 10\% for emissions in the perturbative region. The sensitivity of the distributions to variations of the renormalization scale in the FSR shower can be understood from the linear dependence of the LJP density on \alpSFSR. This uncertainty band shrinks monotonically at low \kt, where nonperturbative effects are more important and are effectively decoupled from the hard perturbative shower. In the region dominated by the parton shower, the predictions with the FSR down variation are in better agreement with the measured data. This is equivalent to choosing a larger value of \alpSFSR. The improvement in the description of jet substructure data with a larger value of \alpSFSR has been observed in other jet substructure measurements~\cite{CMS:2022kqg,ATLAS:2023jdw}. The correlated variations of the FSR and ISR scales have the same effect as the variations of the FSR scales alone. The predictions using the ISR up and down variations are consistent with the predictions calculated with the nominal scale values.

The predictions based on \PYTHIA{8} with the Monash, CUEP8M1, and CP2 tunes shown in Fig.~\ref{fig:LPslicesak8_pythia8tunes} are in better agreement with the data than the predictions from \PYTHIA{8} CP5 in the bulk of the LJP. The main difference of these tunes with respect to the CP5 tune is the larger value of \alpSFSR ($\alpSFSR = 0.1365$ for the Monash and CUEP8M1 tune and 0.130 for CP2). This is consistent with the better description of \PYTHIA{8} CP5 with the FSR down variation mentioned in the previous paragraph. The CP5 tune was introduced to have an improved description of data with high jet multiplicities relative to the other \PYTHIA{8} CP tunes while potentially compromising the mismodeling of some jet substructure observables~\cite{ CMS:2019csb}. The difference between the predictions generated with the CP tunes of \PYTHIA{8} and the previous CUEP8M1 and Monash tunes highlight the complementarity of jet substructure data relative to the minimum bias data that is typically used for tuning. Figure~\ref{fig:LPslicesak8_dipole} shows a comparison of the data to predictions based on different dipole shower implementations. The \PYTHIA{8}+\DIRE and \PYTHIA{8}+\VINCIA predictions tend to overestimate the density of emissions at low \kt by about 10--20\%. \HERWIG{7} with its dipole shower overestimates the density of emissions for collinear emissions at low \kt by about 30--40\%, whereas \SHERPA{2} is able to describe most of the \DeltaR range at low \kt. The \PYTHIA{8}+\VINCIA, \SHERPA{2}, and \HERWIG{7} dipole shower predictions underestimate the density of emissions at large \kt and $\DeltaR \sim 0.2$--$0.4$ by about 10--15\%, whereas \PYTHIA{8}+\DIRE performs better in that region. For hard and collinear emissions, the dipole showers generally underestimate the density of emissions for \kt of about $5\GeV$.

\begin{figure}
  \centering
  \includegraphics[width=0.49\textwidth]{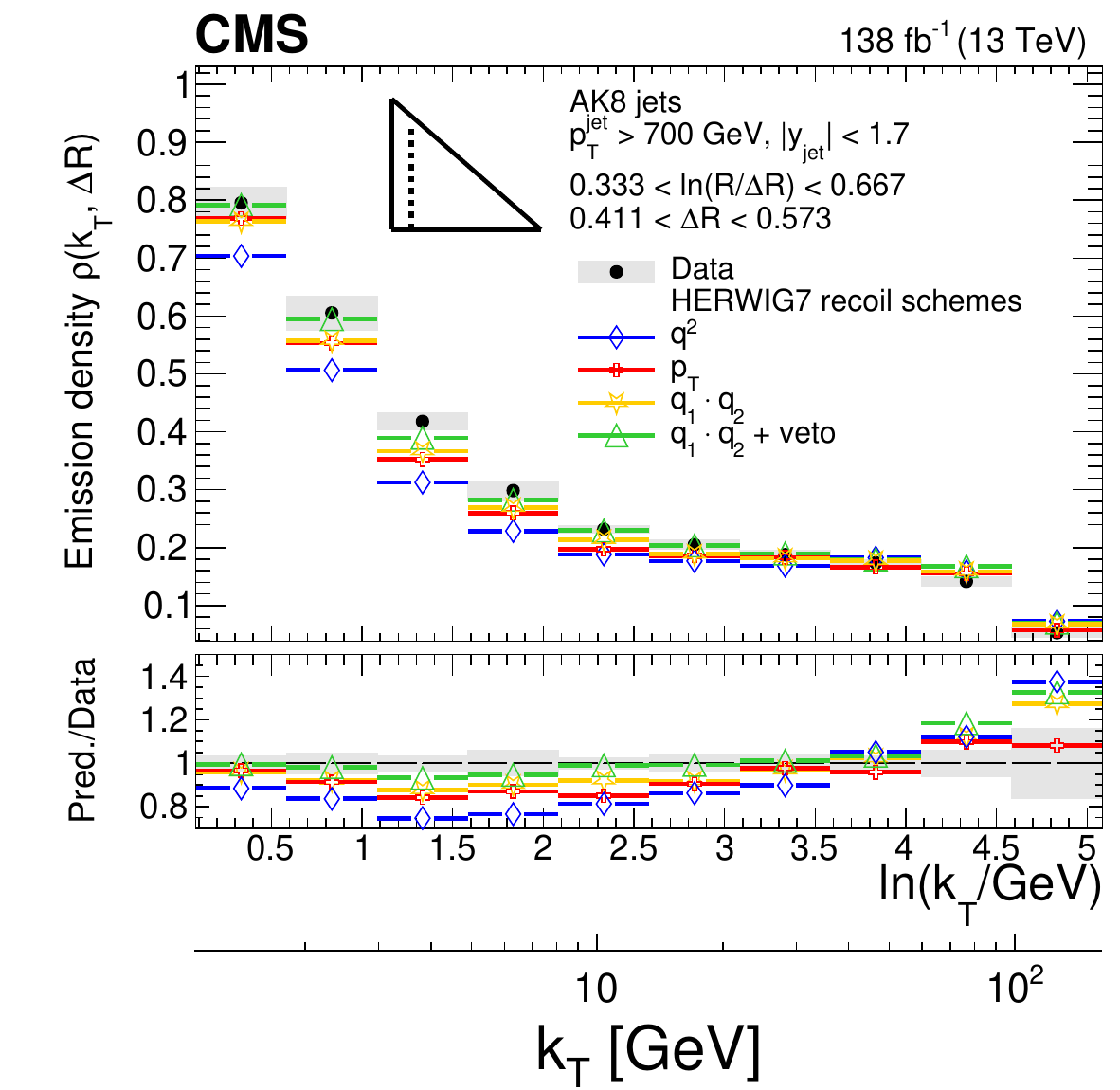}
  \includegraphics[width=0.49\textwidth]{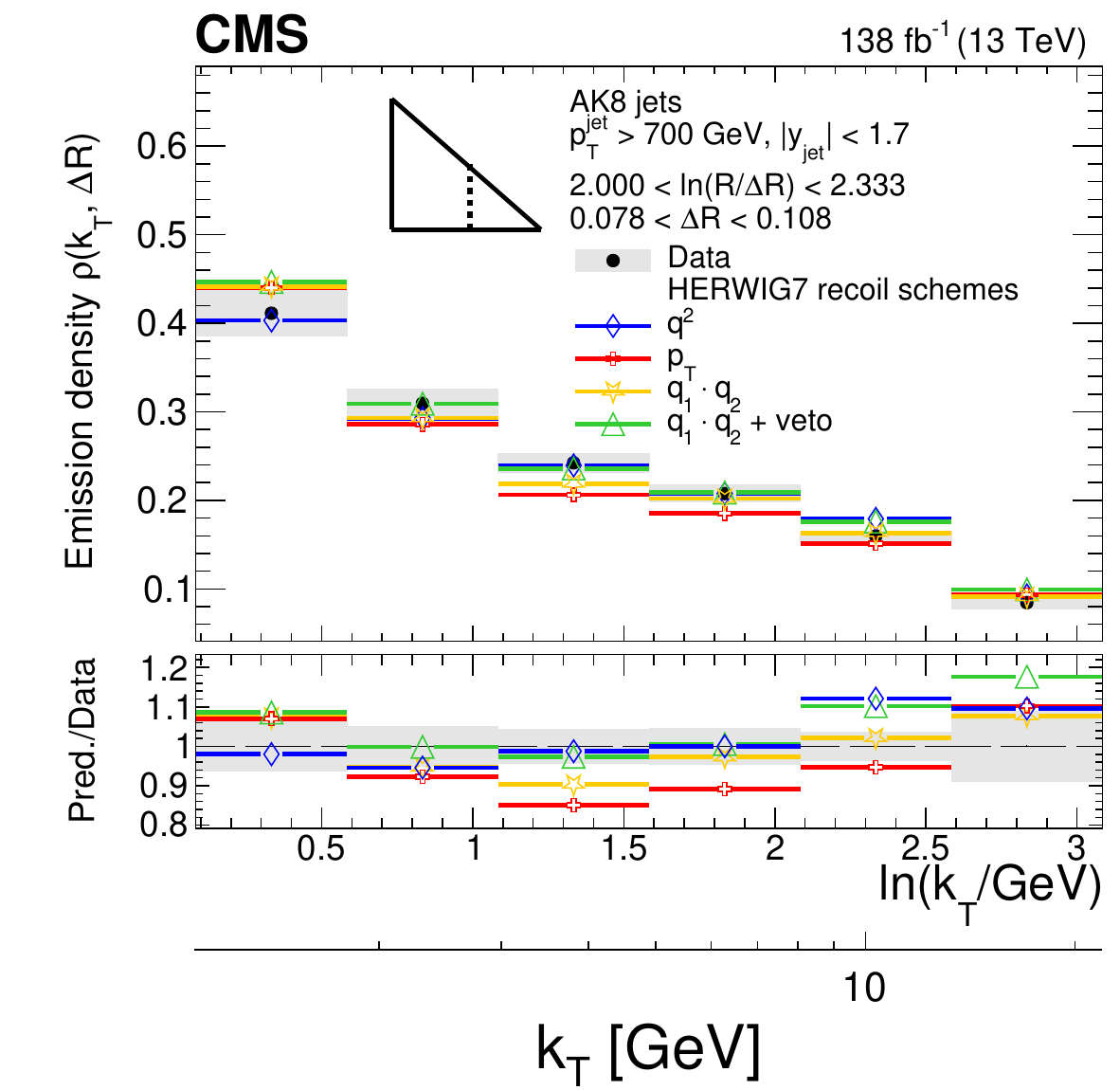}
  \includegraphics[width=0.49\textwidth]{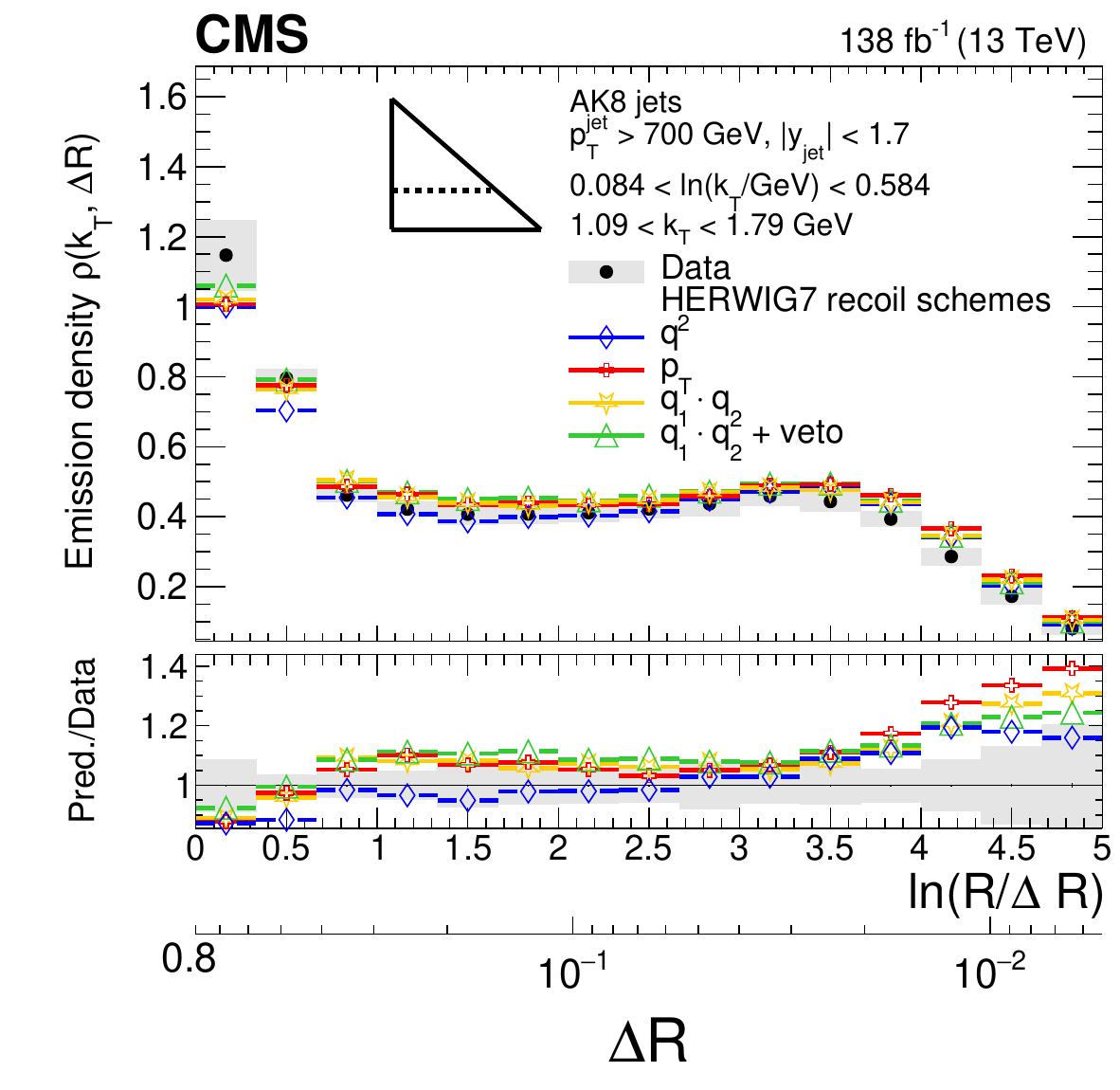}
  \includegraphics[width=0.49\textwidth]{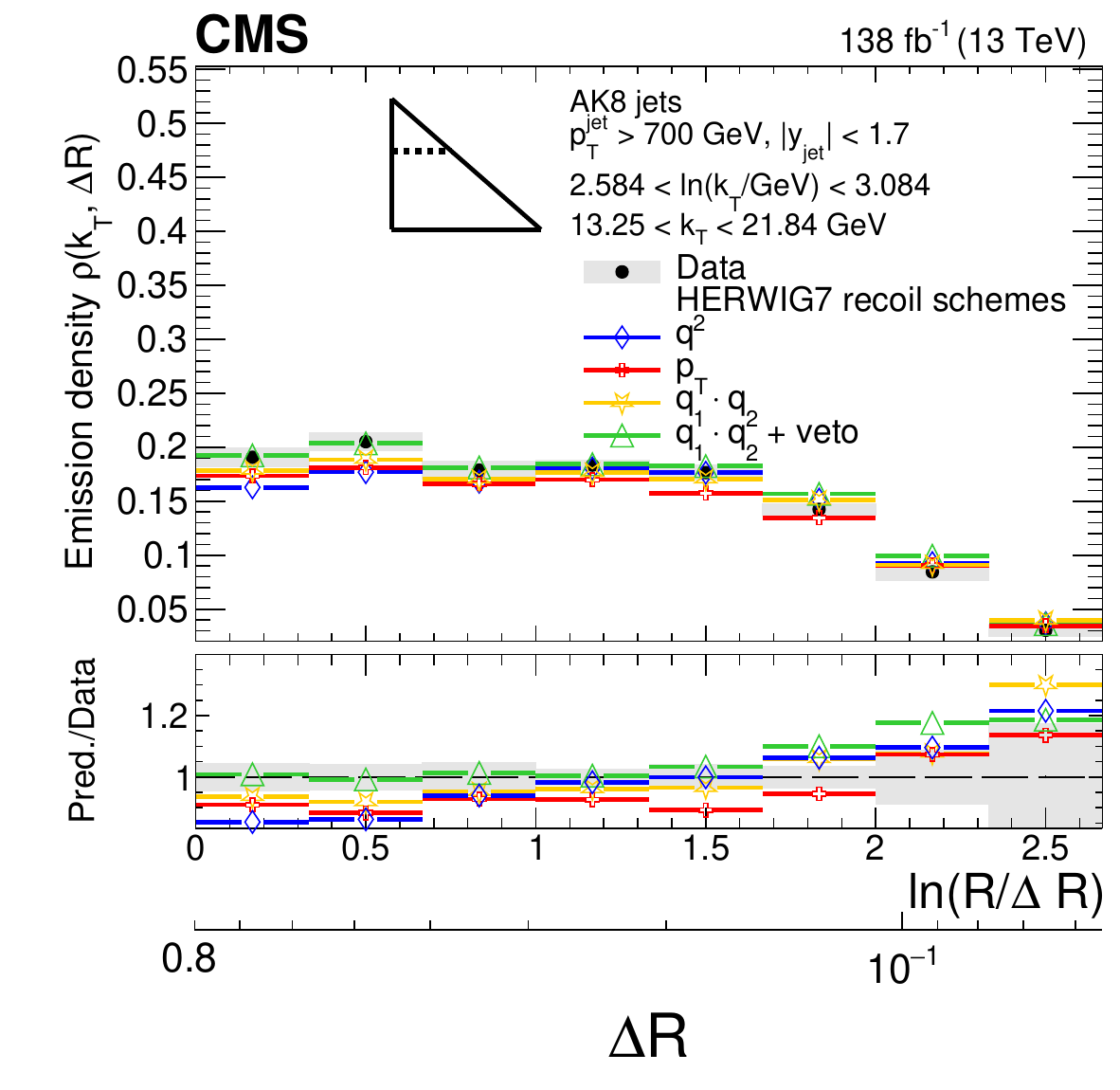}
  \caption{Four different slices of the of the charged-particle primary LJP density of AK8 jets compared with predictions based on different choices of the recoil scheme of the angular-ordered shower of \HERWIG{7}. Each recoil scheme achieves a different degree of logarithmic accuracy, up to NLL for certain observables, as described in the text. The band represents the total experimental uncertainty. The upper two plots correspond to vertical slices of the LJP for fixed \lnDeltaR (large angles on upper-left, small angles on upper-right). The lower two plots correspond to two different horizontal slices for fixed \kt interval: the lower-left plot corresponds to low-\kt splittings and spans the full range in \lnDeltaR, whereas the lower-right plot corresponds to high-\kt splittings, which populate mostly wide-angle radiation. Statistical uncertainties in data and MC-simulated events are represented by vertical bars, which are smaller than the markers in most of the bins.}
  \label{fig:LPslicesak8_recoilChoice}
\end{figure}

Figure~\ref{fig:LPslicesak8_recoilChoice} compares the measured LJP density with predictions generated with \HERWIG{7} using different choices of the recoil scheme of its angular-ordered shower. The recoil scheme, related to the momentum recoil redistribution of the parton shower, affects the logarithmic accuracy of the parton shower, and it can reach NLL accuracy for certain classes of global observables~\cite{Bewick:2019rbu}. The $q^2$ scheme has the largest discrepancy with data, particularly in the high \kt tails. The \pt scheme has a better agreement with the data than the predictions based on the $q^2$ scheme. This was noted in Ref.~\cite{Bewick:2019rbu} for other jet-based observables at the LHC. The $q_1\cdot q_2$ scheme has a better description of the data than the $q^2$ or \pt schemes. A better description of the data in the bulk of the primary LJP is achieved with the $q_1 \cdot q_2+\text{veto}$ scheme, primarily in the perturbative region $\kt > 5 \GeV$ and at large \DeltaR. The $q^2$ scheme predictions are in better agreement with the data in the nonperturbative region $\kt  \approx 1 \GeV$, but the predicted density of emissions is smaller than the data for hard, wide-angle emissions. Similar trends are observed in measurements of hadronic event shape variables in $\EE$ collisions at $\sqrt{s} = 91.2\GeV$ at LEP, where the $q_1\cdot q_2+\text{veto}$ scheme provides a better global description of the data~\cite{Bewick:2019rbu, Bellm:2019zci}. The LJP density can be used as a complementary handle at the LHC to benchmark the choice of the recoil scheme in angular-ordered parton showers in a region where both quark and gluon fragmentation play an important role.

\begin{figure}
  \centering
  \includegraphics[width=0.49\textwidth]{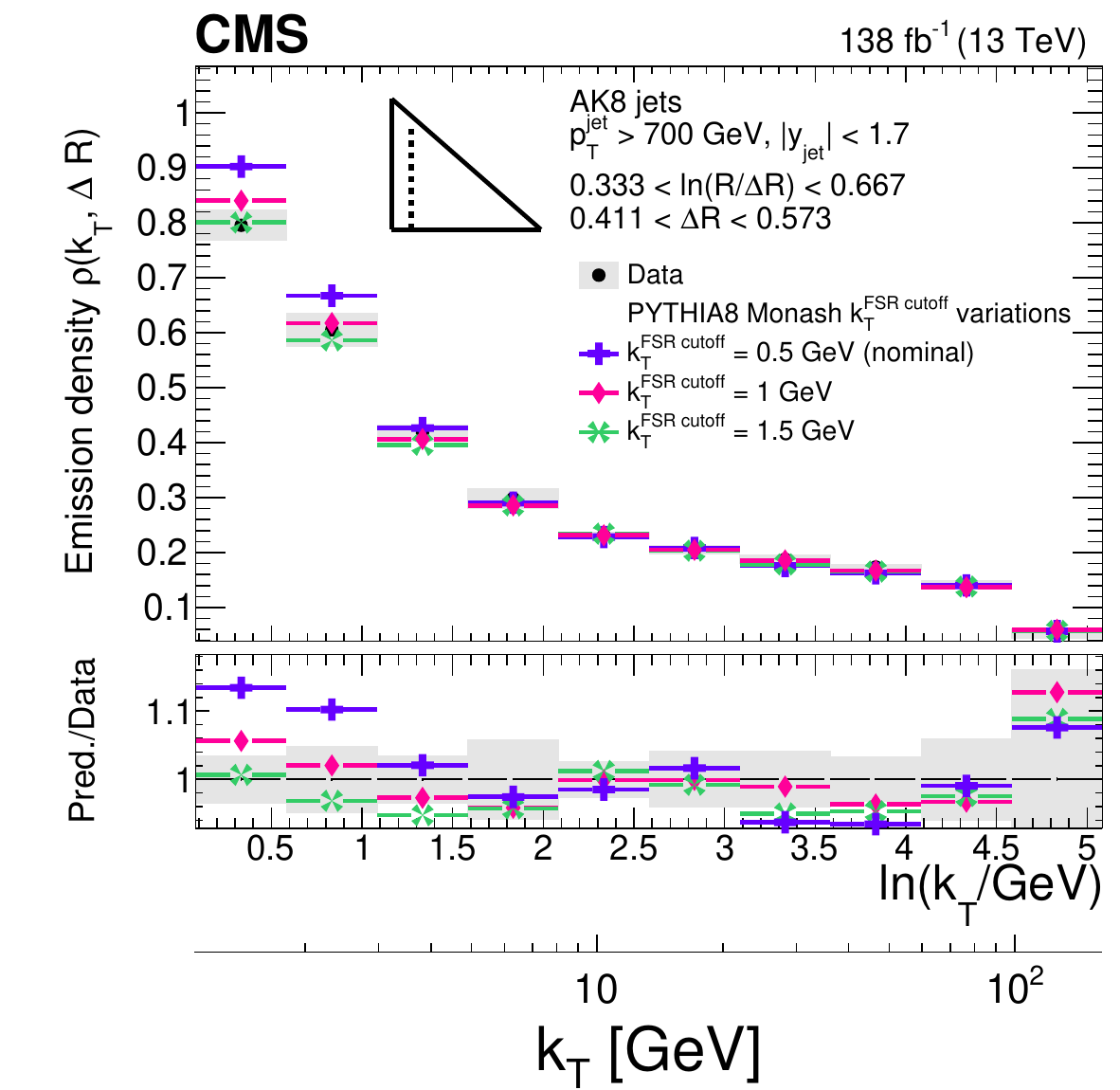}
  \includegraphics[width=0.49\textwidth]{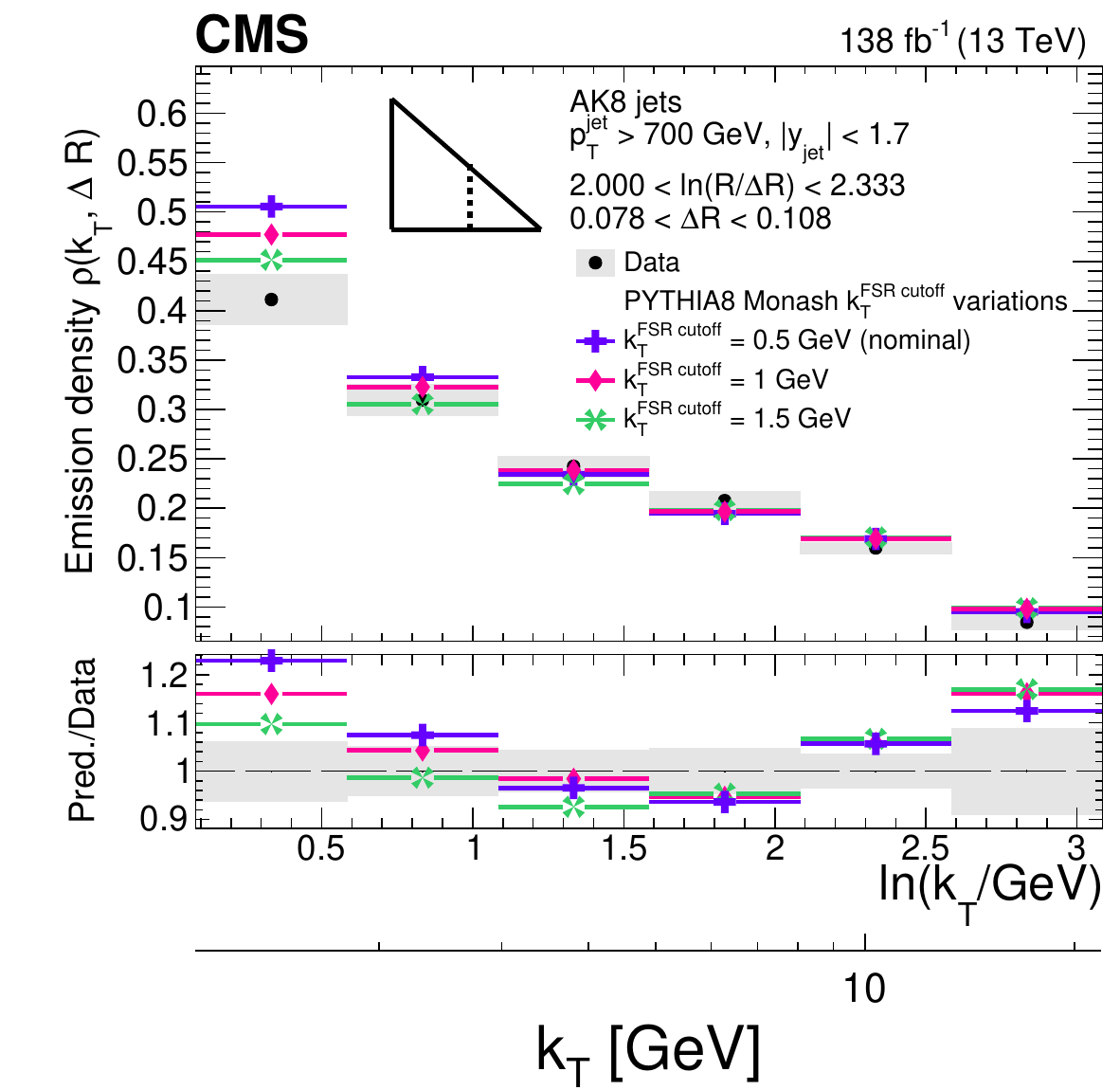}
  \includegraphics[width=0.49\textwidth]{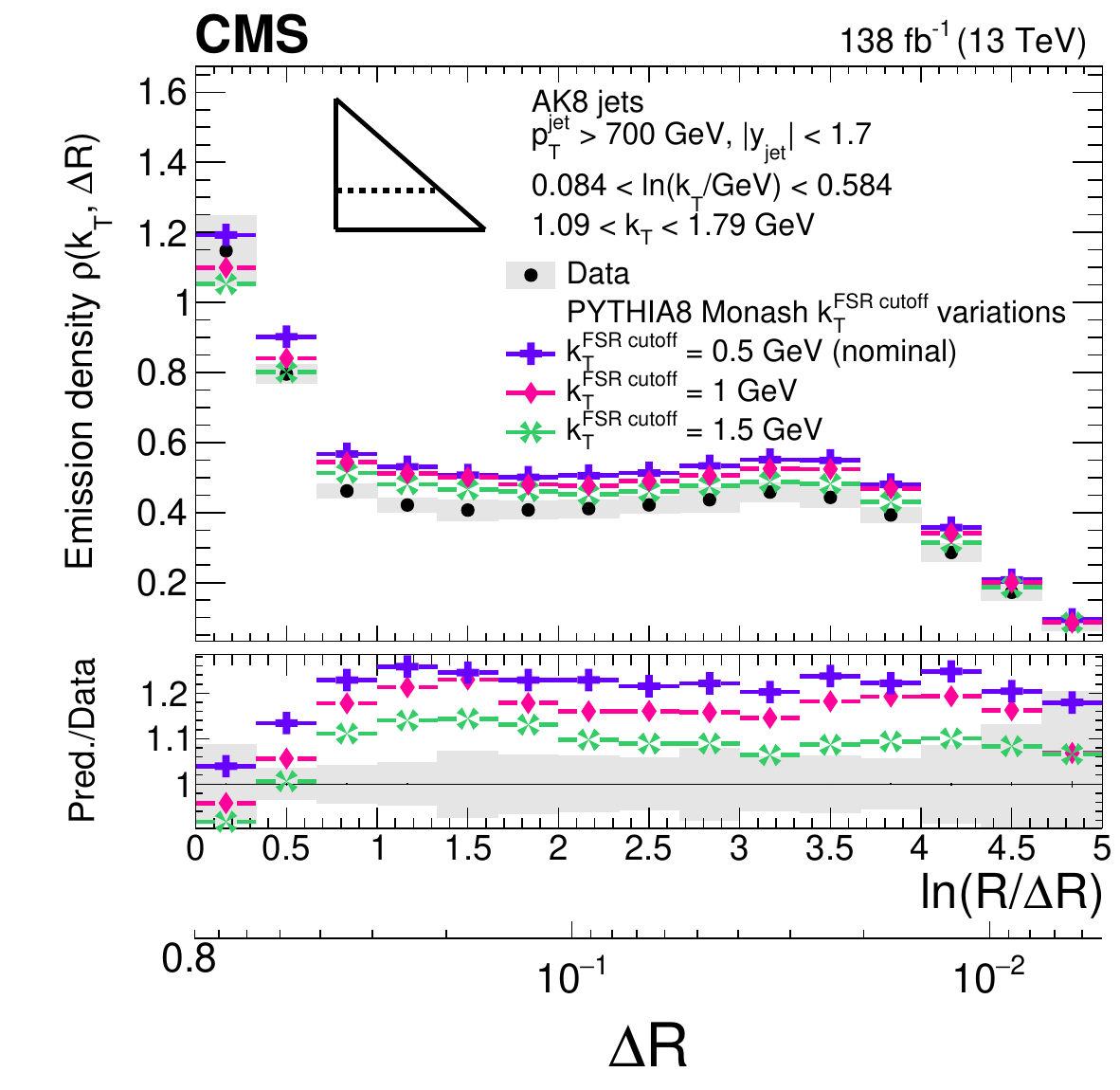}
  \includegraphics[width=0.49\textwidth]{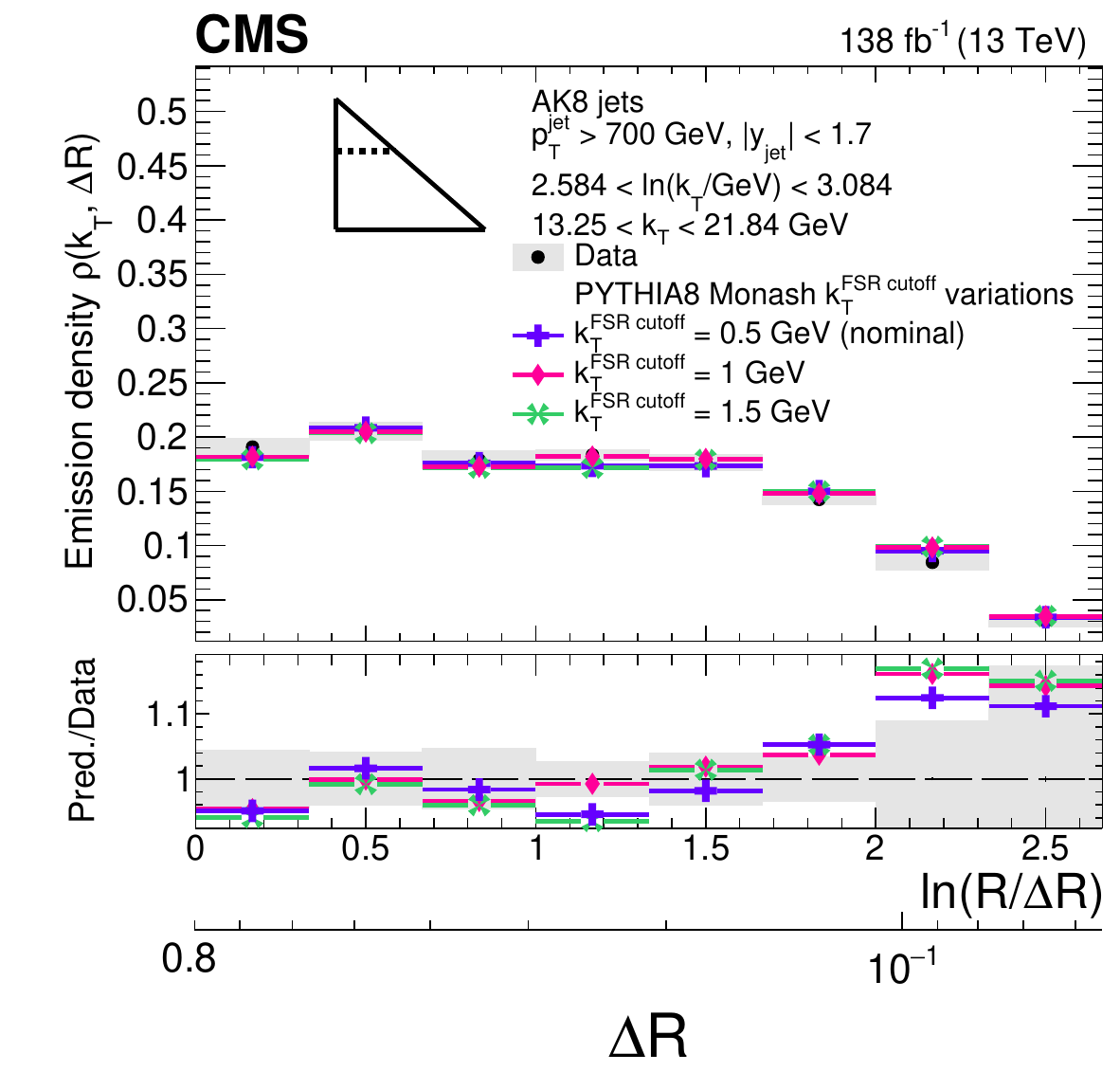}
  \caption{Four different slices of the charged-particle primary LJP density of AK8 jets compared with predictions based on different values of the transverse momentum cutoff used for FSR ($\ktFSRcutoff$) in \PYTHIA{8} with the Monash tune. The larger $\ktFSRcutoff$ value yields a better agreement with the data at low \kt. The lower two plots correspond to two different horizontal slices for fixed \kt interval: the lower-left plot corresponds to low-\kt splittings and spans the full range in \lnDeltaR, whereas the lower-right plot corresponds to high-\kt splittings, which populate mostly wide-angle radiation. Statistical uncertainties in data and MC-simulated events are represented by vertical bars, which are smaller than the markers in most of the bins.}
  \label{fig:LPslicesak8_pTcutoff}
\end{figure}

Another difference between the \HERWIG{7}, \SHERPA{2}, and \PYTHIA{8} generators is the \kt cutoff value used to terminate the FSR evolution and initiate hadronization. For \PYTHIA{8} with the tunes used in this paper, the \kt cutoff for FSR is at 0.5\GeV. For \HERWIG{7} and \SHERPA{2}, the cutoff is at 1\GeV. A perturbative cascade evolving to very low momentum followed by hadronization effects might lead to a larger amount of emissions in the LJP at low \kt. Figure~\ref{fig:LPslicesak8_pTcutoff} shows a variation of \PYTHIA{8} Monash predictions with two different values of the FSR \kt cutoff, $\ktFSRcutoff = 1.0$ and 1.5\GeV; the latter yields a better agreement with the data. These variations do not have a strong effect for emissions with $\kt \gtrsim 5\GeV$, further confirming the ability of the LJP to factorize effects. This indicates that the effect of the $\ktFSRcutoff$ can be effectively decoupled in the primary LJP for event generator tuning.

\begin{figure}
  \centering
  \includegraphics[width=0.95\textwidth]{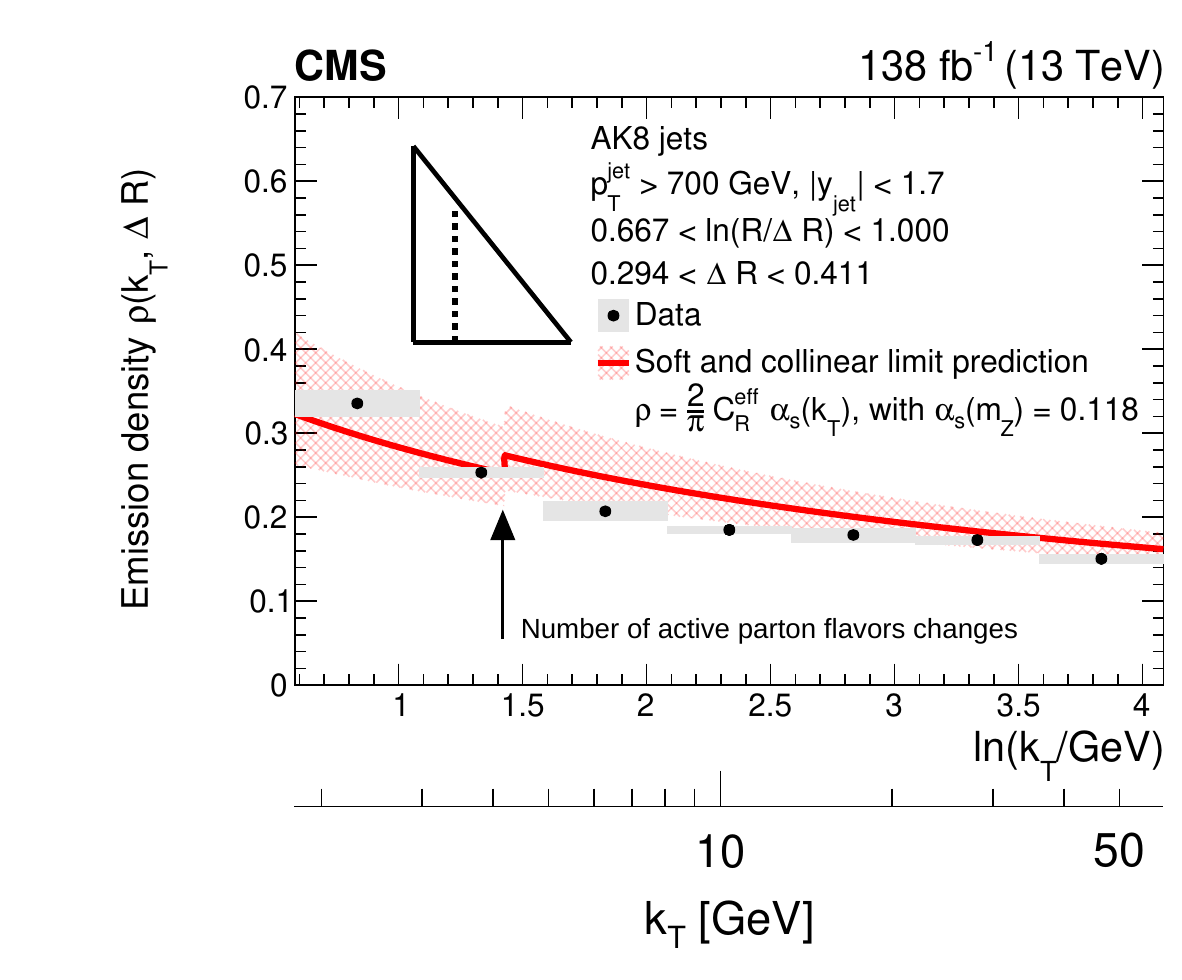}
  \caption{Measured LJP distribution for AK8 jets, compared with the leading-order perturbative-QCD asymptotic prediction in the soft and collinear limit. The grey boxes represent the total experimental uncertainty from the measured data. For the prediction, an effective color factor of $\CReff~=~0.59 ~\CF+0.41~\CA \approx 2$ is assumed, as described in the text. The strong coupling \alpS evolves with \kt using the one-loop $\beta$ function with $\alpS (m_\PZ) = 0.118$. The theoretical uncertainty band is calculated with variations of the renormalization scale up and down by factors of 2. The discontinuity is due to the change of the number of active flavors when \kt reaches the mass of the bottom quark, which is assumed to be 4.2\GeV.}
  \label{fig:runningCoupling}
\end{figure}

It is instructive to analyze the LJP density in terms of the simplest analytical prediction in perturbation theory to illustrate explicitly the effect of the running of \alpS in the substructure of the jet. In Fig.~\ref{fig:runningCoupling}, we display a slice of the LJP density for AK8 jets and compare it to the soft and collinear limit prediction using the one-loop $\beta$ function for the running of \alpS. A slice for slightly more collinear emissions, $0.294 < \DeltaR < 0.411$, is chosen, since the region of validity of the approximation of the LO pQCD prediction is for the collinear limit. The quark jet (0.59) and gluon jet (0.41) fractions from \PYTHIA{8} CP5 simulated events are determined using the jet flavor definition from Ref.~\cite{CMS-PAS-JME-16-003}. Based on this, an effective color factor of $\CReff~=~0.59~\CF+0.41 ~\CA \approx 2$ is assumed. This asymptotic expression does not take into account parton flavor changes in the jet clustering history, \eg gluon to quark-antiquark splitting. We assume a value of $\alpS(m_{\PZ}) = 0.118$, close to the world-average value \cite{PDG2022}. The soft and collinear limit prediction with these basic assumptions qualitatively describes the shape and normalization of the unfolded distribution in the collinear region, consistent with the expectation that the dominant mechanism responsible for the rise of the LJP density at low \kt is due to the running of \alpS in the jet shower, with \kt as the characteristic energy scale used in the evolution of \alpS.

\begin{figure}
  \centering
  \includegraphics[width=0.49\textwidth]{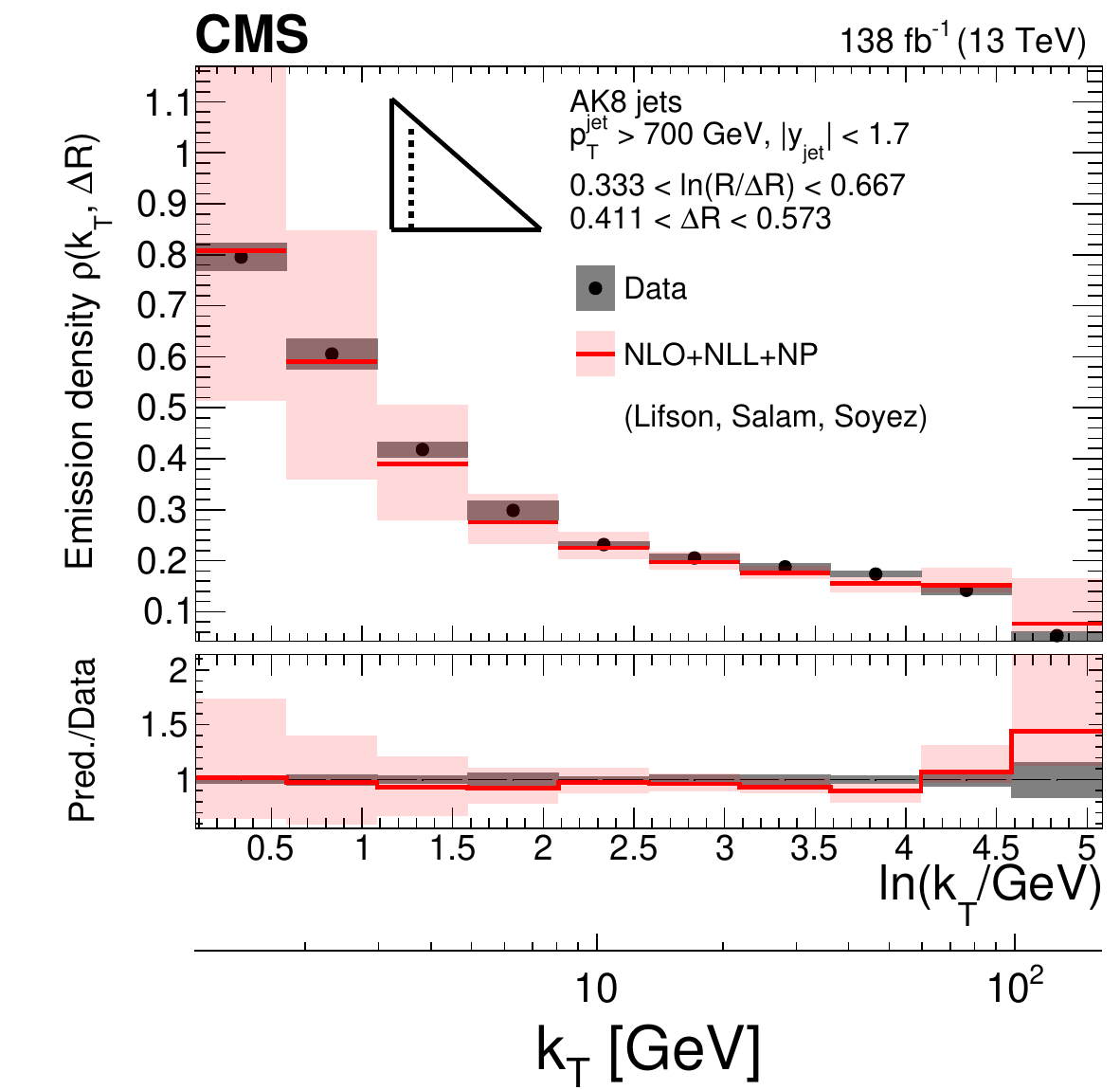}
  \includegraphics[width=0.49\textwidth]{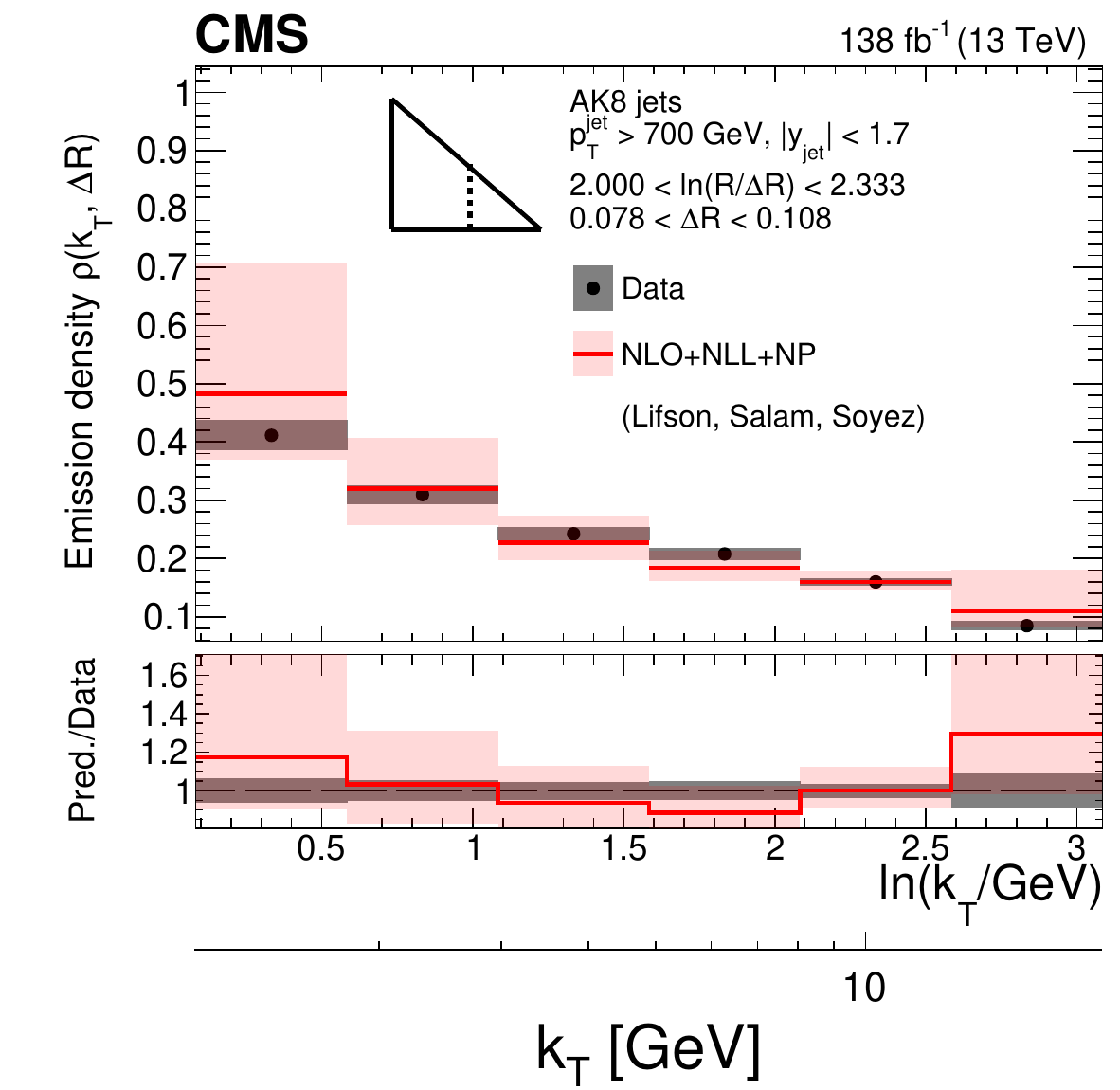}
  \includegraphics[width=0.49\textwidth]{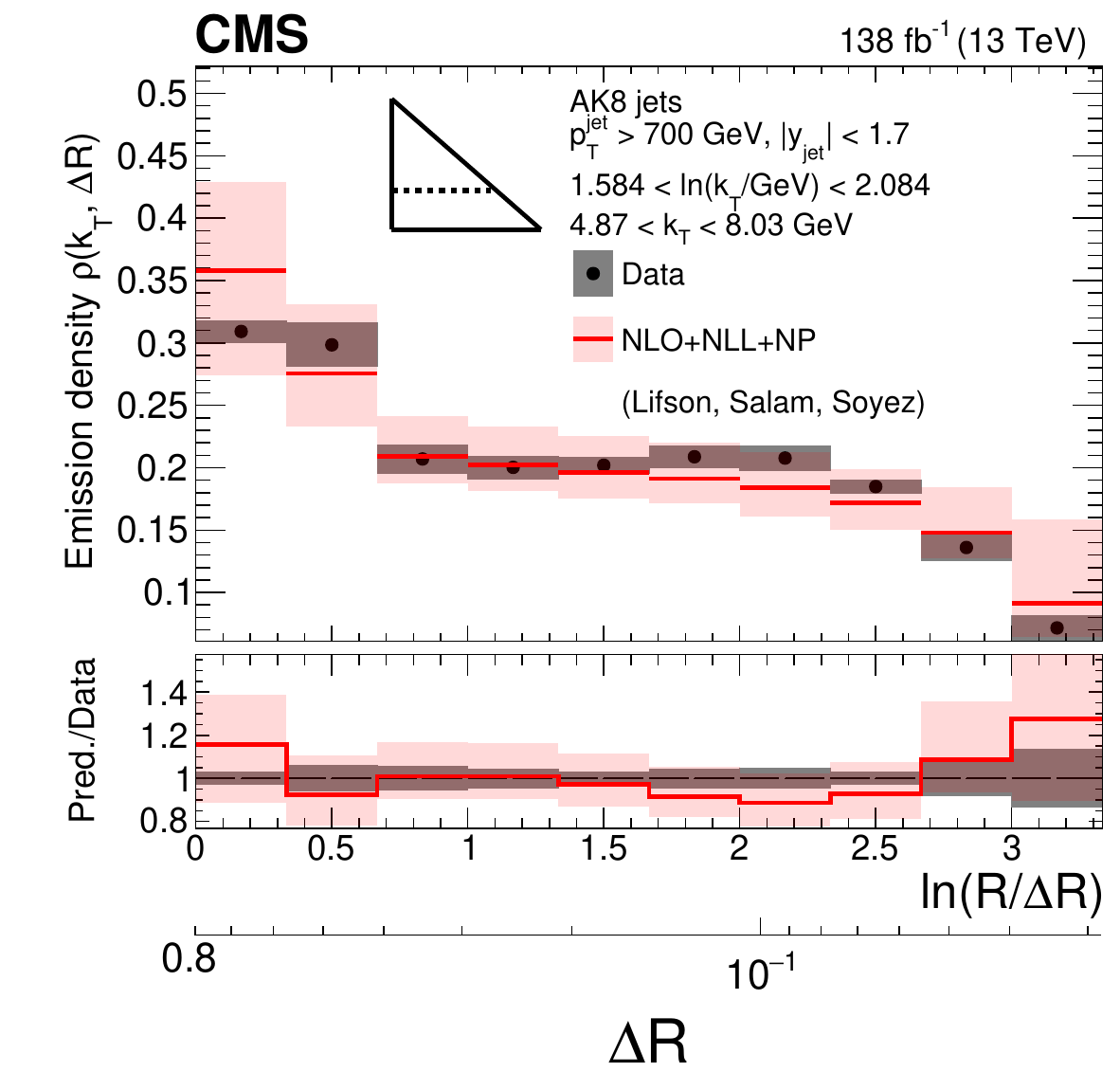}
  \includegraphics[width=0.49\textwidth]{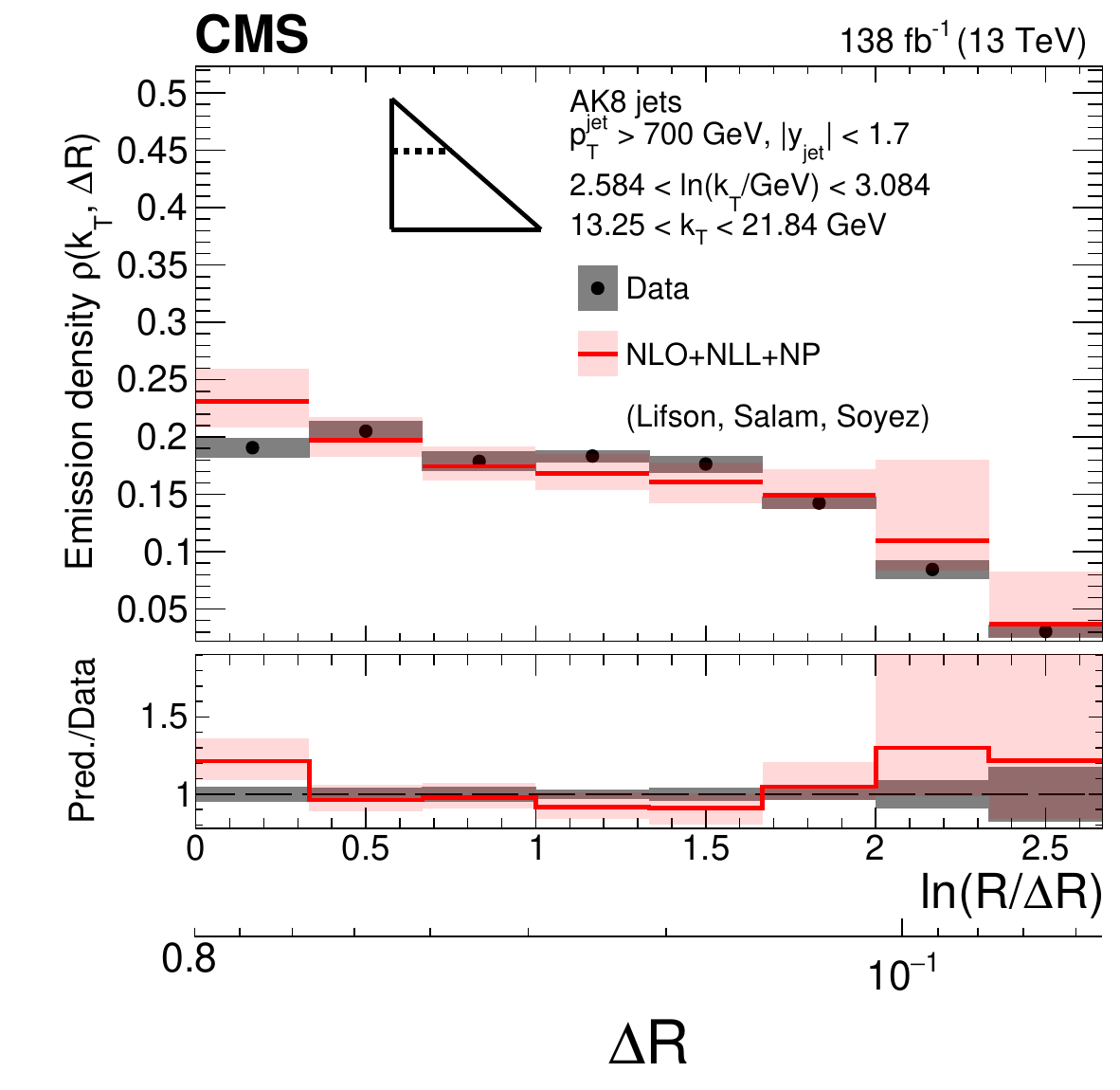}
  \caption{Four different slices of the primary LJP density of AK8 jets compared with perturbation theory calculations by A. Lifson, G. P. Salam, G. Soyez~\cite{Lifson:2020gua}. The calculations include all-orders resummation at next-to-leading logarithmic (NLL) accuracy matched to a next-to-leading order (NLO) fixed-order calculation, and supplemented with nonperturbative (NP) corrections, as described in the text. The band around the theory prediction represents the uncertainty from variations of the renormalization scale uncertainty in the perturbative calculation as well as uncertainties in the NP corrections. The gray band represents the total experimental uncertainty. The upper two plots correspond to vertical slices of the LJP for fixed \lnDeltaR (large angles on upper-left, small angles on upper-right). The lower two plots correspond to two different horizontal slices for fixed \kt interval: the lower-left plot corresponds to low-\kt splittings and spans the full range in \lnDeltaR, whereas the lower-right plot corresponds to high-\kt splittings, which populate mostly wide-angle radiation.}
  \label{fig:Gregory}
\end{figure}

Although the asymptotic formula captures the broad features of the LJP density in terms of the running of \alpS for collinear emissions, higher-order corrections are necessary to better describe the radiation pattern of the jet. Thus, we also present a comparison of the unfolded distributions with the theoretical calculations of the primary LJP density by A. Lifson, G. P. Salam, and G. Soyez using the setup described in Ref.~\cite{Lifson:2020gua}. The calculations are adapted to match the particle-level definition of the jets and their substructure presented in this analysis. The calculations include several pieces associated with the resummation of single-logarithmically enhanced terms in the perturbative expansion in \alpS. The largest contribution comes from the resummation of logarithmic terms from the running of \alpS, which is accounted for using a two-loop $\beta$ function.  A value of $\alpS(m_\PZ) = 0.118$ is used for the pQCD pieces of the calculation. A resummation at NLL accuracy includes contributions that can change the \pt and the flavor of the leading parton. The calculation also includes a resummation of nonglobal logarithms associated to multiple soft-gluon emissions and a resummation of boundary logarithms to account for the fact that the jet is initially clustered with the anti-\kt algorithm and reclustered with the CA algorithm. These pieces modify the LJP density mostly at large \DeltaR. The all-orders resummation is matched to a fixed-order NLO pQCD calculation for jet production. The parton-level calculations are corrected to the hadron level using bin-by-bin corrections derived from MC simulated events. Since the substructure is calculated using the charged particles in the jet, the \kt is scaled by a factor of 0.62 to account for the average charged fraction in a jet. The nonperturbative corrections affect mostly the $\kt < 5\GeV$ region. The resulting theoretical uncertainties include uncertainties from the perturbative calculation, derived from scale variations, and uncertainties in the nonperturbative corrections~\cite{Lifson:2020gua}. Four different slices of the primary LJP density of AK8 jets are shown in Fig.~\ref{fig:Gregory}. For the vertical slices in \DeltaR, one can distinguish the effects from the resummation in \kt. For the comparison with the analytical predictions, we use a slightly larger \kt window for the lower left panel (horizontal \kt slice of $4.87 < \kt <  8.03\GeV$) than what is used in previous plots in this paper. This is a regime of low \kt values where the pQCD calculation is still reliable and hadronization corrections are not as large, and where the phenomenological effect of the resummation of nonglobal logarithms can be observed, which is an increase in the density of emissions at large \DeltaR. The predictions and the unfolded data are consistent with each other within the theoretical and experimental uncertainties, meaning that a first-principles understanding of the radiation pattern of the jet in terms of the primary LJP is achieved.

\clearpage

\section{Summary}\label{sec:summary}

We have presented a measurement of the primary Lund jet plane (LJP) density in inclusive jet production in proton-proton collisions at $\sqrt{s} = 13\TeV$ using data, corresponding to an integrated luminosity of 138\fbinv, collected in Run 2 (2016--2018) with the CMS experiment. The LJP is a two-dimensional representation of the phase space of emissions inside a jet constructed using iterative Cambridge--Aachen declustering. The logarithm of the relative transverse momentum \kt of the emission and the logarithm of the opening angle of the branching \DeltaR are used for the vertical and horizontal axes of the LJP. We analyzed the substructure of jets initially clustered with the anti-\kt algorithm with transverse momentum $\pt > 700\GeV$ and rapidity $\abs{y}<1.7$ clustered with distance parameters $R = 0.4$ or 0.8. The smaller $R = 0.4$ is the standard $R$ for Run 2 analyses. The larger $R = 0.8$, used for the first time in a measurement of the primary LJP density, enables the exploration of a broader kinematical region of the LJP that is inaccessible with the $R =0.4$ parameter value, particularly for wide-angle, hard radiation. Clustering effects associated with the initial anti-\kt clustering have a less strong effect in the collinear region with the larger $R$ value; hence the angular region where the emission density plateaus is wider for $R = 0.8$ jets. The corrected distributions have an experimental uncertainty in a range of 2--7\% in the region away from the kinematical LJP edge and about 15--25\% close to the LJP edge.

We compared the corrected primary LJP density with various particle-level predictions from Monte Carlo (MC) simulated events. The predictions use different implementations of parton showers as well as different models for the underlying-event (UE) activity, beam-beam remnants, hadronization, and color reconnection effects. The aforementioned mechanisms can be effectively factorized in the primary LJP density, which allows for strong constraints in terms of the substructure of jets. At leading-logarithmic (LL) accuracy, the primary LJP density is proportional to the strong coupling $\alpS(\kt)$, so it can be used to tune the value of \alpS evaluated at the \PZ boson mass used for final-state radiation (FSR) in MC event generators, \alpSFSR.

Predictions generated with the CP5 tune of the \PYTHIA{8} generator underestimate the measured density of emissions in the perturbative region ($\kt > 5\GeV$) by about 15\% because of the small value of \alpSFSR used for this tune. Other \PYTHIA{8} tunes or parton shower options tested in the measurement are in better agreement with the data.

The predictions generated with the angular-ordered shower of the \HERWIG{7.2.0} generator are in better agreement with the data than those generated with its alternative dipole shower. The data were also compared with different recoil schemes of the angular-ordered shower of \HERWIG{7}, which allow the parton shower to reach up to next-to-LL (NLL) accuracy for certain global observables. The \HERWIG{7} predictions with the dot-product preserving recoil scheme, together with a veto on high-virtuality partons, are in better agreement with the data in the bulk of the primary LJP in the perturbative region $\kt > 5 \GeV$. The predictions based on the virtuality-preserving scheme describe better the nonperturbative region $\kt \approx 1 \GeV$ than the dot-product preserving scheme, but these do not describe the perturbative region well.

The low-\kt region is dominated by hadronization effects in a wide range of \DeltaR values, with additional contributions from the UE at large $\DeltaR \approx R$. The predictions based on cluster fragmentation models, such as those generated with \HERWIG{7} or \SHERPA{2} generators, are in better agreement with the data at low \kt for a wide range of \DeltaR values than those of \PYTHIA{8}. The \PYTHIA{8} predictions, where hadronization is described with the Lund string fragmentation model, overestimate the LJP density by about 15--20\% for \kt at the \GeV scale for a wide range of \DeltaR values. One possibility to improve the description of the low \kt region is to include the FSR cutoff \kt as a free parameter in future event generator tuning; a larger FSR cutoff \kt value decreases the density of emissions at low \kt in the LJP without affecting the high-\kt region that is dominated by the parton shower.

Finally, the data are also compared with a perturbative QCD calculation with a resummation at NLL accuracy, which is matched to a fixed-order next-to-leading order calculation~\cite{Lifson:2020gua}. To compare with the measured LJP at hadron level, nonperturbative corrections are supplemented to the calculation. The predictions are in agreement with the data within the theoretical and experimental uncertainties. For collinear emissions, the data can be qualitatively described with the running of \alpS with \kt.

These measurements highlight the different aspects of the physics modeling of event generators that should be improved, ranging from the modeling of hadronization and up to the logarithmic accuracy of parton showering algorithms.

\begin{acknowledgments}
We thank Andrew Lifson, Gavin P. Salam, and Gregory Soyez for their theoretical calculations of the primary Lund jet plane density adapted with the selection requirements used in this paper.

We congratulate our colleagues in the CERN accelerator departments for the excellent performance of the LHC and thank the technical and administrative staffs at CERN and at other CMS institutes for their contributions to the success of the CMS effort. In addition, we gratefully acknowledge the computing centers and personnel of the Worldwide LHC Computing Grid and other centers for delivering so effectively the computing infrastructure essential to our analyses. Finally, we acknowledge the enduring support for the construction and operation of the LHC, the CMS detector, and the supporting computing infrastructure provided by the following funding agencies: SC (Armenia), BMBWF and FWF (Austria); FNRS and FWO (Belgium); CNPq, CAPES, FAPERJ, FAPERGS, and FAPESP (Brazil); MES and BNSF (Bulgaria); CERN; CAS, MoST, and NSFC (China); MINCIENCIAS (Colombia); MSES and CSF (Croatia); RIF (Cyprus); SENESCYT (Ecuador); MoER, ERC PUT and ERDF (Estonia); Academy of Finland, MEC, and HIP (Finland); CEA and CNRS/IN2P3 (France); SRNSF (Georgia); BMBF, DFG, and HGF (Germany); GSRI (Greece); NKFIH (Hungary); DAE and DST (India); IPM (Iran); SFI (Ireland); INFN (Italy); MSIP and NRF (Republic of Korea); MES (Latvia); LAS (Lithuania); MOE and UM (Malaysia); BUAP, CINVESTAV, CONACYT, LNS, SEP, and UASLP-FAI (Mexico); MOS (Montenegro); MBIE (New Zealand); PAEC (Pakistan); MES and NSC (Poland); FCT (Portugal);  MESTD (Serbia); MCIN/AEI and PCTI (Spain); MOSTR (Sri Lanka); Swiss Funding Agencies (Switzerland); MST (Taipei); MHESI and NSTDA (Thailand); TUBITAK and TENMAK (Turkey); NASU (Ukraine); STFC (United Kingdom); DOE and NSF (USA).

\hyphenation{Rachada-pisek} Individuals have received support from the Marie-Curie program and the European Research Council and Horizon 2020 Grant, contract Nos.\ 675440, 724704, 752730, 758316, 765710, 824093, 101002207, and COST Action CA16108 (European Union); the Leventis Foundation; the Alfred P.\ Sloan Foundation; the Alexander von Humboldt Foundation; the Science Committee, project no. 22rl-037 (Armenia); the Belgian Federal Science Policy Office; the Fonds pour la Formation \`a la Recherche dans l'Industrie et dans l'Agriculture (FRIA-Belgium); the Agentschap voor Innovatie door Wetenschap en Technologie (IWT-Belgium); the F.R.S.-FNRS and FWO (Belgium) under the ``Excellence of Science -- EOS" -- be.h project n.\ 30820817; the Beijing Municipal Science \& Technology Commission, No. Z191100007219010 and Fundamental Research Funds for the Central Universities (China); the Ministry of Education, Youth and Sports (MEYS) of the Czech Republic; the Shota Rustaveli National Science Foundation, grant FR-22-985 (Georgia); the Deutsche Forschungsgemeinschaft (DFG), under Germany's Excellence Strategy -- EXC 2121 ``Quantum Universe" -- 390833306, and under project number 400140256 - GRK2497; the Hellenic Foundation for Research and Innovation (HFRI), Project Number 2288 (Greece); the Hungarian Academy of Sciences, the New National Excellence Program - \'UNKP, the NKFIH research grants K 124845, K 124850, K 128713, K 128786, K 129058, K 131991, K 133046, K 138136, K 143460, K 143477, 2020-2.2.1-ED-2021-00181, and TKP2021-NKTA-64 (Hungary); the Council of Science and Industrial Research, India; ICSC -- National Research Center for High Performance Computing, Big Data and Quantum Computing, funded by the EU NexGeneration program (Italy); the Latvian Council of Science; the Ministry of Education and Science, project no. 2022/WK/14, and the National Science Center, contracts Opus 2021/41/B/ST2/01369 and 2021/43/B/ST2/01552 (Poland); the Funda\c{c}\~ao para a Ci\^encia e a Tecnologia, grant CEECIND/01334/2018 (Portugal); the National Priorities Research Program by Qatar National Research Fund; MCIN/AEI/10.13039/501100011033, ERDF ``a way of making Europe", and the Programa Estatal de Fomento de la Investigaci{\'o}n Cient{\'i}fica y T{\'e}cnica de Excelencia Mar\'{\i}a de Maeztu, grant MDM-2017-0765 and Programa Severo Ochoa del Principado de Asturias (Spain); the Chulalongkorn Academic into Its 2nd Century Project Advancement Project, and the National Science, Research and Innovation Fund via the Program Management Unit for Human Resources \& Institutional Development, Research and Innovation, grant B37G660013 (Thailand); the Kavli Foundation; the Nvidia Corporation; the SuperMicro Corporation; the Welch Foundation, contract C-1845; and the Weston Havens Foundation (USA).

\end{acknowledgments}
	
\bibliography{auto_generated}
\cleardoublepage \appendix\section{The CMS Collaboration \label{app:collab}}\begin{sloppypar}\hyphenpenalty=5000\widowpenalty=500\clubpenalty=5000
\cmsinstitute{Yerevan Physics Institute, Yerevan, Armenia}
{\tolerance=6000
A.~Hayrapetyan, A.~Tumasyan\cmsAuthorMark{1}\cmsorcid{0009-0000-0684-6742}
\par}
\cmsinstitute{Institut f\"{u}r Hochenergiephysik, Vienna, Austria}
{\tolerance=6000
W.~Adam\cmsorcid{0000-0001-9099-4341}, J.W.~Andrejkovic, T.~Bergauer\cmsorcid{0000-0002-5786-0293}, S.~Chatterjee\cmsorcid{0000-0003-2660-0349}, K.~Damanakis\cmsorcid{0000-0001-5389-2872}, M.~Dragicevic\cmsorcid{0000-0003-1967-6783}, A.~Escalante~Del~Valle\cmsorcid{0000-0002-9702-6359}, P.S.~Hussain\cmsorcid{0000-0002-4825-5278}, M.~Jeitler\cmsAuthorMark{2}\cmsorcid{0000-0002-5141-9560}, N.~Krammer\cmsorcid{0000-0002-0548-0985}, D.~Liko\cmsorcid{0000-0002-3380-473X}, I.~Mikulec\cmsorcid{0000-0003-0385-2746}, J.~Schieck\cmsAuthorMark{2}\cmsorcid{0000-0002-1058-8093}, R.~Sch\"{o}fbeck\cmsorcid{0000-0002-2332-8784}, D.~Schwarz\cmsorcid{0000-0002-3821-7331}, M.~Sonawane\cmsorcid{0000-0003-0510-7010}, S.~Templ\cmsorcid{0000-0003-3137-5692}, W.~Waltenberger\cmsorcid{0000-0002-6215-7228}, C.-E.~Wulz\cmsAuthorMark{2}\cmsorcid{0000-0001-9226-5812}
% ...
\par}
\vskip\cmsinstskip
\dag:~Deceased\\
$^{1}$Also at Yerevan State University, Yerevan, Armenia\\
$^{2}$Also at TU Wien, Vienna, Austria\\
% ...
\end{sloppypar}
\end{document}